\documentclass[aps,pra,twocolumn,groupedaddress,longbibliography]{revtex4-1}

\usepackage{graphicx}

\newcommand{\vort}{\mbox{\boldmath$\omega$}} 
\newcommand{\vel}{\mbox{\boldmath$v$}}

\newcommand{\nrm}{\mbox{\boldmath$n$}}

\newcommand{\Avec}{\mbox{\boldmath$A$}}
\newcommand{\Bvec}{\mbox{\boldmath$B$}}
\newcommand{\Evec}{\mbox{\boldmath$E$}}

\newcommand{\cubr}[1]{\left(#1\right)}

\newcommand*\diff{\mathop{}\!\mathrm{d}}

\newcommand{\atube}{\mathcal{A}}
\newcommand{\ctube}{\mathcal{P}}
\newcommand{\core}{\mathcal{P}_c}
\newcommand{\shell}{\mathcal{P}_s}

\usepackage{color}

\begin{document}


\title{Vortex line topology during vortex tube reconnection}


\author{P.~McGavin and D.~I.~Pontin}
\affiliation{Division of Mathematics, University of Dundee, Dundee, DD1 4HN, UK}


\date{\today}

\begin{abstract}
This paper addresses reconnection of vortex tubes, with particular focus on the topology of the vortex lines (field lines of the vorticity). 
This analysis of vortex line topology reveals previously undiscovered features of the reconnection process, such as the generation of many small flux rings, formed when reconnection occurs in multiple locations in the vortex sheet between the tubes. 
Consideration of three-dimensional reconnection principles leads to a robust measurement of the reconnection rate, even once instabilities break the symmetry. It also allows us to identify internal reconnection of vortex lines within the individual vortex tubes. 
Finally, the introduction of a third vortex tube is shown to render the vortex reconnection process fully three-dimensional, leading to a fundamental change in the topological structure of the process. An additional interesting feature is the generation of vorticity null points. 
\end{abstract}


\maketitle

\section{Introduction}
\subsection{Background}

Vortices are observed on a wide range of scales in fluids and in many different domains, from the meteorological (e.g.~hurricanes, tidal currents) to the man-made (e.g.~from wingtips of aircraft, rotors) and even the natural world, where birds utilise them in flight to maintain a `V' formation. 
While vortex tubes tend to have well-defined identities, under certain circumstances their topology has been observed to change, in a process known as {\it vortex reconnection}.
Studies of vortex reconnection took off following the work of \cite{1970AIAAJ...8.2172C}, who discussed the reconnection of wingtip vortices; \cite{1998AnRFM..30..107S} discussed the relevance of these wingtip vortices for air traffic control. 
The study of vortex reconnection is also motivated by the proposed turbulent cascade induced during the process \citep[e.g.][]{1989PhFlA...1..630K}, that has been implicated in jet noise \citep{2011PhFl...23b1701H}.
Recently, experimental advances have allowed the study of vortex reconnection in more complicated three-dimensional (3D) configurations: \cite{2013NatPh...9..253K} showed a method of 3D printing hydrofoils to generate vortices, including knotted vortices whose evolution was viewed with the help of micro-bubbles. 

Motivated by Crow's work many studies of vortex reconnection have studied the interaction of two anti-parallel vortex tubes subject to a sinusoidal perturbation, starting with the work of \cite{1987PhRvL..58.1636P} who focussed on the approach phase of the perturbed vortex tubes. \cite{1989PhFl....1..633M,iutam1989melander} followed the evolution after the onset of reconnection, and first noted the importance of the `bridging' process of reconnected field lines, that eventually chokes off the reconnection {(this bridging having previously been described in the reconnection of a trefoil knot vortex by \cite{kida1987})}. \cite{1995JFM...304...47V} studied the effects of compressibility, noting that it reduces the stretching in the bridges, leading to a reduction in the peak vorticity. \cite{2011PhFl...23b1701H} performed a systematic study varying the Reynolds number ($Re$) and noted amongst other things the onset of a Kelvin-Helmholtz instability that breaks the symmetry of the process. 
\cite{2012PhFl...24g5105V} performed high-$Re$ simulations of reconnecting anti-parallel vortex tubes with and without axial flow. Other studied configurations  include the collision of two vortex rings  \citep{1987PhRvL..58.1632A,1989PhFlA...1..630K} and more recently the self-reconnection of a trefoil knot \citep{2013NatPh...9..253K}. 

The aim of this paper is to provide a new perspective on these elementary vortex reconnections by invoking a body of theory developed over the last 20 years or so to understand reconnection of magnetic fields. We will measure robustly where and how quickly the topology of the vorticity field is changed by the reconnection process, and how this evolution of the topology influences the dynamics. This is compared with the classical analyses of vortex tube reconnection described above. 

Our analysis, which focusses on topological properties of vortex lines during the interaction, leads to the discovery of new features to the reconnection process. These features were not discovered by previous studies which examined the reconnection using only isosurfaces of vorticity -- they become apparent only when vortex line topology is considered. One such feature is internal vortex line reconnection within the tubes and its relation to the helicity, that has recently been highlighted in an experimental study \citep{scheeler2017}.
One of the major recent breakthroughs in magnetic reconnection theory is the realisation that two-dimensional (2D) and three-dimensional (3D) reconnection (that is, reconnection for which the magnetic field is locally 2D or 3D in the immediate vicinity of the reconnection site) are fundamentally different \cite{2003JGRA..108.1285P,birn2007}. As such we first study the interaction of an isolated pair of anti-parallel vortex tubes -- which turn out to reconnect `in a 2D manner' -- before considering the addition of a third vortex tube that renders the reconnection process fully 3D. Prior to describing our results, we introduce in the following section the necessary theoretical background.

\subsection{Theory of reconnection}\label{sec:rectheory}
In an inviscid barotropic fluid, vorticity field lines (hereafter vortex lines) are material lines, meaning that all fluid elements that initially lie on the same vortex line will remain connected by a vortex line at later time, since
\begin{equation}\label{eq:ideal}
\frac{D}{Dt}\left(\frac{\vort}{\rho}\right)=\left(\frac{\vort}{\rho}\cdot\nabla\right)\vel,
\end{equation}
which can be compared to the evolution equation of a material line element \citep{1978magnetic,1994AnRFM..26..169K}.
However, when viscosity is introduced to the system this `frozen in' condition may break down. There exists a direct analogue to the process in high magnetic Reynolds number $R_m$ plasmas. For a perfectly conducting plasma the magnetic field is frozen into the plasma (obeying an equation identical to \ref{eq:ideal}, where $\vort$ is replaced by the magnetic field $\Bvec$), while a large but finite conductivity permits a breakdown of the frozen-in condition. The parallels between vortex reconnection and magnetic reconnection in plasmas are described in detail by \cite{2001LNP...571..373H}. 
The parallels between magnetic and vortex reconnection can be understood by comparing the Navier-Stokes equation for a barotropic fluid in the form
\begin{equation}
-\frac{\partial\vel}{\partial t}-\nabla\cubr{\tilde{p}+\frac{\vel^2}{2}-\frac{4}{3}\nu\nabla\cdot\vel}+\vel\times\vort=\nu\nabla\times\vort, \label{eq:dvdt}
\end{equation}
where $\nabla\tilde{p}=(1/\rho)\nabla p$, with the following plasma equation obtained from Maxwell's equations and Ohm's law:
\begin{equation}
-\frac{\partial \Avec}{\partial t} - \nabla\phi +\vel\times \Bvec = \frac{1}{\mu_0\sigma}\nabla\times \Bvec,
\end{equation}
where $\Bvec=\nabla\times\Avec$, $\phi$ is a gauge potential for the electric field, and $\vel$ is the plasma velocity. The magnetic permeability $\mu_0$ and electrical conductivity $\sigma$ can be combined into the magnetic diffusivity, $\eta=1/(\mu_0\sigma)$.
Comparing these two equations leads to the following associations:
\begin{eqnarray}\label{eq:analog}
&\Avec~\leftrightarrow~\vel\qquad \Bvec~\leftrightarrow~\vort, \nonumber\\
&\phi~\leftrightarrow~\tilde{p}+\frac{\vel^2}{2}-\frac{4}{3}\nu(\nabla\cdot\vel)\qquad \eta=\frac{1}{\mu\sigma}~\leftrightarrow~\nu.
\end{eqnarray} 
Using these parallels, we seek to understand the structure of vortex reconnection. However, a fundamental difference between vortex and magnetic reconnection should be noted: while the vorticity and velocity fields are directly dependent in hydrodynamics, the magnetic field and bulk velocity in a plasma are not \citep[for further discussion see e.g.][]{1993PhFlB...5.2355G}. 

Since different conventions appear is the literature, we begin by defining what we mean by vortex reconnection. As pointed out by \cite{kida1991,takaoka1996} the reconnection of vortex lines is critically different from the `reconnection' of vorticity isosurfaces: it is only the former that is prohibited in an inviscid fluid. 
{In other words, vortex \emph{line} reconnection, which is prohibited in an ideal evolution, is distinct from vortex \emph{tube} reconnection, which may occur under ideal deformations (the vortex tubes being defined by isosurfaces of $|\vort|$).}
Herein \emph{we define vortex reconnection as a change in the topology of the vorticity (vector) field}, {i.e.~vortex line reconnection}.
Note that two vorticity fields are topologically equivalent if and only if one field can be transformed into the other by means of a smooth (continuously differentiable) deformation. Equivalently, some smooth ideal evolution (flow) can transform one field into the other. Such an evolution between topologically equivalent fields preserves all linkages or knottedness of field lines within the volume as well as all connections of vortex lines between co-moving boundary points. With these definitions we follow the \emph{general magnetic reconnection} framework of \cite{1988JGR....93.5547S} in defining reconnection as the breakdown of field line conservation -- i.e. the connection between fluid elements by vorticity field lines -- this being equivalent to a change of the topology of the vorticity field. Note that in order to be defined as reconnection, this change of topology must be due to a local non-ideal evolution (as opposed to e.g.~a global diffusion). The relationship between  topology and reconnection in this sense is discussed in more detail for the magnetic case by \cite{hornig1996}. A number of caveats should be noted. First, this definition is not equivalent to the definition of \cite{melander1994}, who restrict their attention to axisymmetric flows, and end up with a definition of reconnection that is not synonymous with change of vortex line connectivity. Note further that while a strictly isolated non-ideal region is typically realised for magnetic reconnection in high-$R_m$ plasmas,  this may not always be the case for vortex reconnection, for example in vortex rings with swirl as noted by \cite{melander1994}.
Finally,  the general magnetic/vorticity reconnection framework that we adopt relies on laminar fields -- the extension of the notion to turbulent magnetic/vorticity fields was presented by \cite{eyink2015}. 

When categorising reconnection processes, the first distinction that needs to be made is between {\it 2D reconnection} -- in which the vortex lines lie locally in a plane -- and {\it 3D reconnection} where all three components of the vorticity field are non-zero (except perhaps at isolated points). We also consider further below the possibility of vortex line {\it annihilation}, which occurs when vortex lines that are one dimensional are brought together (either at a zero plane of vorticity or at an `O-point') -- this is not true reconnection but involves a loss of vorticity flux \citep{1989JFM...205..263B}.

2D reconnection occurs at X-type null points of the vorticity field.   
This type of reconnection occurs in many simulations of vortex tube reconnection, either due to the initial anti-parallel geometry of the tubes, or because the dynamics of the tubes as they approach conspire to bring them together locally anti-parallel {\citep[e.g.][]{kida1988,1992PhFlA...4..581B}}. Sample vortex lines showing the X-point structure are shown in Figure \ref{fig:xptandloop} from our simulations of reconnecting anti-parallel vortex tubes (see below).
Field lines break at the X-point, and connect with partners in the opposite tube, leading to the formation of two differently connected vortex tubes. This cut-and-connect of individual pairs of field lines in 2D reconnection is a manifestation of the fact that the field lines can be considered as being frozen into a virtual flow that is singular at the X-point where the cut and connect occurs \citep{greene1993,melander1994}.

It can be demonstrated as follows that the rate at which vorticity flux is reconnected can be measured by integrating $(\nabla\times\vort)$ along the extension of this X-type null into three-dimensions, sometimes termed the `X-line'.
Consider the rate of change of vorticity flux through the surface, $S$, whose boundary is the yellow-green loop in Figure \ref{fig:xptandloop}(b):
\begin{eqnarray}
\frac{\partial}{\partial t}\int_S\vort\cdot \nrm\diff S &=&  \int_S \frac{\partial \vort}{\partial t}\cdot \nrm\diff S \nonumber \\
&= & \int_S[\nabla\times(\vel \times \vort) - \nu \nabla\times(\nabla\times\vort)]\cdot \nrm \diff  S, \nonumber\\
&= &  -\nu\oint_{\partial S}(\nabla\times\vort)\cdot\diff {\bf l}
\end{eqnarray}
where the final equality follows by applying Stokes' Theorem and noting that $\vort={\bf 0}$ along the integration path. Thus the rate at which vorticity flux is converted from threads (red vortex lines in Figure \ref{fig:xptandloop}) to bridges (cyan) can be measured by integrating the component of $(\nabla\times\vort)$ along the X-line (yellow) -- since $\vort=\nabla\times\vort={\bf 0}$ on the green portion of the loop. 

\begin{figure}
(a)\includegraphics[width=0.46\textwidth]{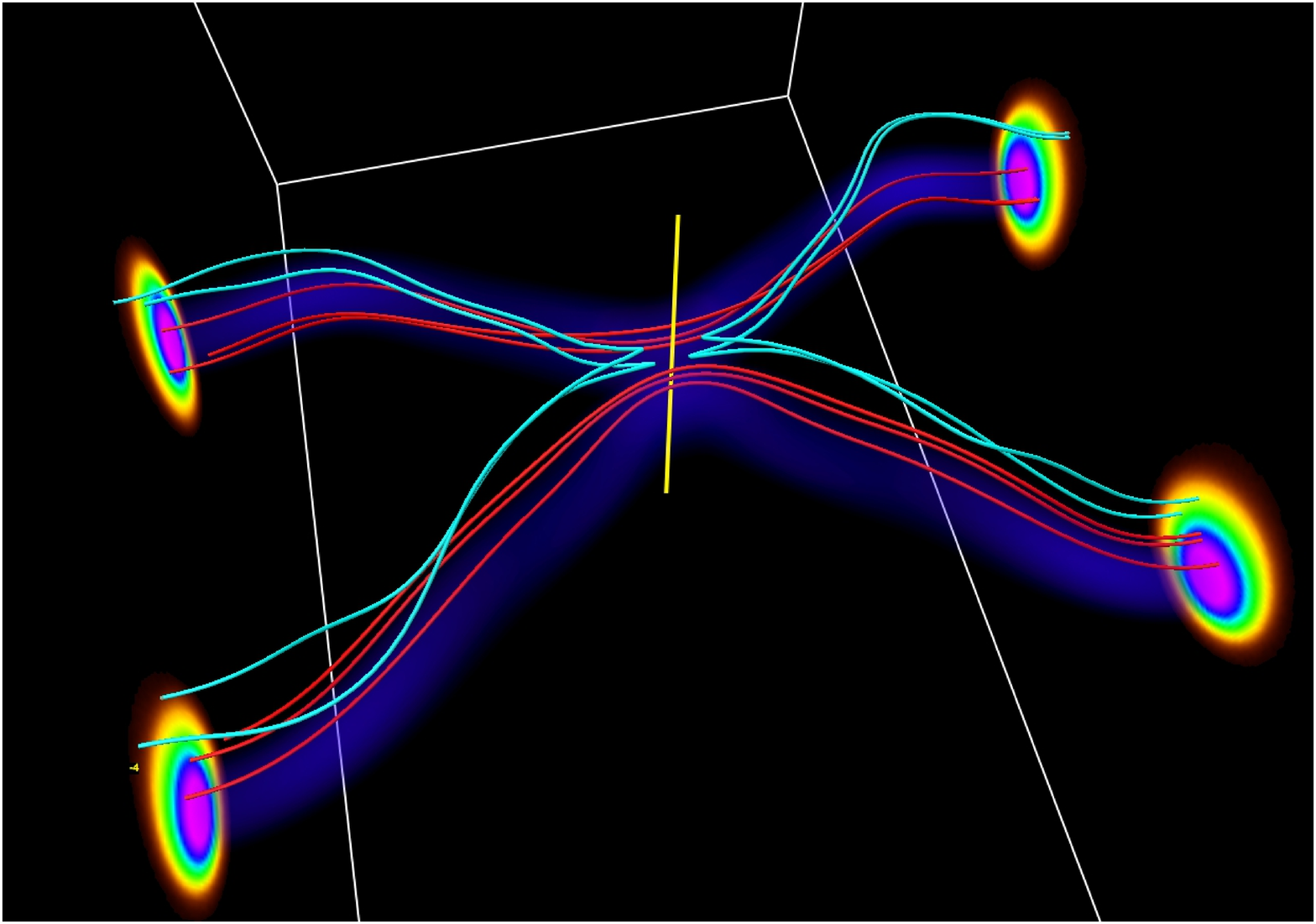}
(b)\includegraphics[width=0.46\textwidth]{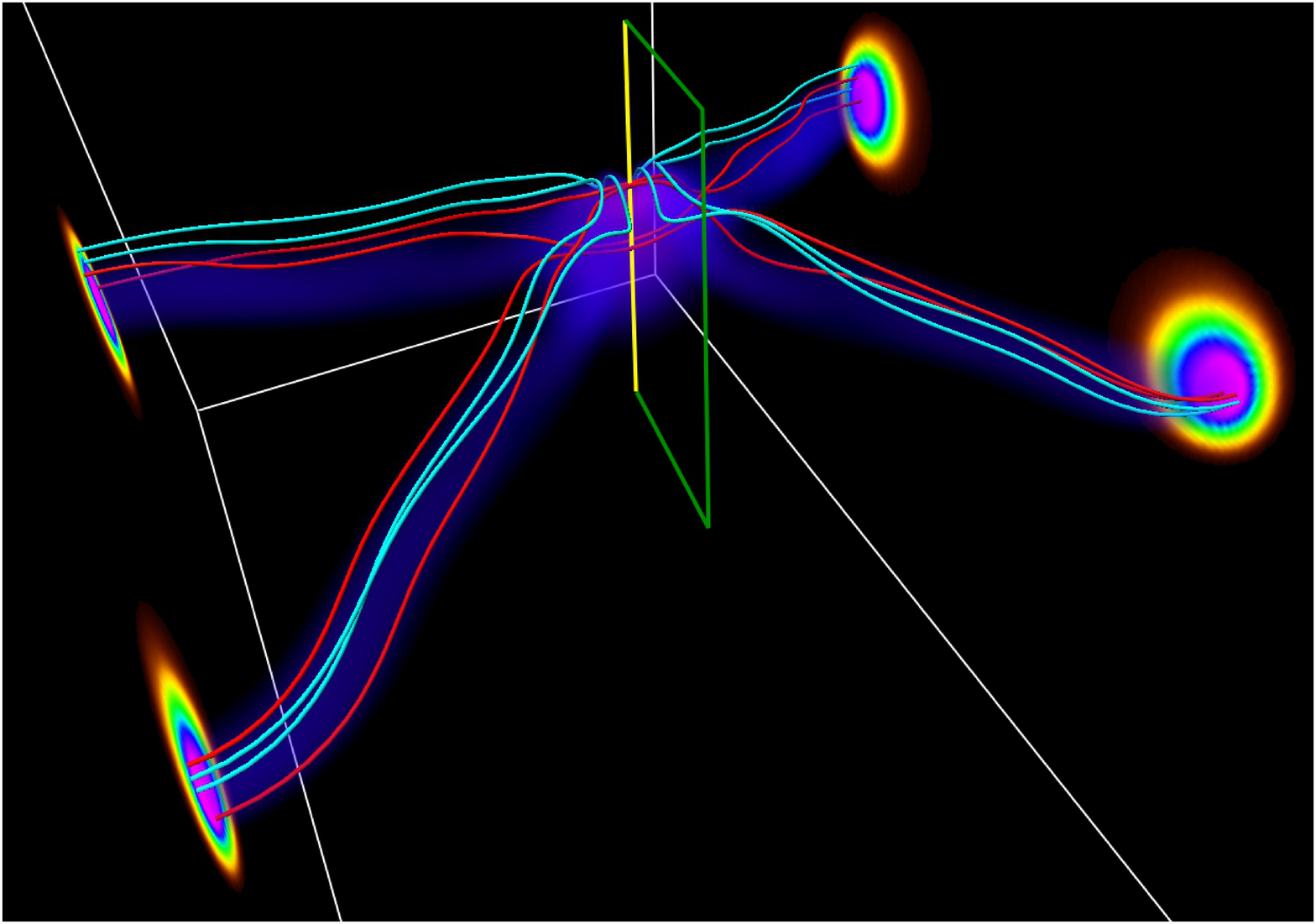}
\caption{Selected vortex lines during the interaction of anti-parallel vortex tubes (red, threads; cyan, reconnected bridges). Shading on the end planes and in the volume shows $|\vort|$. Taken from the simulation described in Section \ref{sec:apsetup} at (a) $t=30$, (b) $t=45$.}
\label{fig:xptandloop}
\end{figure}

\begin{figure}
\begin{center}
(a)\includegraphics[width=0.45\textwidth]{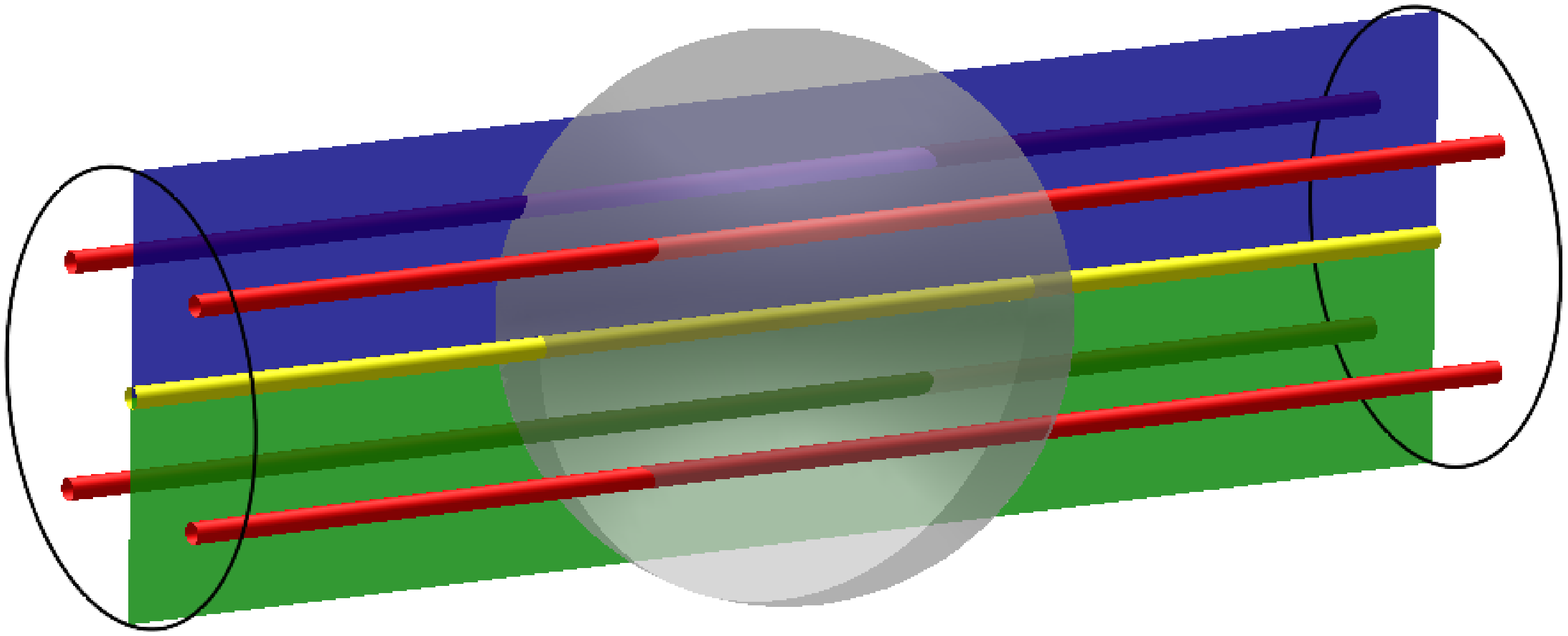}
(b)\includegraphics[width=0.45\textwidth]{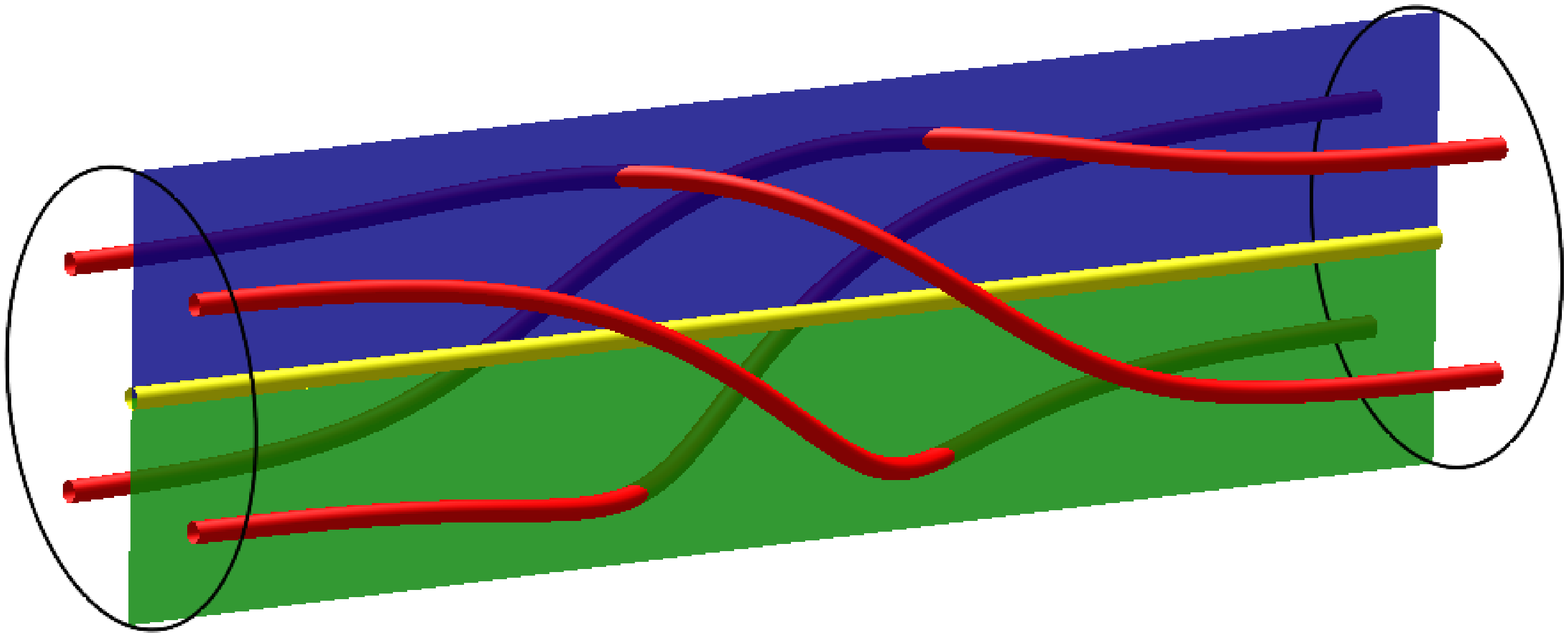}\\
\caption{Vorticity fieldlines (a) before and (b) after 3-D reconnection. The shaded sphere in (a) represents the {\it non-ideal} region, within which $(\nabla\times\vort)\cdot\vort\neq 0$.}
\label{fig:intrectype3drec}
\end{center}
\end{figure}

Turning now to reconnection in a fully 3D vorticity or magnetic field, Schindler {\it et al.} \cite{1988JGR....93.5547S} demonstrated in their {\it general magnetic reconnection} framework that this process requires the existence of a localised region within which the electric field ${\bf E}$ has a non-zero component parallel to the magnetic field, denoted $E_\|$. The rate of change of connectivity between plasma elements is then measured by finding the supremum of this quantity over all field lines passing through the region in which $\Evec\cdot\Bvec\neq0$:
$
\left(\int \Evec_\parallel dl\right)_{max}. 
$
Noting that ${\bf E}=-\partial \Avec/\partial t-\nabla\phi$ and using the analogies drawn in Equation (\ref{eq:analog}), the 3D vortex reconnection rate is therefore given by
\begin{equation}\label{eqn:vortrecrate}
\left(\nu\int\nabla\times\vort \cdot {\bf dl}\right)_{max},
\end{equation}
the integral being performed with respect to arc length along vortex lines, and the supremum being taken over all vortex lines threading a localised reconnection region, i.e.~a region in which $(\nabla\times\vort)\cdot\vort\neq 0$.
{The detailed theoretical development describing how this integral quantifies the rate of change of flux connectivity, making use of an Euler potential representation, is presented by \cite{1988JGR....93.5547S,1988JGR....93.5559H}.}
3D reconnection in the absence of vorticity nulls is illustrated in Figure~\ref{fig:intrectype3drec}. Consider a single vortex tube, within which exists a localised region of $(\nabla\times\vort)\cdot\vort\neq 0$ (marked as a grey sphere). The existence of this non-ideal region implies a rotational `slipping' of field lines that are integrated from either side of the non-ideal region \citep{1988JGR....93.5559H,hornig2003,2005ApJ...631.1227H}. The reconnection acts to change the flux through the green and blue surfaces in the Figure, despite there being no relative rotation between the two end planes. A key feature is that field line connectivity change no longer occurs at a single point or line, but throughout the region in which $(\nabla\times\vort)\cdot\vort\neq 0$ \citep{2003JGRA..108.1285P}. 
For a detailed review of 2D and 3D magnetic reconnection theory, the reader is referred to \cite{pontin2011b}, and references therein.


\begin{figure}
\begin{center}
(a)\includegraphics[width=0.32\textwidth]{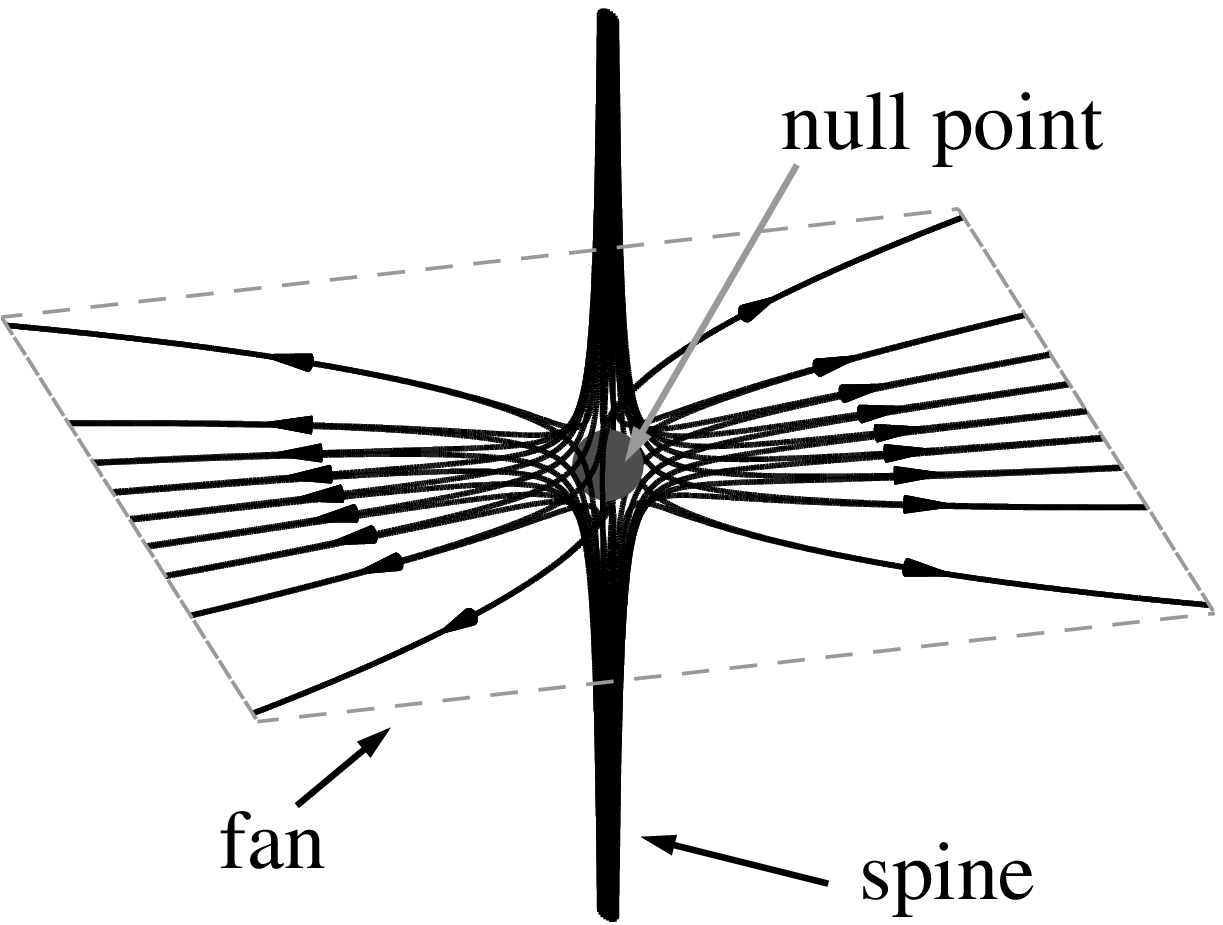}
~~(b)\includegraphics[width=0.22\textwidth]{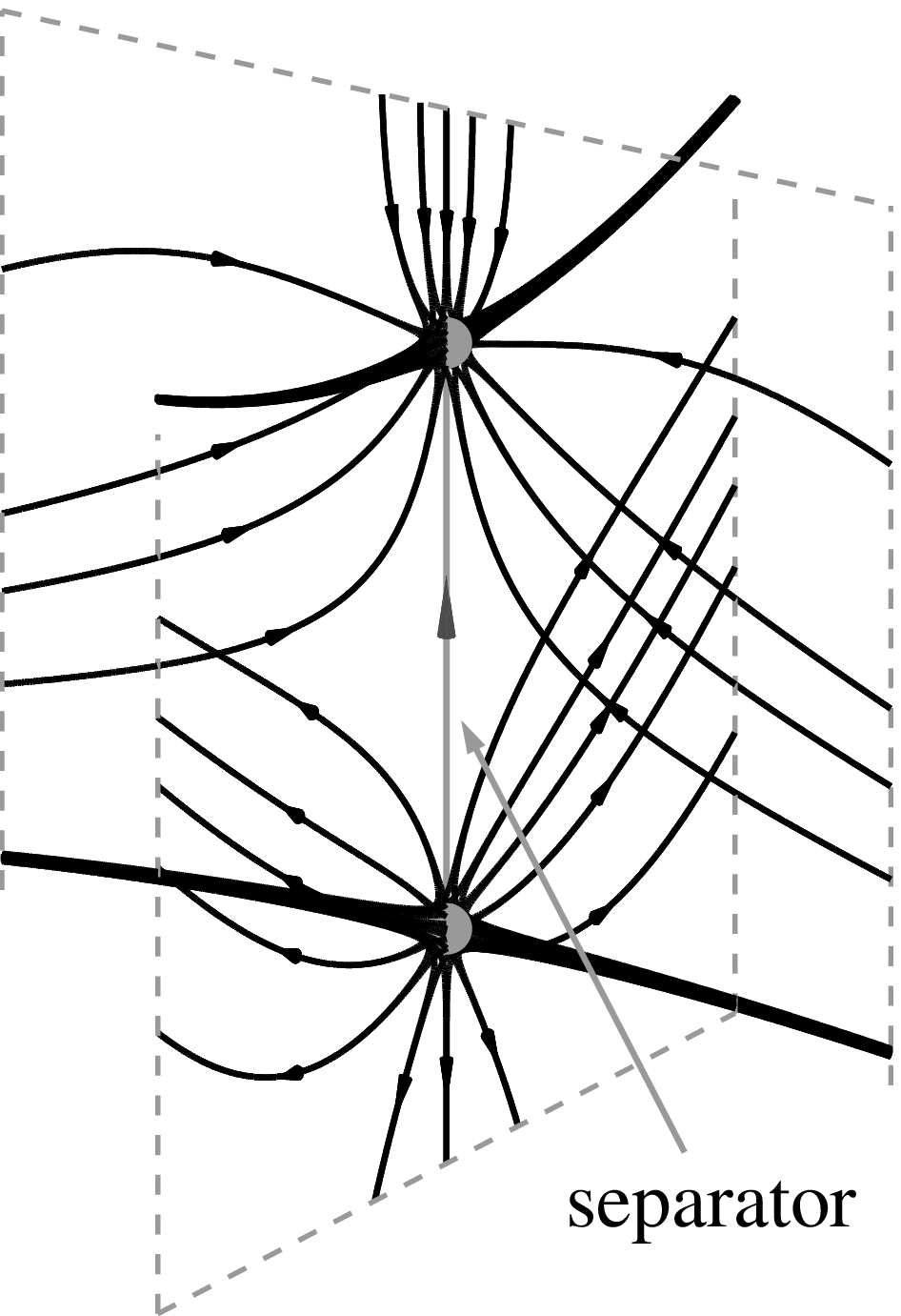}
\caption{Vortex lines in the vicinity of (a) 3D null point, and (b) a separator. After \cite{pontin2011b}.}
\label{fig:intrectypesep}
\end{center}
\end{figure}

To complete our review of the properties of reconnection, we note that one feature of a 3D magnetic field known to be a likely site for reconnection is a 3D magnetic null point. The rate at which flux is reconnected at a 3D null can be evaluated -- by extension of \cite{1988JGR....93.5547S} -- using Equation (\ref{eqn:vortrecrate}), as shown by \cite{pontin2005a}. As we shall see below, in some of our simulations pairs of 3D vorticity null points (isolated points at which $\vort={\bf 0}$) are created. 3D vorticity null points are always created in pairs with opposite topological degree \citep[e.g.][]{1993PhFlB...5.2355G}. The topology of the field lines in their vicinity is described by a `spine' curve and a separatrix (or `fan') surface, as shown in Figure \ref{fig:intrectypesep}(a). The orientations of these structures can be found by calculating the eigenvectors of the Jacobian of the vorticity field at the null point \citep{1993PhFlB...5.2355G,parnell1996}. If the separatrices of two null points intersect one another, the field line determining this intersection is known as a `separator', and such separators are also proposed to be likely sites for magnetic reconnection. The presence of separators in the vorticity field during a vortex reconnection event is noted below.

\section{Simulation setup}\label{sec:apsetup}

The initial configuration that we consider comprises two straight anti-parallel vortex tubes aligned to the $x$-axis, with their axes located at $y=\pm 1$, $z=-9$. 
Specifically, in cylindrical coordinates ($r$,$\phi$,$z'$) centred on the tube axis we take
$
\vel=v_\phi{\bf e}_\phi=\omega_0{\tanh(18r^2)}/{16r}\,{\bf e}_\phi,
$
which leads to a vorticity distribution
\begin{equation}\label{eq:tubevort}
\vort=\omega_{z'}{\bf e}_{z'} = -\frac{\omega_0}{\cosh^2(8r^2)}{\bf e}_{z'}.
\end{equation}
We set $\omega_0=\pm 1$ for the two tubes, such that each vortex tube has a circulation of $\pi/8$.


A perturbation is applied that affects a displacement of the vortex tubes in the positive-$z$ direction, localised around the $x=0$ plane -- see Figure~\ref{fig:lines_and_iso}(a). The vorticity distribution  within  the tubes (\ref{eq:tubevort}) is chosen to ensure that in the initial state the vorticity flux connecting between the tubes is negligible. The curvature  introduced to the vortex tubes induces a velocity that is locally directed along the binormal vector of the vortex lines.
The geometry of the perturbation therefore means that the tubes will rotate and press against each other, eventually forming a vortex sheet. The perturbation is achieved by applying a deformation of the form  $z\to z + \cos^6\!\!\left(\frac{\pi x}{6} \right), ~x\in[-3,3]$, and then calculating the pull-back on the 1-form $\vel$ \citep{frankel2004geometry}. This ensures that the vorticity field lines are also deformed as per the perturbation (though note that the velocity field is no longer divergence-free).
Generating the perturbation in this way allows us to preserve the exact divergence-free nature of the vorticity field, avoiding the complications described by, e.g., \cite{1989PhFl....1..633M}. We note further that this Gaussian initial condition for $\vort$ -- together with the use of a finite difference code -- avoids, by construction, the issues of numerical noise {in the initial condition} discussed by \cite{2008PhyD..237.1912B} that hampered previous studies that initialised the vortex tubes with compact support.

 When discussing the reconnection process in the next section we will refer to certain planes to aid discussion. The $x=0$ plane will be referred to as the `symmetry plane' due to the symmetry of the velocity field about this plane. The $y=0$ plane will be referred to as the `dividing plane' as prior to reconnection it divides the flux of the two vortex tubes.  The primary reconnection process will involve a transfer of vorticity flux from the symmetry plane to the dividing plane.

For computational expediency and to isolate the flow within the simulated domain we choose to employ periodic boundaries  that are closed to the flow at $x=\pm 3$, $y=\pm6$, $z=\pm12$. Thus an additional pair of vortex tubes is positioned  at $z=9$, $y=\pm 1$, with vorticity sign and perturbation anti-symmetric about the $z=0$ plane to those of the tubes located at $z=-9$. The periodic perturbation satisfies the boundary condition in $x$, and to obtain an initial condition periodic in $y$ and $z$ a $9\times 9$ array of image vortices is constructed. Here we restrict our study to the pair of tubes contained within the sub-domain $x\in[-3,3]$, $y\in[-6,6]$, $z\in[-12,0]$. The simulation is terminated before the evolution  is significantly affected by any of the image vortex tubes -- this was verified by repeating selected simulations in larger domains. 

The simulations are conducted using a 3D code developed and thoroughly tested for hydrodynamic and magnetohydrodynamic problems \citep{nordlund19953d,galsgaard1996}. This is a high-order finite difference code using staggered grids to maintain conservation of physical quantities. The derivative operators are sixth-order in space -- meaning that numerical diffusion is minimised -- while the interpolation operators are fifth-order. The solution is advanced in time using a third-order explicit predictor-corrector method. We solve the equations
\begin{eqnarray}
\frac{\partial(\rho \vel)}{\partial t} &=& -\mathbf{\nabla}\cdot(\rho \vel\vel) -{\nabla}p + \mu 
\left( \nabla^2 \vel + \frac{1}{3} \nabla (\nabla \cdot \vel) \right) \label{momentum}\\
\frac{\partial \rho}{\partial t}&=& -{\nabla} \cdot (\rho \vel) \label{mass}\\
\frac{\partial e}{\partial t}&=& -{\nabla}\cdot(e\vel)-p\mathbf{\nabla}\cdot\vel \nonumber\\
&&+ \mu\left(  \frac{{\partial v_i}}{\partial x_j}\frac{{\partial v_i}}{\partial x_j}+\frac{{\partial v_j}}{\partial x_i}\frac{{\partial v_i}}{\partial x_j}-\frac{2}{3}(\nabla\cdot\vel)^2   \right) \label{energy}
\end{eqnarray}
where $\vel$ is the fluid velocity, $\rho$ the density, $e$ the thermal energy, $p=(\gamma-1)e=2e/3$ the gas pressure, $\mu$ the viscosity, and summation over repeated indices is assumed. {We emphasise that the equation of motion is written in the form (\ref{momentum}) -- different to Equation (\ref{eq:dvdt}) -- to ensure conservation of momentum on the staggered grid for the employed numerical scheme \cite{nordlund19953d}. If a barotropic fluid is assumed then Equation (\ref{momentum}) is equivalent to Equations (\ref{eq:dvdt}), (\ref{mass}), up to vector identities. We have performed the simulations both using the full energy equation (\ref{energy}) and an adiabatic equation of state, and find both the qualitative and quantitative differences to be negligible. Here we present the results of the simulations using the full system (\ref{momentum}--\ref{energy}).}

The viscosity is set explicitly to a constant value throughout the volume. At $t=0$ we set $\rho=0.1$, $e=0.09$, uniform in the domain, giving a sound speed of 0.77 in the non-dimensional code units. 
A grid resolution of $[120,600,480]$ is chosen for the sub-domain $x\in[-3,3]$, $y\in[-6,6]$, $z\in[-12,0]$. The grid is uniformly-spaced in $x$ and $z$, but is stretched along $y$ so as to increase the density of points in the vicinity of $y=0$, in order to better resolve the vortex sheet that forms there. 
In line with previous studies we define the Reynolds number of our simulations to be $\Gamma/(\mu/\rho)$, where $\Gamma$ is the tube circulation at $t=0$.

\section{Qualitative description of the reconnection process}

\begin{figure*}
\centering
\includegraphics[width=\textwidth]{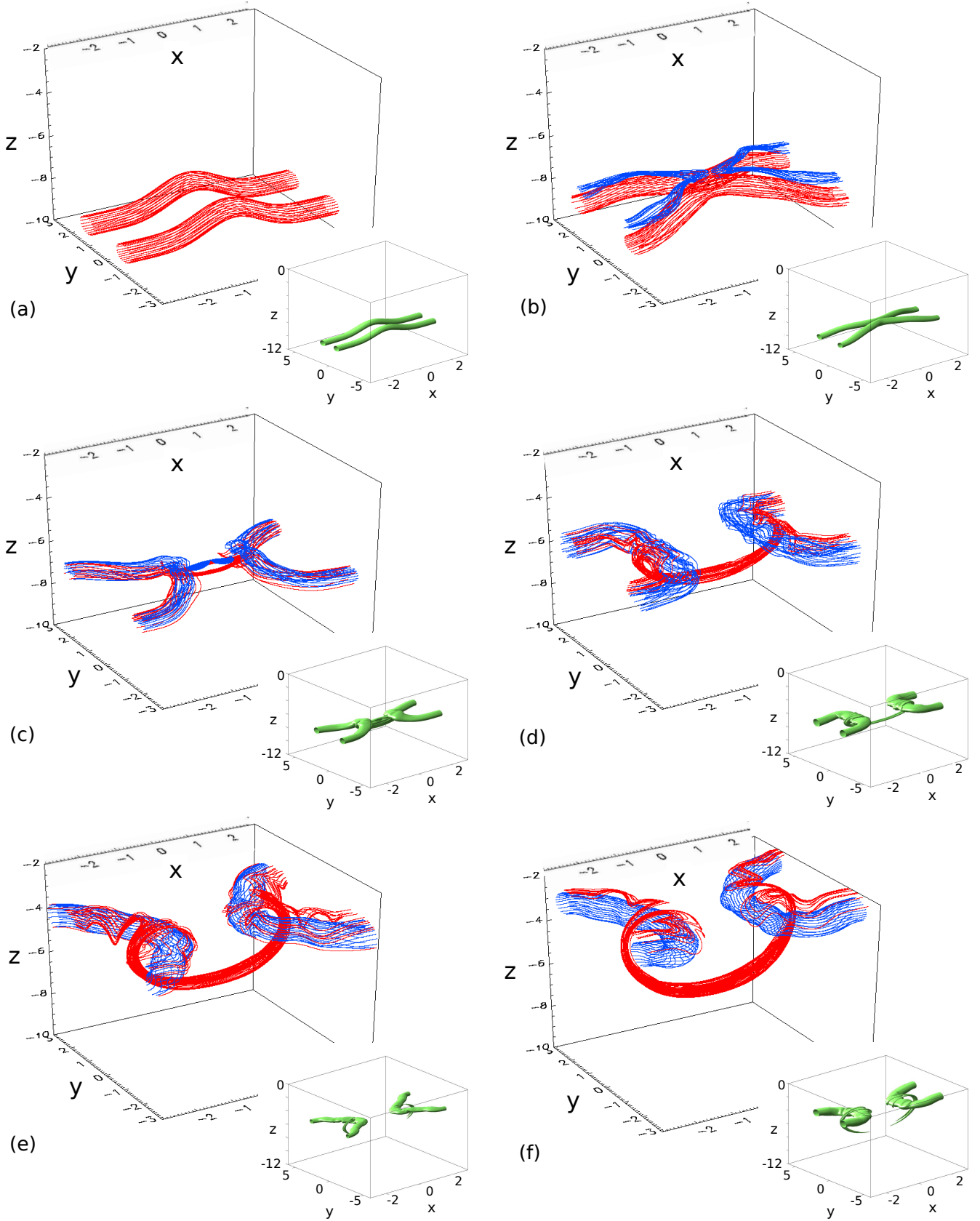}
\caption{Vorticity fieldlines plotted from $30\%$ $|\vort|$ contours at $x=0$ (red) and $y=0$ (blue) at (a) $t=0$, (b) $t=30$, (c) $t=60$, (d) $t=90$, (e) $t=120$ and (f) $t=150$. Inset: $|{\vort}|$ isosurface of $30\%$ maximum $|\vort|$ at the $x=-3$ boundary, plotted over the full numerical domain.}
\label{fig:lines_and_iso}
\end{figure*}

\begin{figure}
\centering
(a)\includegraphics[width=0.4\textwidth]{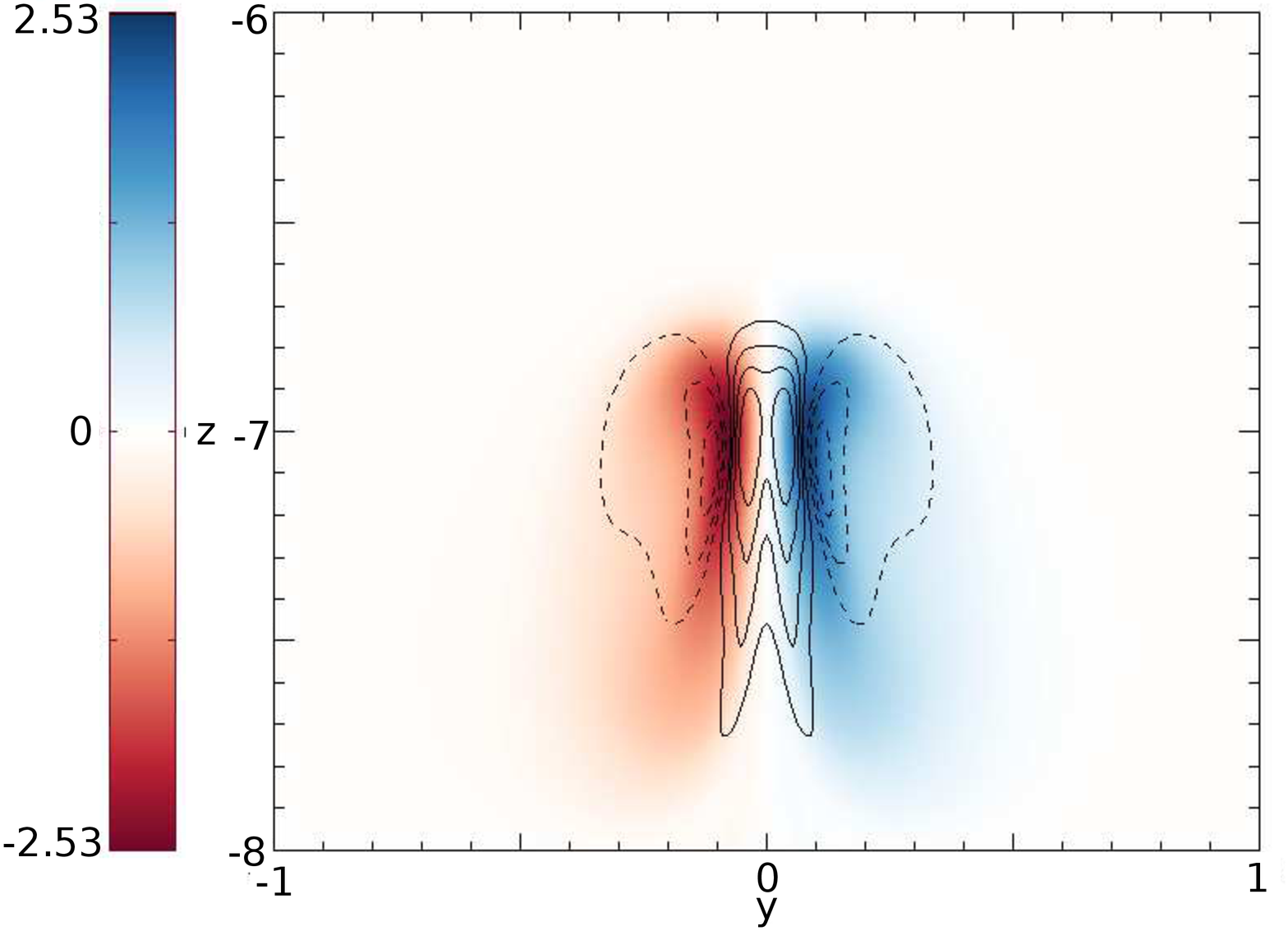}
(b)\includegraphics[width=0.4\textwidth]{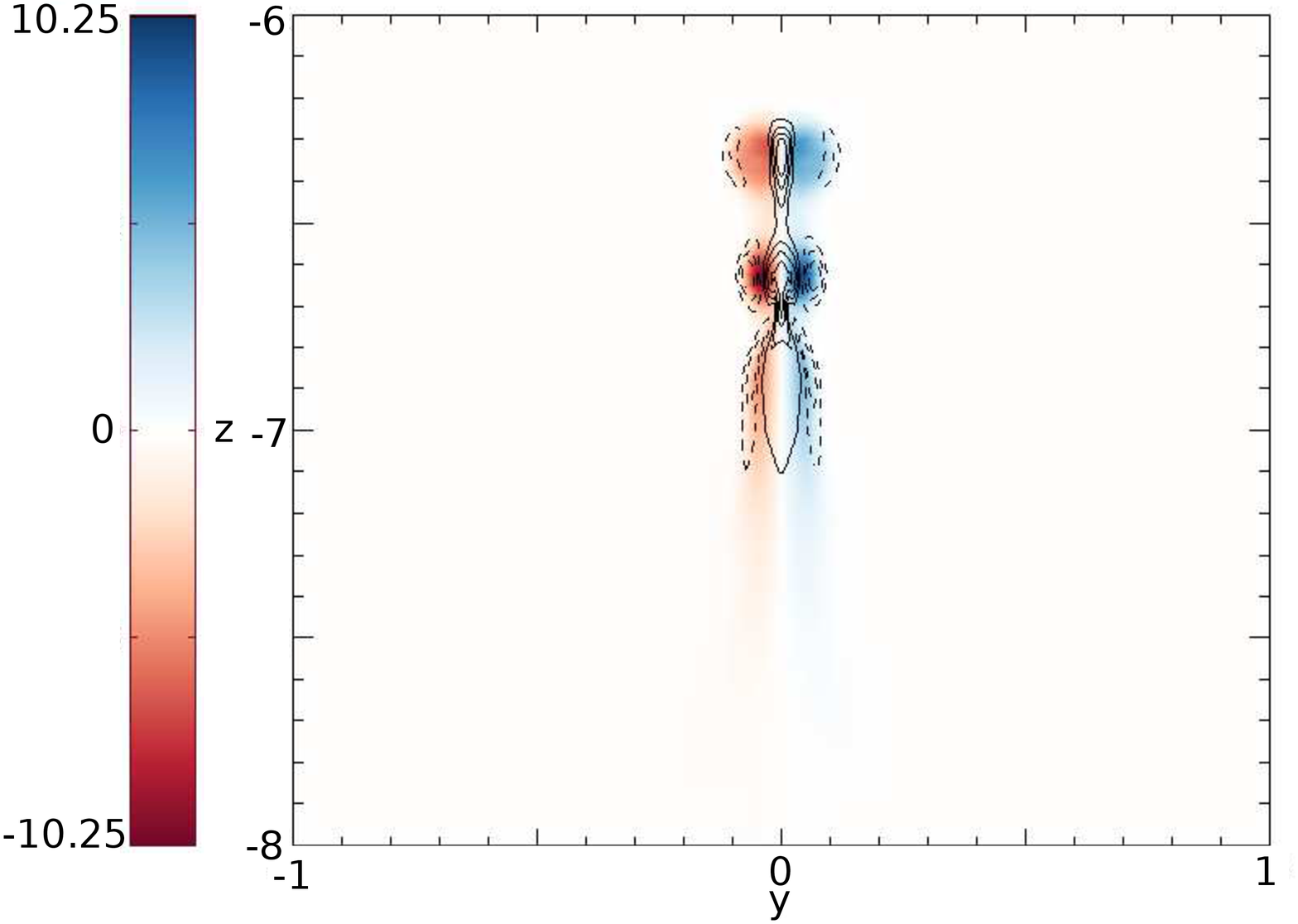}\\
(c)\includegraphics[width=0.4\textwidth]{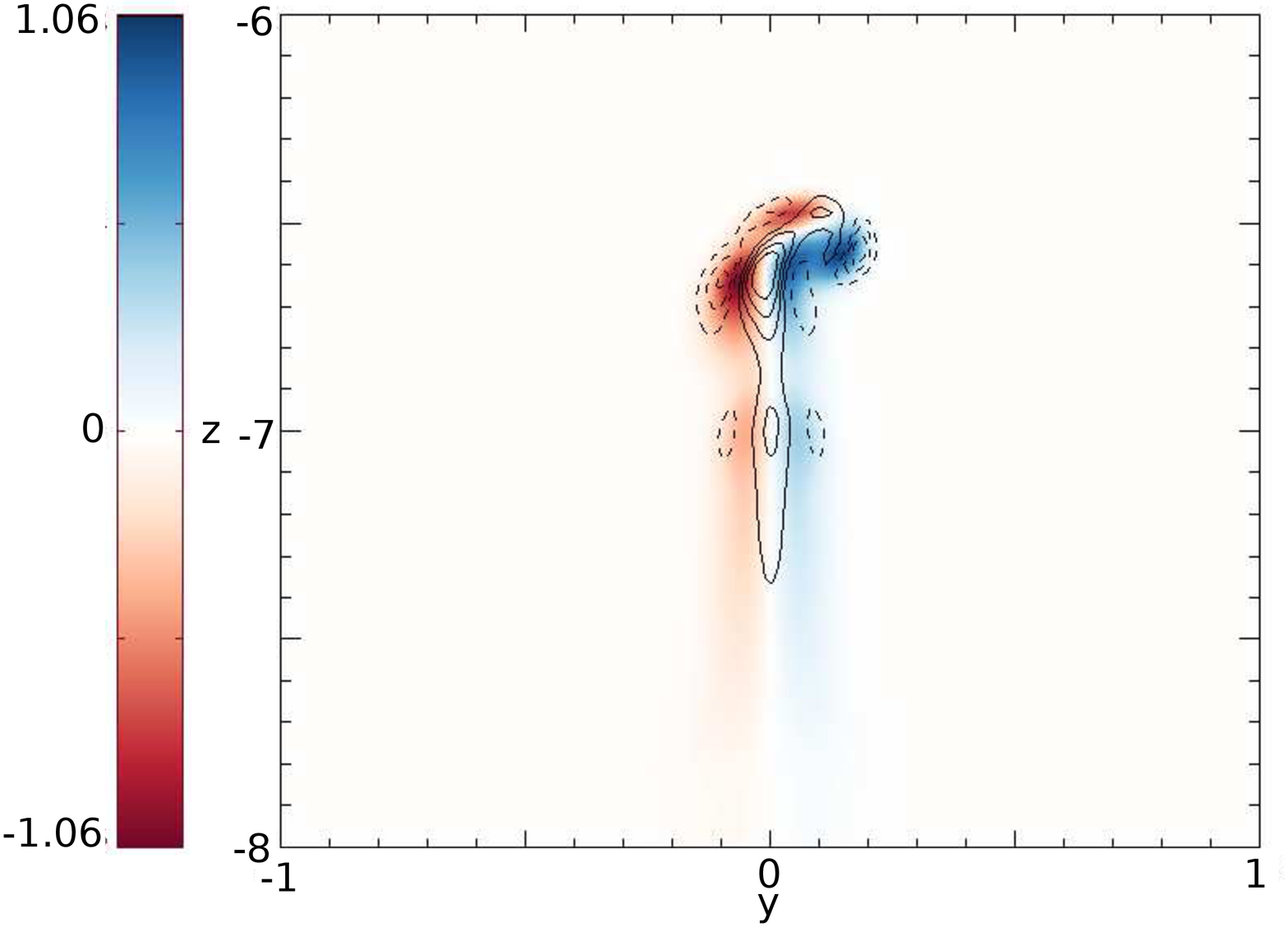}
(d)\includegraphics[width=0.4\textwidth]{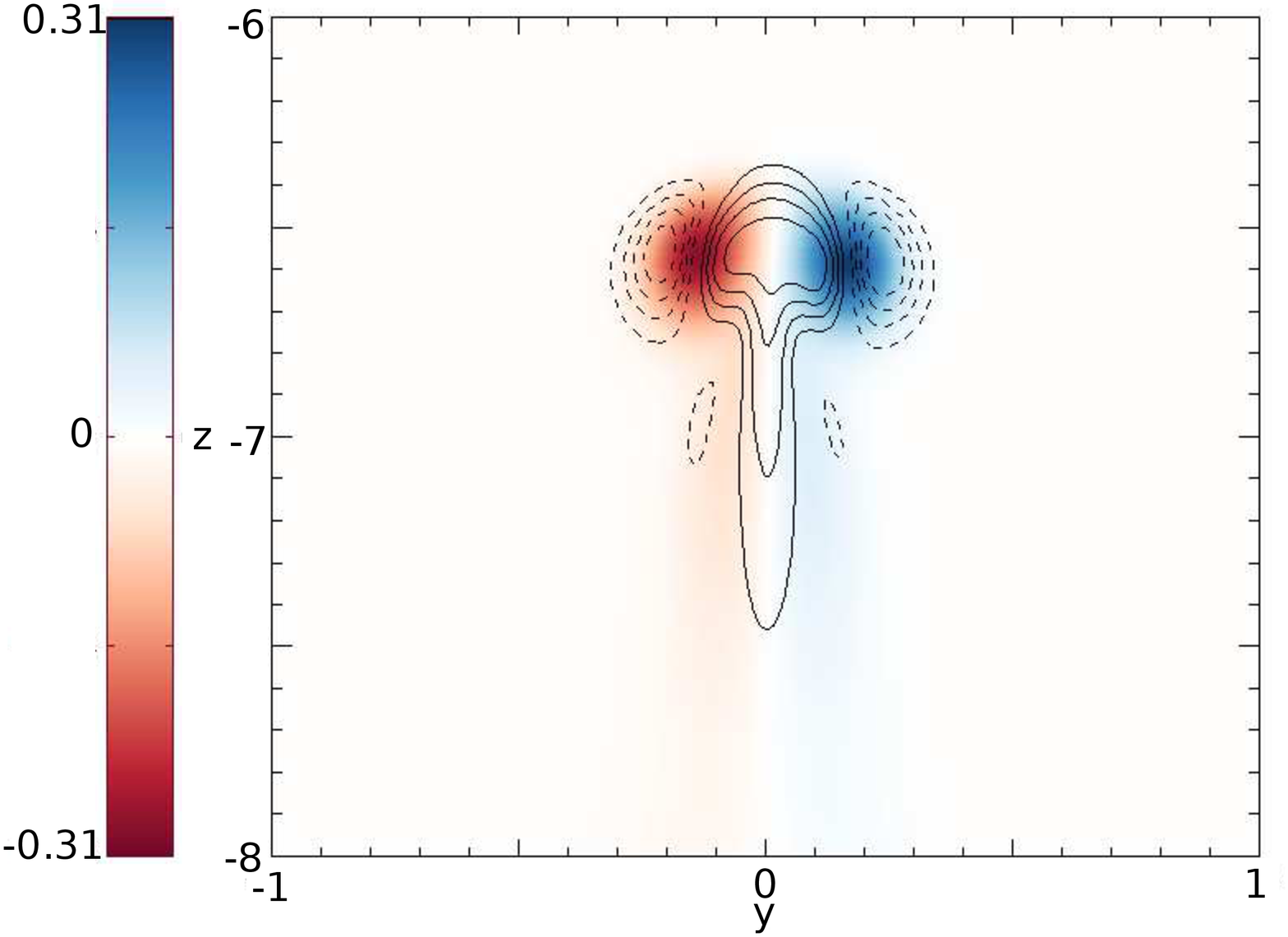}
\caption{Contour plots of $\omega_x$ (shaded, filled contours) and $(\nabla\times\vort)_z$ (unfilled contours; solid positive, dashed negative) in the $x=0$ plane, at (a) $t=35$, (b) $t=51$, (c) $t=75$, (d) $t=120$.}
\label{fig:apresdvs15}
\end{figure}

The simulations have been run for a series of Reynolds numbers. We focus principally on a simulation with $Re=4000$, while notable variations of the results with $Re$ are mentioned. Qualitative properties of the evolution can be seen in the inset images in Figure \ref{fig:lines_and_iso}. The perturbation applied to the vortex tubes at $t=0$ leads to a rotation of the two tube segments towards one another, leading them to collide. As the vortex tubes approach one another in the vicinity of $y=0$, their cross-sections each become stretched in the $z$-direction and squeezed in the $y$-direction to form what we describe here as a `vortex sheet' geometry \citep{1987PhRvL..58.1636P}. As these vortex sheets approach one another they form a double vortex sheet  at the centre of which is an intense concentration of $(\nabla \times \vort)_z$, as shown in Figure \ref{fig:apresdvs15}. This is the quantity responsible for determining the reconnection rate, as discussed in Section \ref{sec:rectheory}.

As the double vortex sheet thins and intensifies, its motion in the $z$-direction accelerates.
At this value of $Re$ the double-vortex sheet reaches a sufficiently large aspect ratio near the plane $y=0$ that it undergoes a Kelvin-Helmholtz instability \citep{opac-b1090797}, forming
a `head-tail' structure as described by \cite{iutam1989melander} -- see Figure \ref{fig:apresdvs15}(b). Eventually the symmetry about the $y=0$ plane is broken by small (numerical) fluctuations, as previously observed by \cite{2011PhFl...23b1701H} and shown in Figure \ref{fig:apresdvs15}(c).
It is worth noting that this instability leads to a non-zero vorticity flux through the dividing plane even in the absence of reconnection. For this reason care must be taken when attributing measured flux changes to the reconnection process.
Post-reconnection, the bridges form elliptical vortex rings (recall the periodic boundary conditions in $x$) that exhibit Kelvin-wave oscillations as they travel upward in $z$ \citep[as discussed in detail by, e.g.][]{2012PhFl...24g5105V}. 
Note that herein we use `vortex ring' to describe a structure characterised by closed vortex lines.
Note also that while the simulation is compressible, we find that in practice density fluctuations are small (maximum of 2-3\%), and so the compressibility has a minimal effect on the dynamics.


\section{Topological analysis of the reconnection process}\label{topolsec}

\subsection{Visualising the Reconnection Process - Vorticity Fieldlines}\label{subsec:apresfl}

%

To achieve a more detailed understanding of the reconnection process vorticity fieldlines are plotted in Figure~\ref{fig:lines_and_iso}, allowing analysis of the topological changes in the vorticity field. At each time we integrate 50 fieldlines from seed points in the symmetry plane (red lines; threads) and dividing plane (blue lines; reconnected bridge fieldlines). Specifically, the vortex lines are initiated from starting points that are equally spaced along contour lines of $|\vort|$ in the plane in question, at $30\%$ of the maximum vorticity in the plane. This permits the identification of new features of the reconnection process, as described below. From Figure~\ref{fig:lines_and_iso} we observe the rotation of the vortex tubes and evidence of reconnection. The reconnection process begins at the leading edge of the vortex tubes in $z$, in the relatively weak `head' of the double vortex sheet -- Figure~\ref{fig:lines_and_iso}(b). This occurs due to the shape of the double vortex sheet, with the stronger vorticity in the `head' moving faster along $z$ than weaker vorticity of the `tail' \citep{iutam1989melander}. The higher vorticity at the leading edge leads to a higher $(\nabla \times \vort)_z$ which induces reconnection. 

As the bridges evolve the threads begin to wrap around them, and the curvature of the thread vortex lines changes -- Figure~\ref{fig:lines_and_iso}(c-e). This new curvature means the thread vortex lines begin to separate, slowing the reconnection process and ultimately preventing it from being `complete' -- i.e.~preventing all thread flux from being converted to bridge flux \citep{1989PhFl....1..633M}.
{The geometry of the field lines post reconnection at early times shows a pronounced cusp shape (Figure \ref{fig:lines_and_iso}b), which gradually smooths out as the field lines retract from the reconnection site \citep[for more details see][]{mcgavin2017}. At later times, the Kelvin waves develop as part of the process, as shown in Figure \ref{fig:lines_and_iso}(d-f)}.

\subsection{Flux Evolution}\label{subsec:apresflux}
\subsubsection{Reconnection at the symmetry plane}

\begin{figure}
\centering
(a)\includegraphics[width=0.45\textwidth]{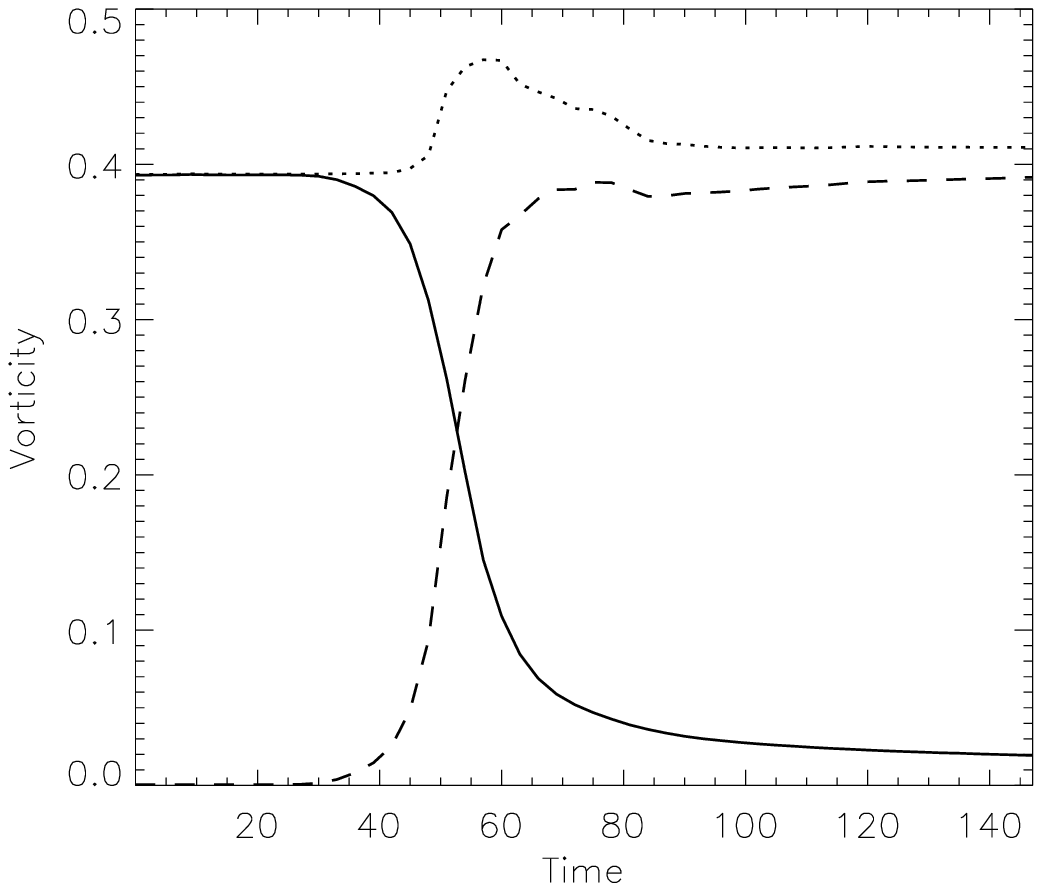}
(b)\includegraphics[width=0.45\textwidth]{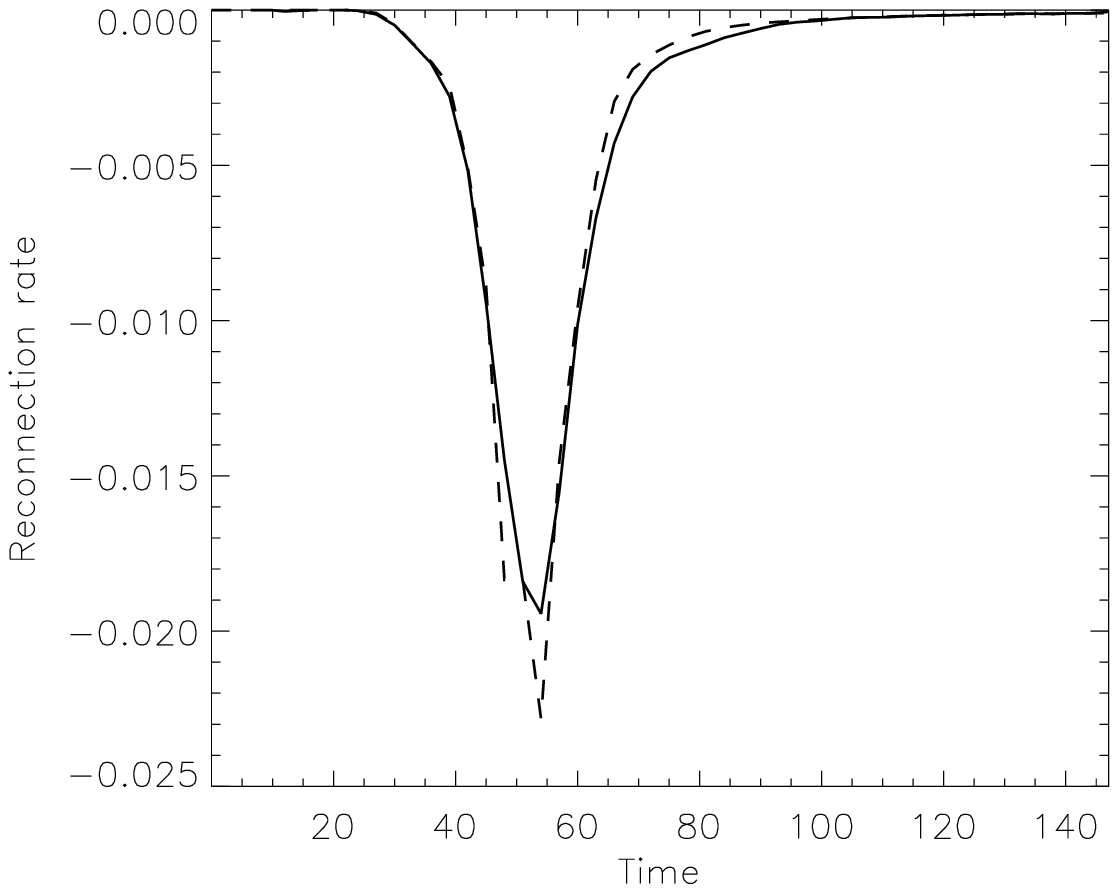}
\caption{(a) Vorticity flux measured at $x=0$ (solid), $y=0$ (dashed) and total of both (dotted) as a function of time (b) Reconnection rate measured from rate of change of vorticity flux  through the symmetry and dividing planes (solid) and by integrating $(\nabla\times\vort)_z$ along the central axis (dashed).}
\label{fig:apresflux}
\end{figure}

Here we quantify the rate of reconnection of vorticity flux by two methods. The first involves invoking symmetry and measuring the fluxes through the symmetry and dividing planes. The second involves integrating $(\nabla \times \vort)\cdot d{\bf l}$ along an appropriate path and invoking the theory presented in Section \ref{sec:rectheory}.
Considering the first method, the vorticity flux in both the symmetry and dividing planes is plotted in Figure~\ref{fig:apresflux}(a). The sum of these two fluxes is plotted as the dotted line -- in a simple symmetric reconnection process (obtained for lower Reynolds number) this sum is expected to be constant; the reason that it is not is discussed below. In Figure~\ref{fig:apresflux}(b) we see that the reconnection rate increases rapidly to its maximum, after which it begins to stall as the newly reconnected bridges inhibit the reconnection of subsequent fieldlines. After $t\approx 65$, the rate tails off, but remains non-zero as the elongated threads reconnect after the main event. 

\begin{figure}
\centering
(a)\includegraphics[width=0.46\textwidth]{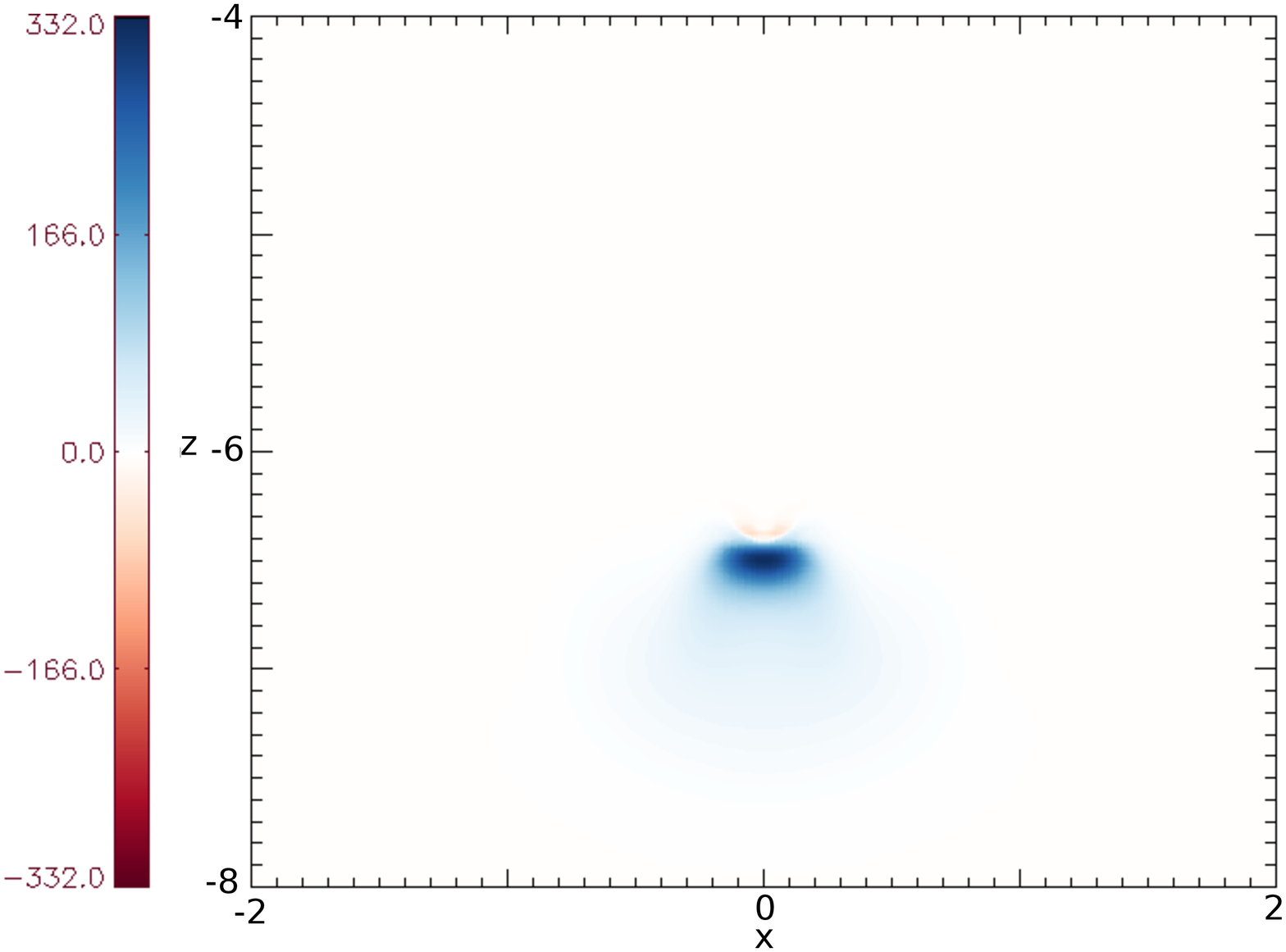}
(b)\includegraphics[width=0.46\textwidth]{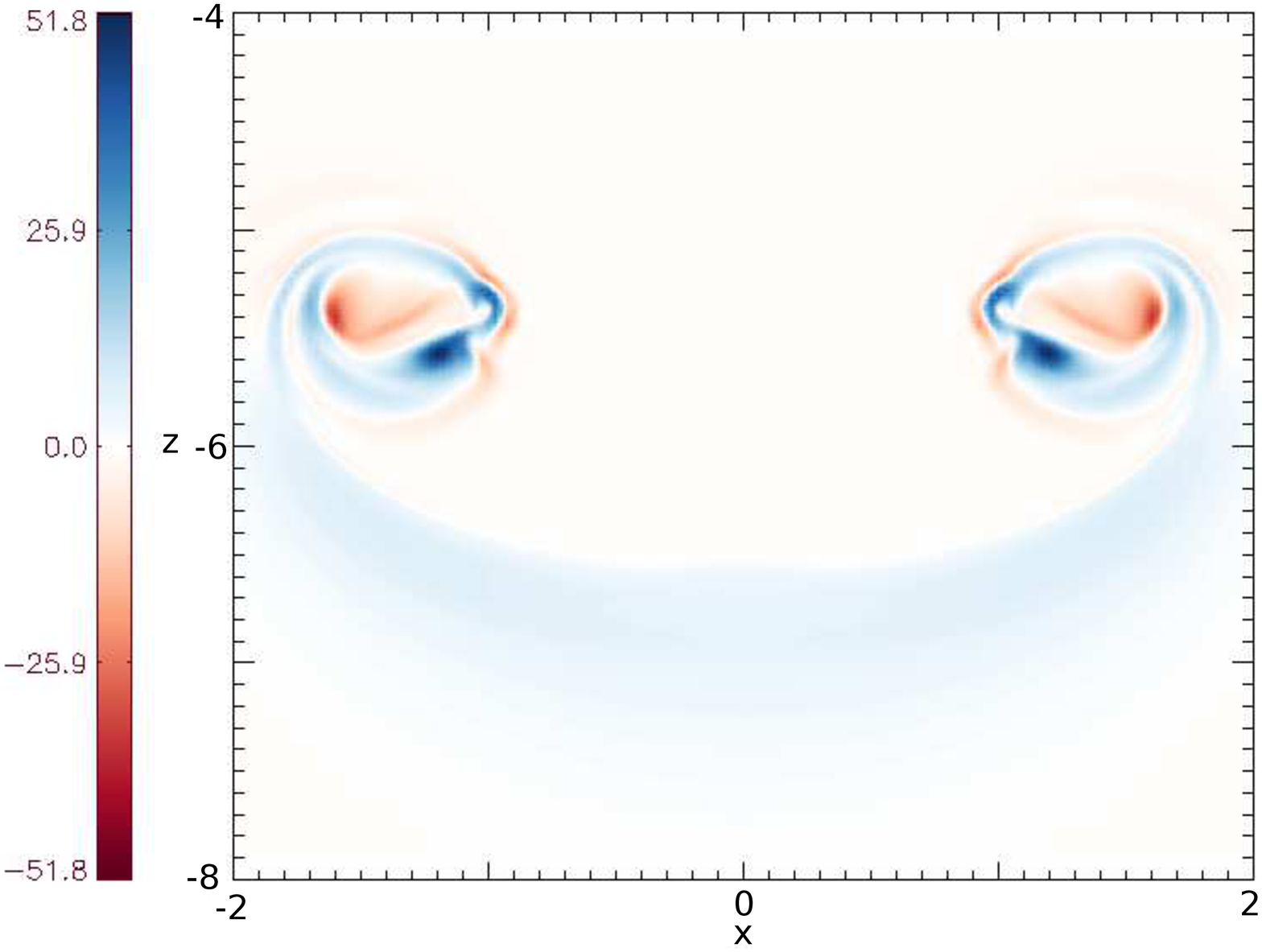}
\caption{$(\nabla \times \vort)_z$ contour plot in the dividing plane ($y=0$) at $t=45$ and $t=90$.}
\label{fig:aprescwdw15}
\end{figure}

Consider now method 2 for calculating the rate of reconnection. The layer within which the reconnection takes place can be identified by the region of enhanced $(\nabla \times \vort)\cdot\vort$. Within this layer lies the central axis of the box (the $z$-axis) where the vorticity is zero by symmetry (at least at early times), and where the vortex lines are locally planar as seen in Figure \ref{fig:xptandloop}. Thus the reconnection occurs exactly along the X-line, along which $|\vort|=0$ and $(\nabla \times \vort)_z\neq 0$, and the reconnection rate can thus be measured as discussed in Subsection~\ref{sec:rectheory} by integrating along the X-line. When the vortex tubes first press together, $(\nabla \times \vort)\cdot\vort$ is concentrated in a well localised layer in the centre of the box (Figure \ref{fig:aprescwdw15}a). However, at later times this layer stretches quite far along the tube (as seen by the contours in the dividing plane -- Figure~\ref{fig:aprescwdw15}b). 
Integrating $\nu(\nabla \times \vort)_z$ along the central axis {(dashed curve, Figure~\ref{fig:apresflux}(b))} gives a measure of the reconnection rate that closely matches the brute-force flux measurements {(solid curve, Figure~\ref{fig:apresflux}(b))}. We note however that both methods rely on the assumption of a symmetric reconnection process, while this symmetry is broken at higher $Re$ by the Kelvin-Helmholtz instability as well as the formation of additional vortex rings (see below). The difference is that method 2 can in principle be extended to take account of this breaking of the symmetry: one only has to first identify the path of the X-line(s) in the domain and integrate along them, and in this way the changes in flux connectivity are measured.

%

\subsubsection{Additional Vortex Rings}\label{subsec:apreycascade}

\begin{figure}
\centering
(a)\includegraphics[width=6.5cm]{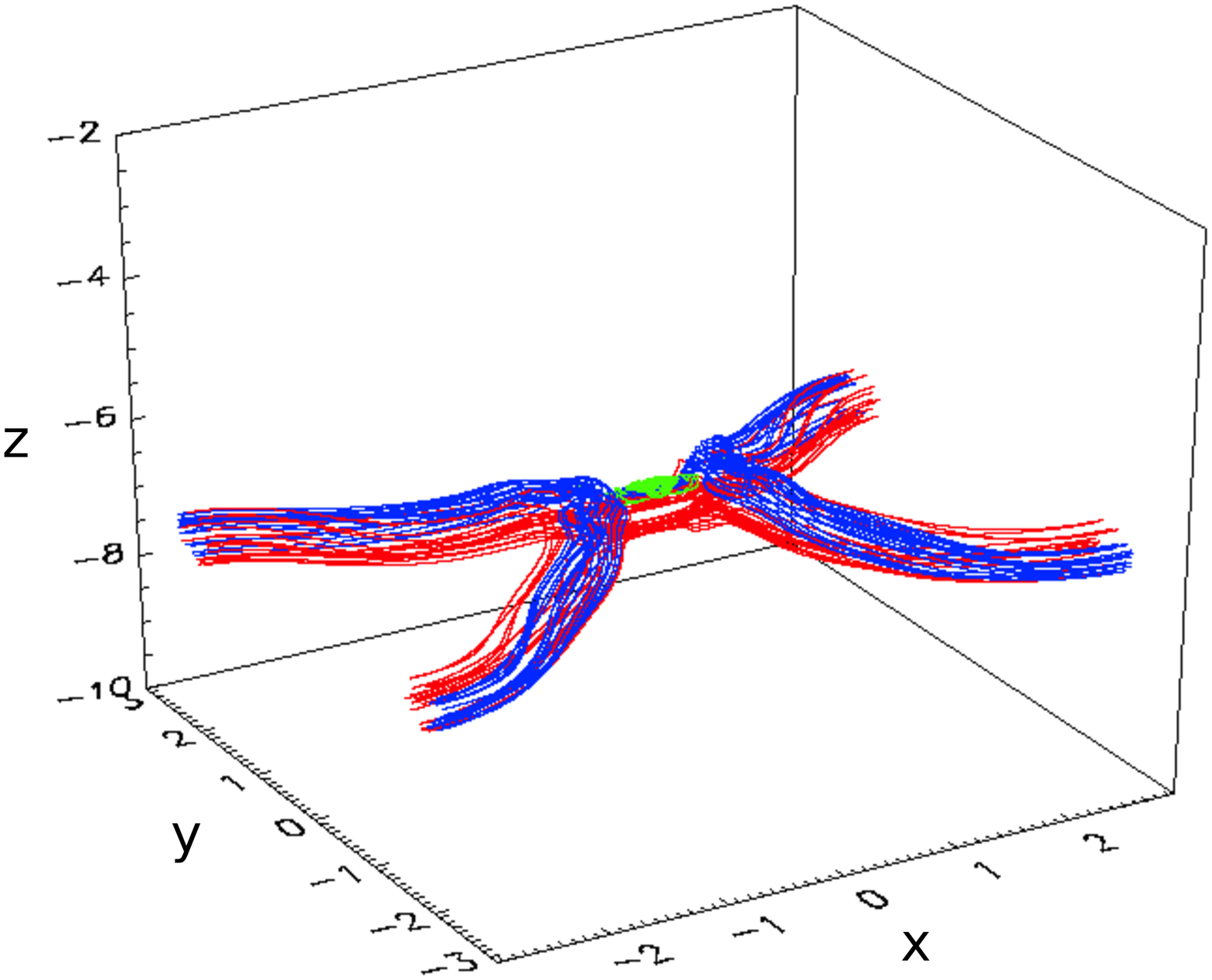}
(b)\includegraphics[width=6.5cm]{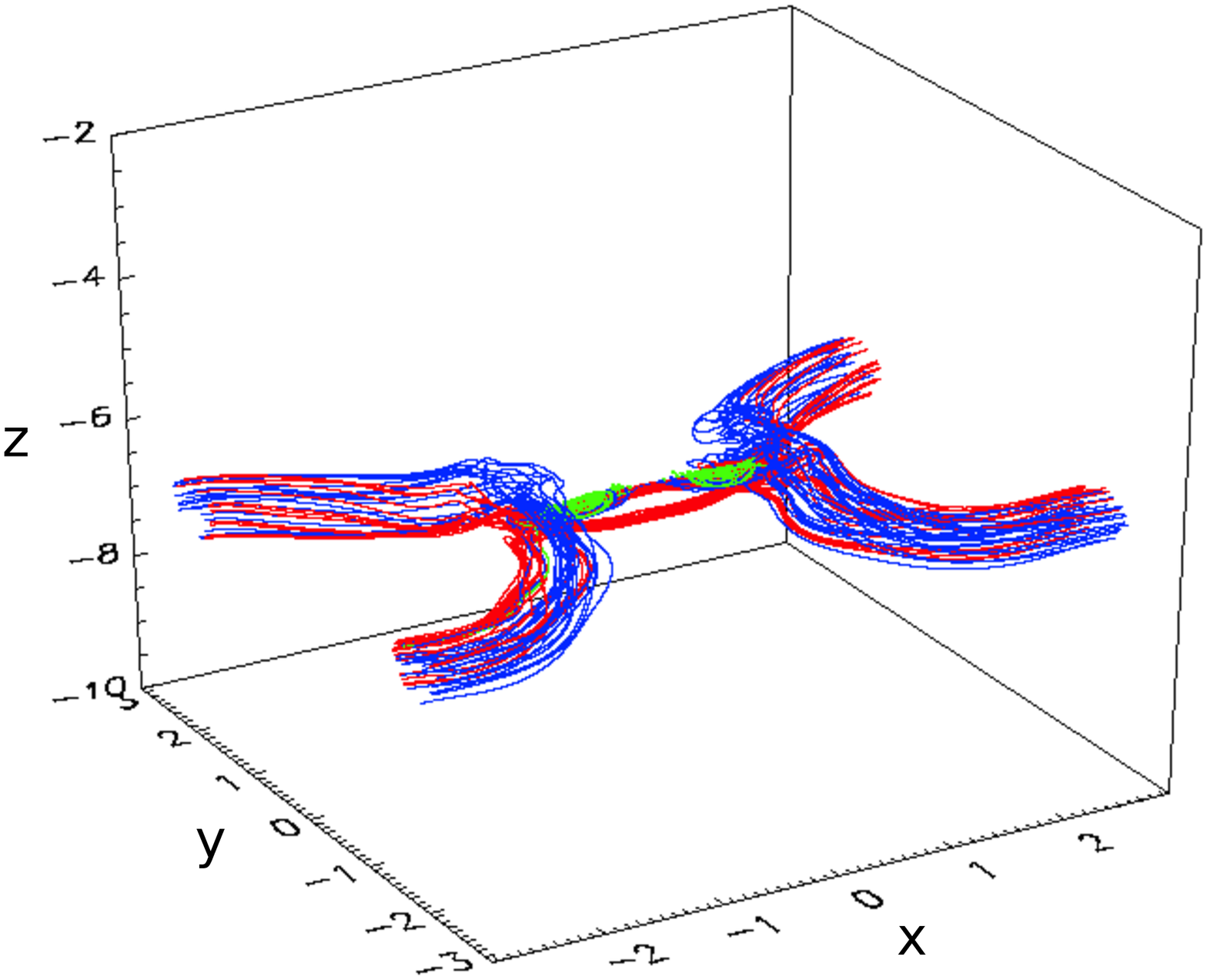}
(c)\includegraphics[width=6.5cm]{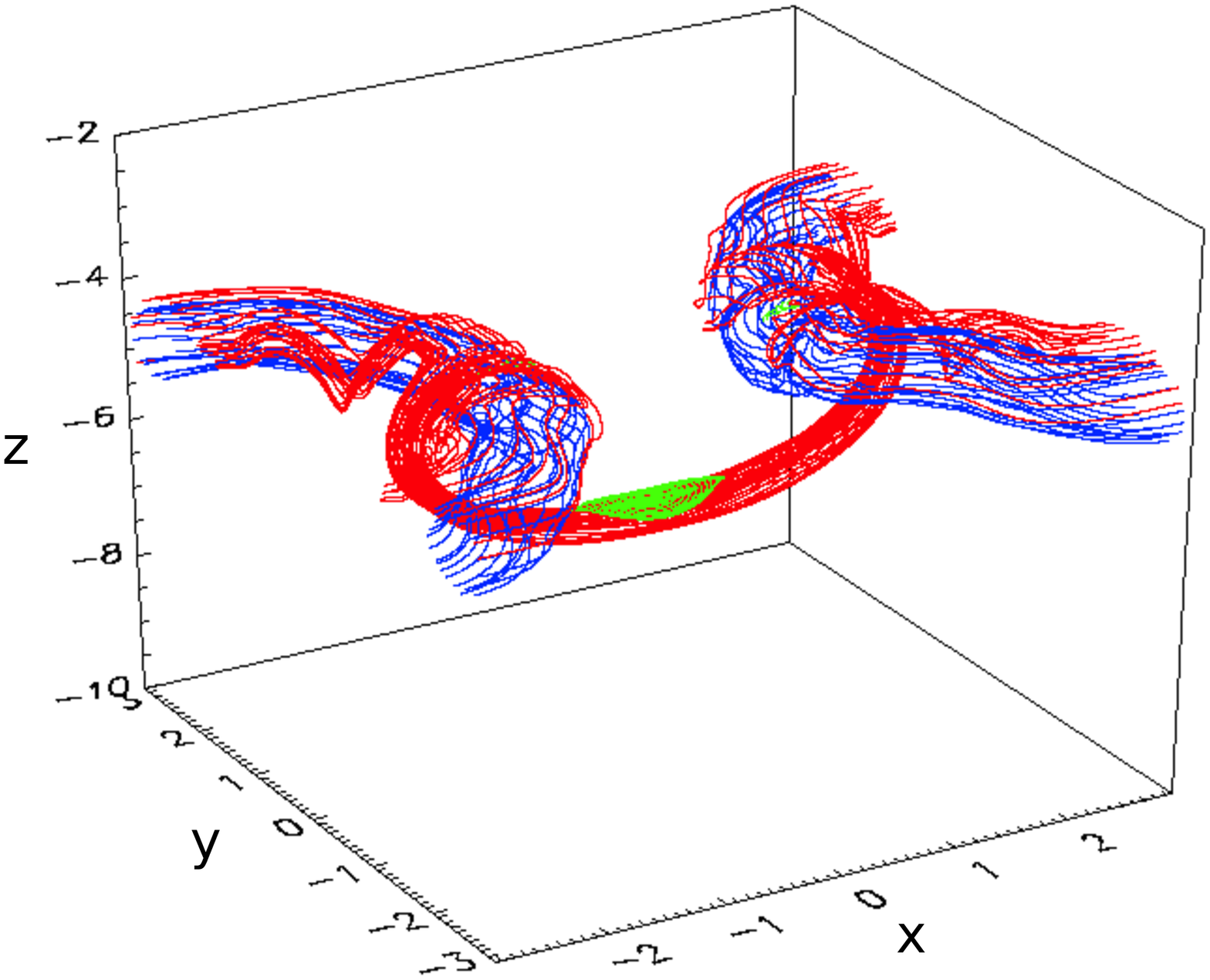}
(d)\includegraphics[width=6.5cm]{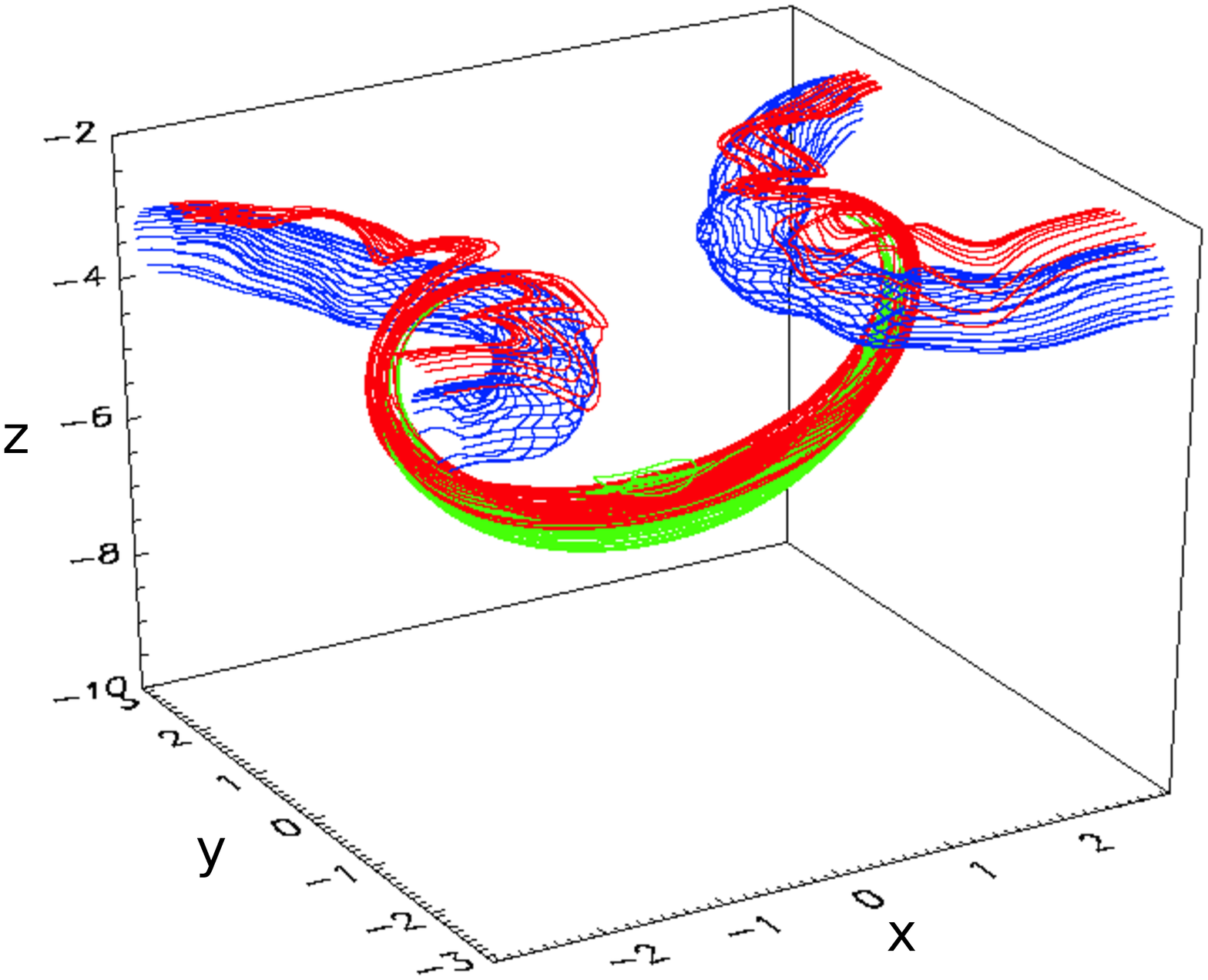}
\caption{Threads (red) plotted from $30\%$ maximum $\omega_x$ contours at $x=0$, bridges (blue) plotted from $30\%$ maximum $\omega_y$ contours at $y=0$, additional vortex ring fieldlines (green) plotted from $30\%$ maximum $\omega_y$ (of opposite sign to bridges) contours at $y=0$, $Re=4000$ at (a) $t=54$, (b) $t=63$, (c) $t=111$ and (d) $t=138$.}
\label{fig:apreyallfl4000}
\end{figure}

\begin{figure}
\centering
\includegraphics[width=0.45\textwidth]{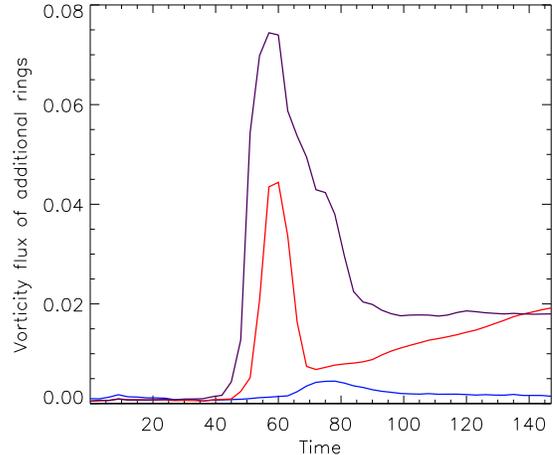}
\caption{Vorticity flux in the dividing plane due to additional vortex rings as a function of time. For $Re=4000$ (purple), 2000 (red) and 800 (blue).}
\label{fig:apreyadditionalflux}
\end{figure}

For $Re \gtrsim 800$, analysis of the vortex lines reveals reconnection events out of the symmetry plane where the main reconnection occurs. This additional reconnection creates new vortex rings; {some examples from the simulation with ${\rm Re}=4000$ are} shown by green field lines in Figure~\ref{fig:apreyallfl4000}. These rings constitute additional flux through the dividing plane, and are the reason for the `bump' on the dotted curve in Figure~\ref{fig:apresflux}(a). They are associated with the formation of extra null-lines close to the dividing plane -- both X-lines (between the rings) and O-lines (at their centres). In Figure~\ref{fig:apreyadditionalflux} we plot {a lower bound to} the flux in these additional rings, and we observe a sharp peak followed by a rapid dissipation {between $t\approx 60$ and $t\approx 80$}. Since the threads are very close together when they reconnect the flux rings formed are very thin (in $y$) and annihilate shortly after forming. Small remnants of the rings are left over after this annihilation, and at later times the elongated threads reconnect at multiple locations, creating high aspect ratio vortex rings -- see Figure~\ref{fig:apreyallfl4000}(d). At higher values of $Re$, progressively more of these rings are created and then annihilated (Figure~\ref{fig:apreyadditionalflux}).  {We note however, that at higher $Re$ the rings become increasingly difficult to identify computationally, as they become progressively longer and thinner, while at certain times we observe a `cascade' of rings forming away from the symmetry plane ($x=0$; see e.g.~Figure~\ref{fig:apreyallfl4000}b). At $Re=2000$ (red curve in Figure~\ref{fig:apreyadditionalflux}) all additional rings appear to be centred on the $z$-axis, and do not overlap/wrap into the bridges -- these are readily identifiable with our algorithm and thus the plot gives a relatively accurate indication of the flux in the rings. However, the purple curve in the figure representing the flux in the rings for $Re=4000$ definitely provides an underestimate, since it misses both the cascade and the rings that wrap into the bridges (Figure~\ref{fig:apreyallfl4000}d). This is the reason why the red curve overtakes the purple one at late times.}
We note further that these additional vortex rings could be responsible for the `curved vortex belts' of \cite{1995AcMSn..11..209W}. Importantly, the vortex rings do not show up clearly in isosurface plots and were therefore missed by many previous studies -- they are revealed only by plotting the vortex lines.

\subsection{Field line helicity and `internal' reconnection}\label{subsec:apreshelicity}

\begin{figure}
\centering
(a)\includegraphics[width=0.4\textwidth]{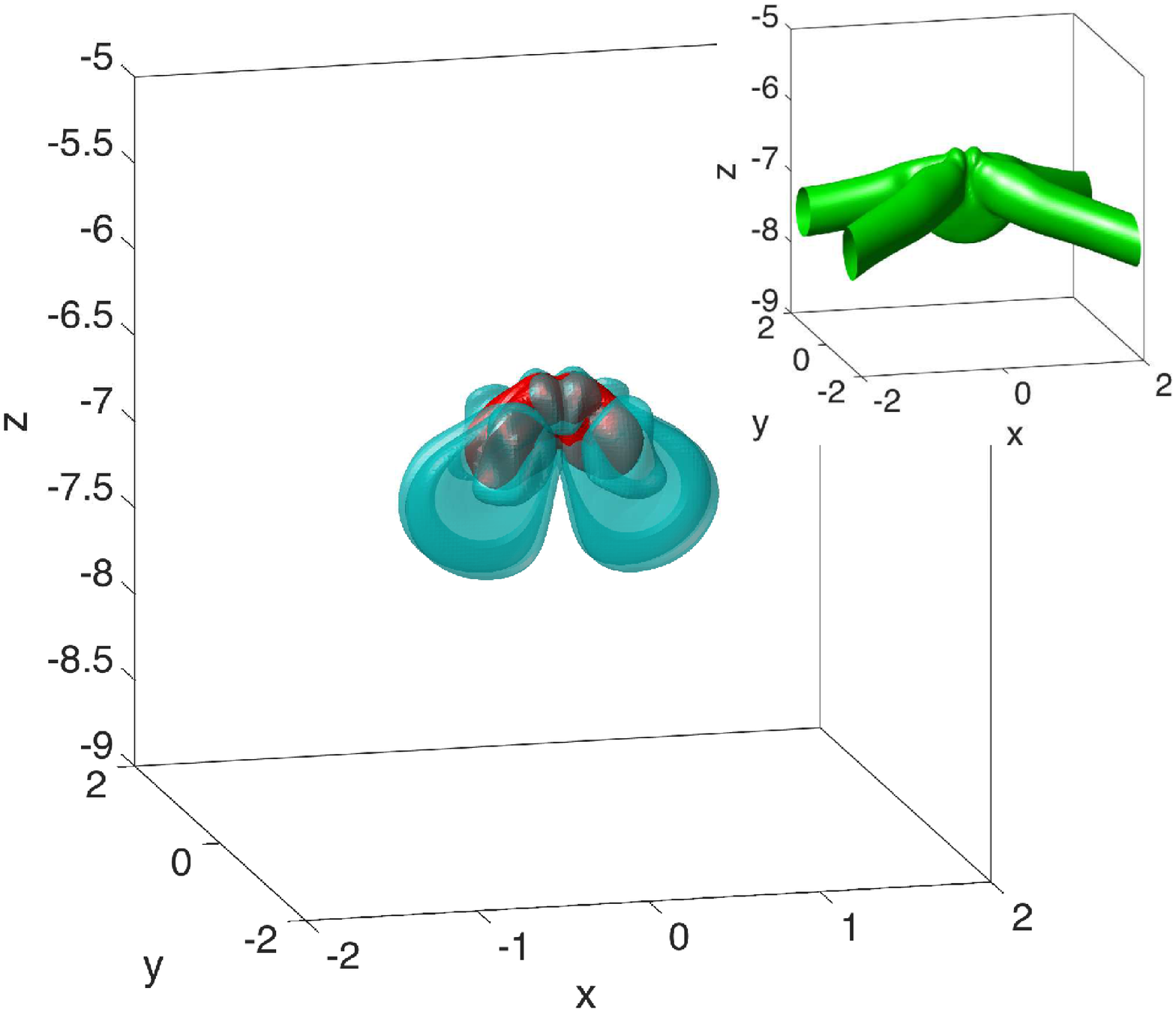}
(b)\includegraphics[width=0.4\textwidth]{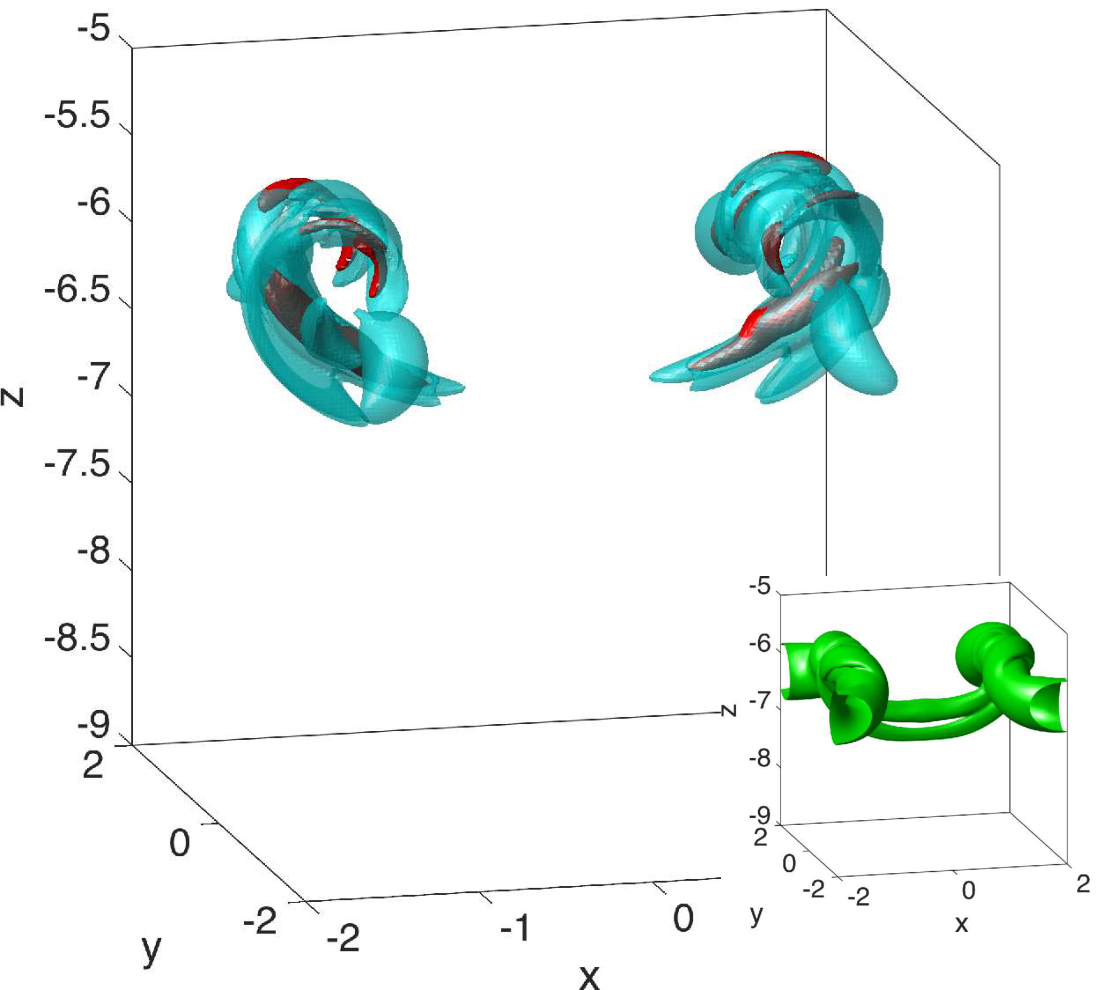}
\caption{Isosurfaces of $|\vel\cdot\vort|$ (cyan), $|\vort\cdot(\nabla \times\vort)|$ (red), and $|\vort|$ (inset, green), in each case at 10\% of the domain maximum, for (a) $t=45$, and (b) $t=75$.}
\label{fig:apreshelicisosurf}
\end{figure}

In addition to the reconnection of vortex lines between the two anti-parallel tubes -- that creates the primary and secondary vortex rings as described above -- we also find a reconnection of vortex lines within each individual tube that is previously unexplored. In order to examine this we require to understand the internal topological structure of vortex lines within each tube. `Internal' reconnection within the tubes is expected to occur when a significant twist is imparted to field lines, and takes the form of a rotational slippage in the connectivity of fluid elements by vortex lines as shown in Figure~\ref{fig:intrectype3drec}(a,b) \citep[e.g.][]{hornig2003}.

A powerful measure of the topology of the vorticity field is the {\it kinetic helicity}
\begin{equation}
H=\int_V \vel\cdot\vort\, dV,
\end{equation}
that measures the tangling -- or more precisely the net linkage -- of vortex lines within the domain $V$ \citep{moffatt1969}. This gives a single value for the whole domain, while more detailed information can be obtained by evaluating the {\it field line helicity}, 
\begin{equation}
h({\bf x}_0)=\int_{F({\bf x}_0)}\vel\cdot\vort\diff l,
\end{equation}
which measures the net winding of all vortex lines  in the domain (weighted by their flux) with the vortex line $F({\bf x}_0)$ through the point ${\bf x}_0$ \citep{berger1988}. There is an intimate link between the helicity and the reconnection process that can permit a change in the field line tangling. Recall that $(\nabla\times\vort)\cdot\vort$  -- necessary for 3D reconnection -- is a source term in the evolution equation for helicity \citep{1978magnetic}, and indeed 3D reconnection is known to redistribute the helicity density within the domain. Generation of helicity in 3D vortex reconnection was discussed by \cite{takaoka1996}.

For our initial condition, a single perturbed tube has zero helicity, since each vortex line lies in a plane and therefore has no self-twist \citep{1965JFM....22..471B}. Once the tube pair is introduced a small total unsigned helicity of $\sim 1.3\times 10^{-3}$ is present due to the tube curvature, that leads to a mutual helicity between the pair (of equal and opposite sign in the two halves of each tube at $x>0$ and $x<0$ by symmetry). 
{It is important to note that the helicity density $\vel\cdot\vort$ is not a Galilean invariant (although $H$ is).
 As such, care must be taken in ascribing physical significance to its value. The reader is referred to discussions in \cite{kida1991b,takaoka1996}, and references therein. Importantly, $H$ is an inviscid invariant, while $\vel\cdot\vort$ is not. Nevertheless, in a given reference frame, changes in the spatial distribution of the helicity density can still give a clue to the nature of the reconnection process. Here we calculate $\vel\cdot\vort$ using the native reference frame of the simulations, a natural choice since it minimises the unsigned helicity density prior to perturbation of the tubes, and respects the symmetry of the vortex tubes. An alternative choice is suggested in section 7.2 of \cite{kida1991b}. We have verified that calculating the helicity density using this alternative choice of frame maintains all of the qualitative properties presented here, with peak value of $\vel\cdot\vort$ changing by $<10\%$.}

{With the above caveats in mind}, as the tubes evolve we note the development of significant local concentrations of helicity density $\vel\cdot\vort$ -- see Figure~\ref{fig:apreshelicisosurf} \citep[and also the discussion of][]{iutam1989melander}. 
We hypothesise that the concentrations of $|\vel\cdot\vort|$ are generated because the fluid elements at the symmetry plane rotate faster than those at the $x=\pm3$ boundaries, twisting the vortex lines around the tube axes and leading to a concentration of $|\vel\cdot\vort|$ in the vortex sheet (Figure~\ref{fig:apreshelicisosurf}a). This is suggested by the enhanced intensity of $|\vort|$ in the mid-plane compared with at $x=\pm3$, though quantifying this relative twist precisely is difficult since the vortex line evolution is not ideal close to the reconnection site, but rather there is some slippage between the field lines and flow. Later in the simulation the helicity density is most strongly concentrated within the threads as shown in Figure~\ref{fig:apreshelicisosurf}(b). This helicity is associated with the wrapping of the threads around the reconnected bridges.
We note from the Figure that the Galilean invariant $\vort\cdot(\nabla\times\vort)$ is focussed in the same regions as the kinetic helicity, though is even more localised in space. {The distribution of both $\vel\cdot\vort$ and $\vort\cdot(\nabla\times\vort)$ is reminiscent of that observed during reconnection of vortex rings \cite{kida1991b}.}

As twist is introduced within the vortex tubes, we also observe the generation of regions of non-zero $(\nabla\times\vort)_\parallel$.  
The dynamics in these regions allows the vortex lines to slip (breaking the connections between opposite points on the $x=\pm 3$ boundaries), and the twist is dissipated. This type of reconnection is very different conceptually from the flux exchange between the tubes; nevertheless its importance in magnetised plasmas is now appreciated. Experimental evidence of this internal change of twist within vortex tubes and the relation to changes of helicity has recently been reported by Scheeler {\it et al.} \cite{scheeler2017}.

To examine this behaviour in detail we calculate the field line helicity $h$ for field line segments between the boundary and either the symmetry or dividing plane  (by symmetry $h$ is always zero for the full field line within the domain). We also calculate 
\begin{equation}
\Psi=\int_{F({\bf x}_0)}(\nabla\times\vort)\cdot \vort\, dl
\end{equation} 
along the same segments of the vortex lines -- which we call the `slipping rate' due to its relation with the reconnection rate in 3D -- and compare the distribution of the two quantities. 

\begin{figure*}
\centering
\includegraphics[width=0.8\textwidth]{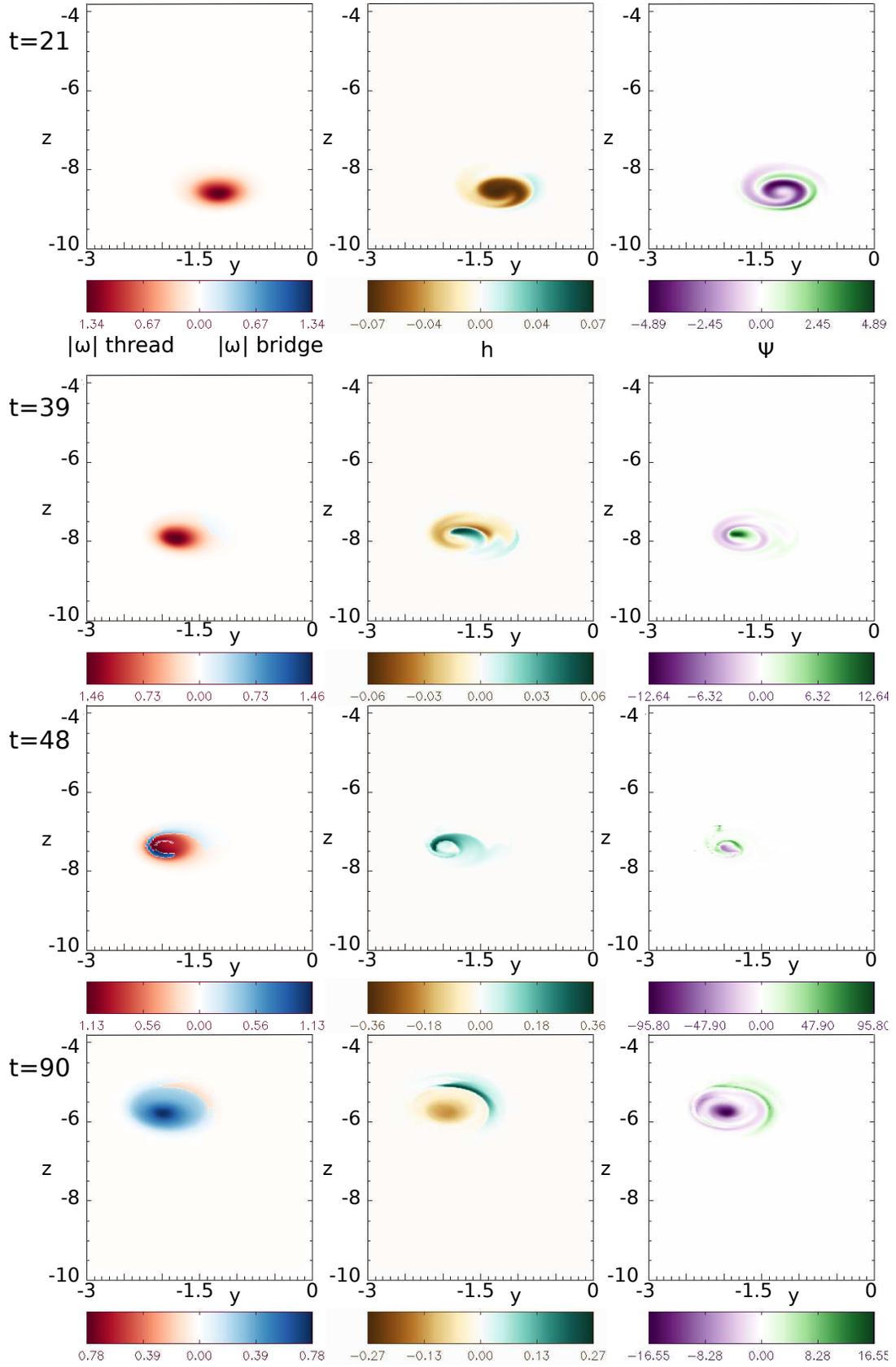}
\caption{Contour plots, from left to right, of vorticity magnitude (red, threads; blue, bridges) field line helicity $h$, and field line `slipping' rate $\Psi$, in the plane $x=-3$. {Threads and bridges are distinguished in the left frame by tracing vortex lines from a grid of points and determining whether they connect to $x=3$ or back to $x=-3$.}}
\label{fig:apreshelplot}
\end{figure*}

%


Figure~\ref{fig:apreshelplot} shows {a strong similarity between the patterns of} the slipping rate $\Psi$ and the field line helicity $h$ {over field lines}. 
The presence of both positive and negative regions of $\Psi$ indicates that reconnection is leading to a twisting/untwisting in both a right- and left-handed sense within the tube, correlated with the sign of $h$. 
{Calculating a per-field-line linear Pearson correlation coefficient for the two quantities yields values 0.66, 0.58, 0.38, and 0.74 for the times displayed in Figure \ref{fig:apreshelplot}, {indicating a moderate correlation between $h$ and $\Psi$.}
The left panels in the Figure show how the main reconnection between the tubes proceeds. The spatial distribution of reconnected field lines at a given time within the tube is  more complex than might initially be expected. In particular, we see at $t=48$ that the footprint of the reconnected field lines on the $x=-3$ plane exhibits a spiral pattern, due to the twisting of the field lines in the tube. {Examining $h$ at the same time we see that it is strongly concentrated in the vicinity of these reconnected vortex lines.} Examining the plots at $t=90$ we observe that at later times the twist (as measured by $h$) becomes concentrated along the threads as they are wrapped around the bridges. {We emphasise again that $h$ is not Galilean invariant, and so changes in its value are not sufficient to characterise a change of vortex line structure. However, the relative distribution of $h$ gives some insights into the local structure during the reconnection process.}

The contour plots of $\Psi$ allow us to estimate the vorticity flux that is reconnecting within each tube, by seeking local maxima and minima of $\Psi$ -- as described by \cite{2015PhPl...22d2117W}. These maxima and minima of $\Psi$ can be combined to give different measures of the reconnected flux, which can be interpreted as the net and gross reconnection rates. {The net reconnection rate takes into account only the difference between the global maxima and minima of $\Psi$ (dashed line, Figure \ref{fig:whcalc}), while the total, or gross, reconnection rate is obtained by examining adjacent local maxima and minima of $\Psi$ -- for a full discussion see \citep{2015PhPl...22d2117W}.} Applying those techniques we find that the minimum measure of the flux reconnected is approximately equal to the total flux of a single vortex tube ({note that this flux is 0.4 in non-dimensional code units}). Therefore on average each field line is reconnected once `internally' within the tube during the whole evolution. We find that the temporal maximum of $\Psi$ occurs after the main reconnection, probably due to the oscillations observed on the main reconnected vortex rings.

\begin{figure}
\centering
\includegraphics[width=0.45\textwidth]{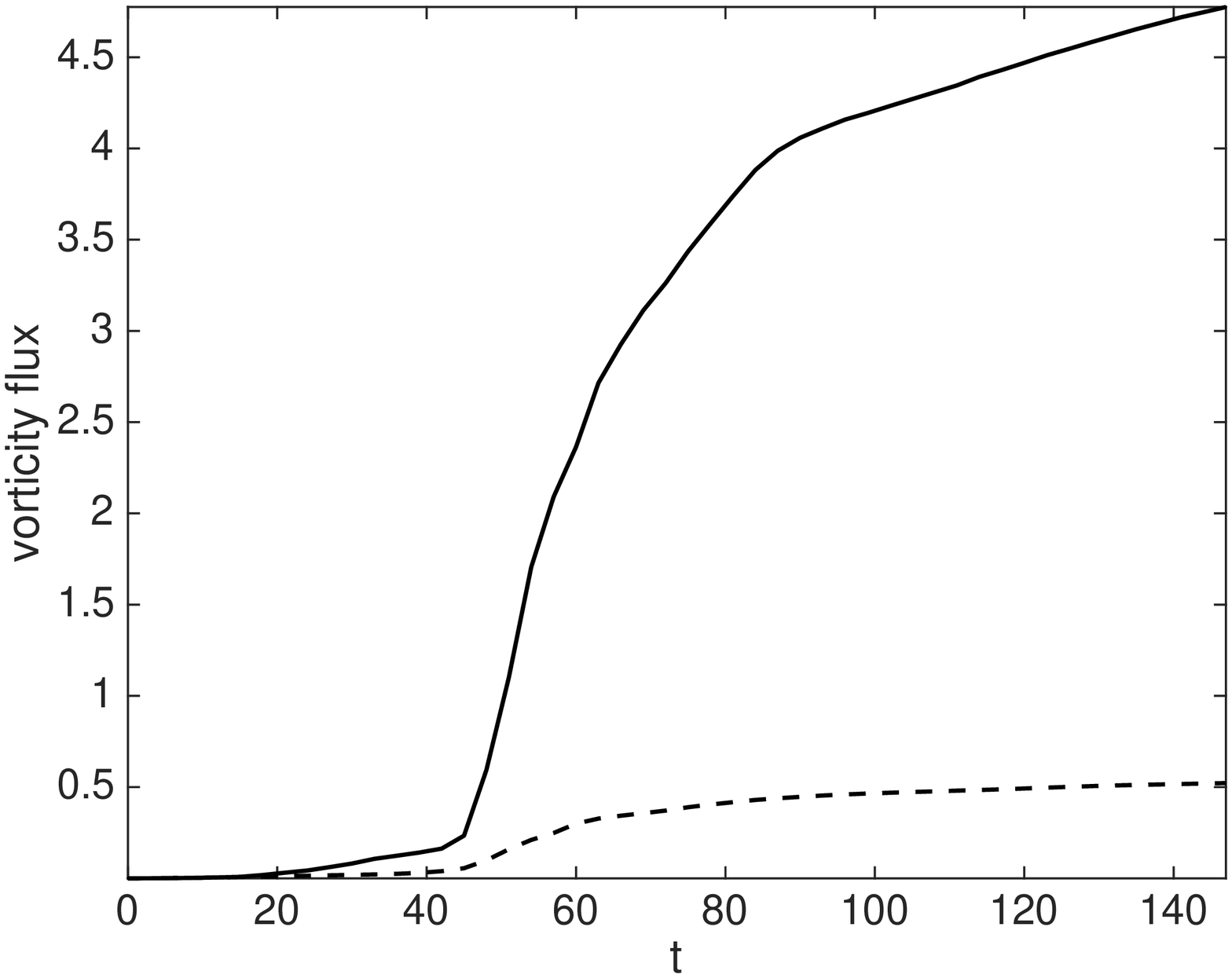}
\caption{Estimates of the cumulative flux reconnected internally within the tubes. Solid curve is the maximum measure of total flux reconnected (See Equation (21) of Wyper and Hesse \cite{2015PhPl...22d2117W}); dashed curve is the minimal measure or `net' flux reconnected (see Equation (19) of \cite{2015PhPl...22d2117W}).}
\label{fig:whcalc}
\end{figure}

\section{Reconnection between three vortex tubes}
\subsection{Simulation setup}

Thus far we have considered the reconnection of two anti-parallel vortex tubes, in which the main reconnection process itself is found to be locally two-dimensional. We complete our study of vortex line geometry during vortex reconnection by studying a fully 3D vortex reconnection process -- motivated by the observation that 2D and 3D reconnection are fundamentally different (see Section \ref{sec:rectheory}).
In order to analyse fully 3D reconnection,  we consider the addition of a third vortex tube to the system.
The additional vortex tube is located perpendicular to both anti-parallel tubes, centred on the $z$-axis where the symmetry and dividing planes intersect, see Figure~\ref{fig:pthighres30fieldlines}(a). This geometry is chosen to provide a non-zero component of $\vort$ along the X-line at which reconnection took place in the previously described simulations.

As before we ensure that a negligible amount of flux connects between the tubes at $t=0$. To ensure this, we choose the perpendicular vortex tube to have a radius of $\sim0.3$ (compared to $\sim0.6$ for the anti-parallel tubes). To maintain the periodic boundary conditions in $x$ and $y$ for computational tractability, the vorticity flux associated with the perpendicular tube must be zero (otherwise there would be a net circulation around the $xy$-boundary). Thus the vortex tube consist of a `core' within which $\omega_z$ has one sign, surrounded by an opposite sign `shell'. 
Specifically, it is constructed by taking cylindrical co-ordinates ($r$, $\phi$, $z$) centred on the $z$-axis, and setting $\vel=v_\phi{\bf e}_\phi$, with
\begin{equation}
v_\phi=\frac{3}{4r}\left(\tanh(50r^2)-\tanh(48r^2)\right).
\end{equation}
To aid discussion the anti-parallel vortex tubes will be referred to as $\atube_1$ and $\atube_2$. The perpendicular vortex tube will be referred to as $\ctube$, and its core and outer shell $\core$ and $\shell$, respectively.
Apart from the addition of $\ctube$, the simulations are identical to those described above. We primarily describe the results of a simulation with $Re=2000$.

\begin{figure*}
\begin{center}
(a)\includegraphics[width=7cm]{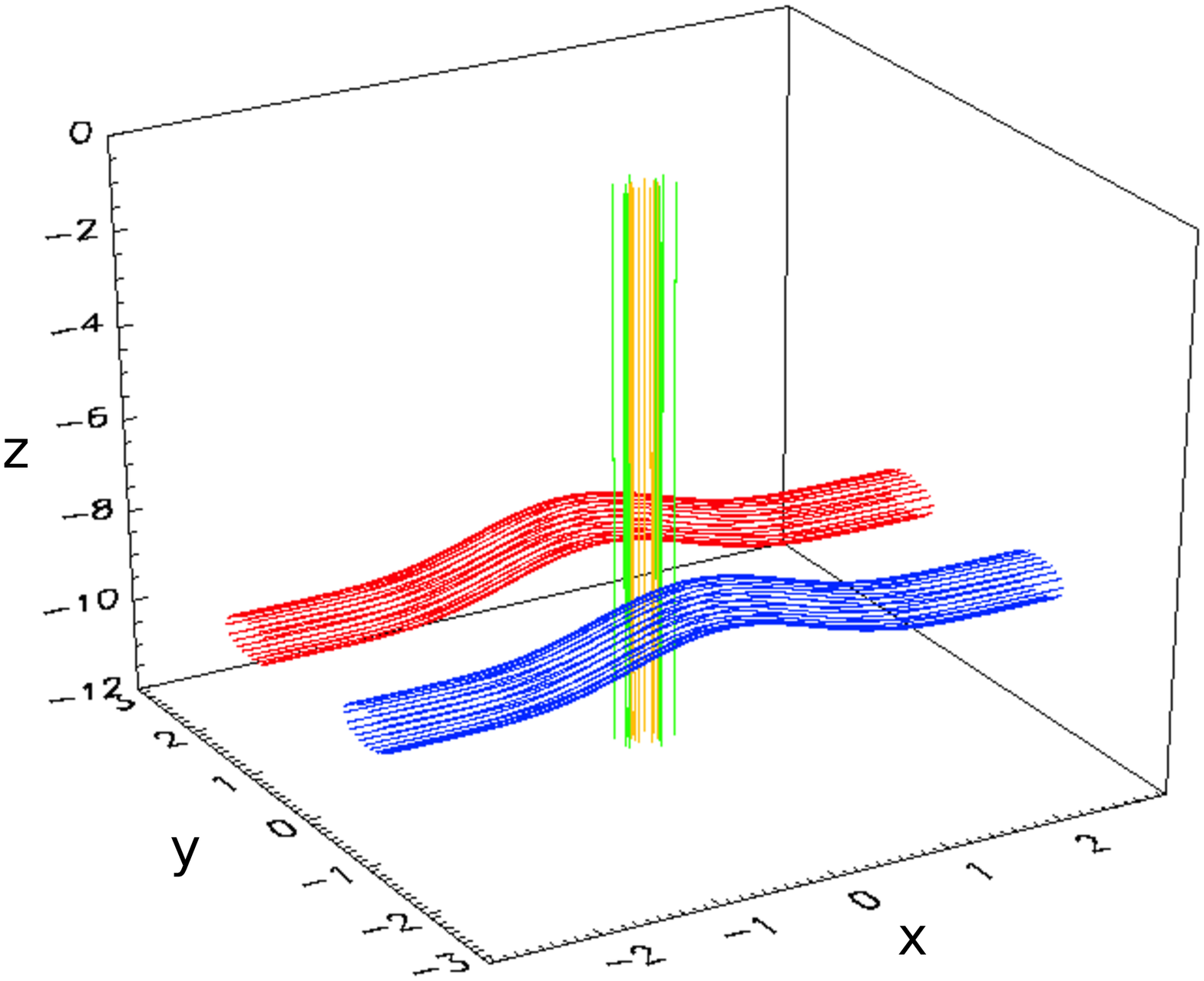}
(b)\includegraphics[width=7cm]{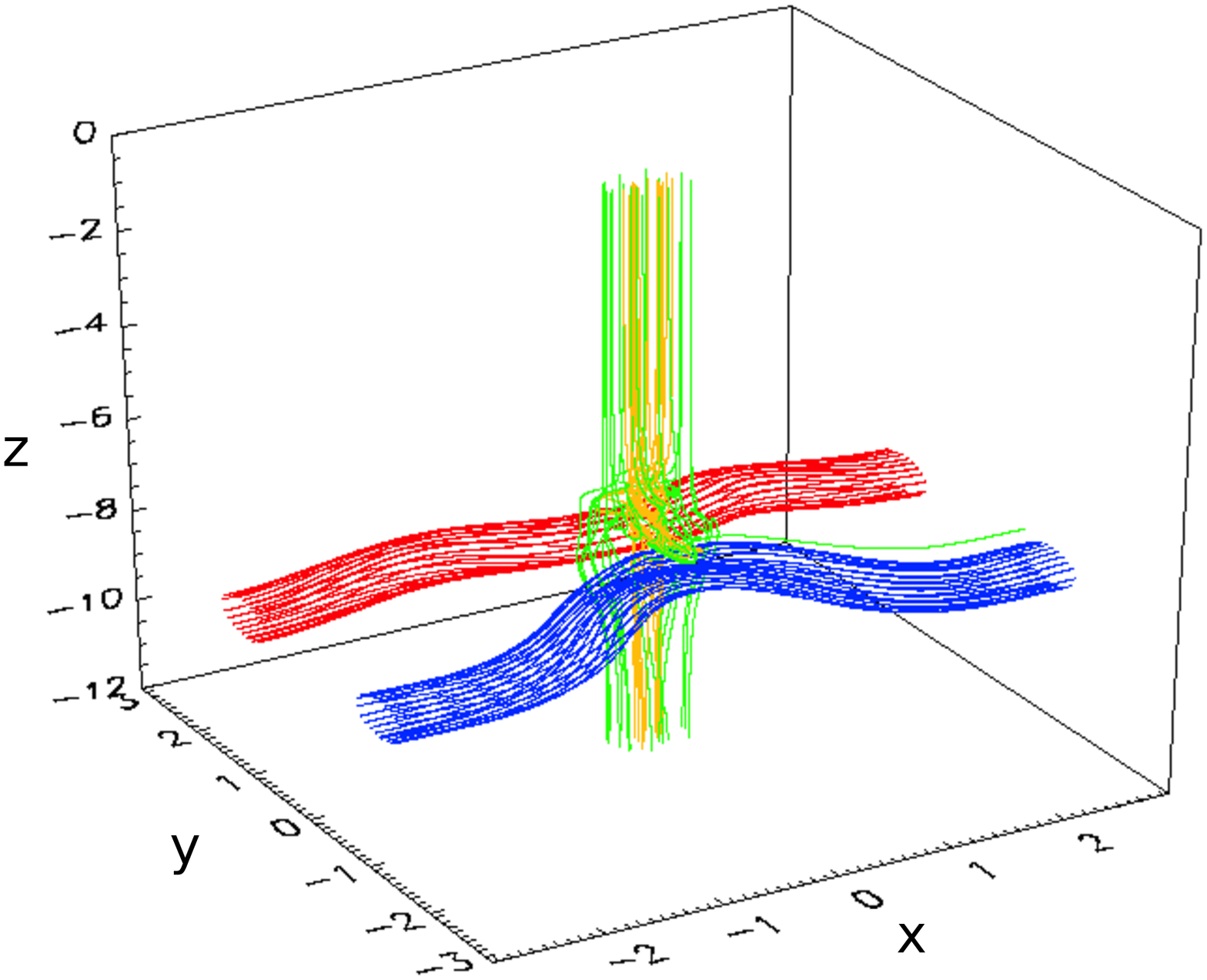}
(c)\includegraphics[width=7cm]{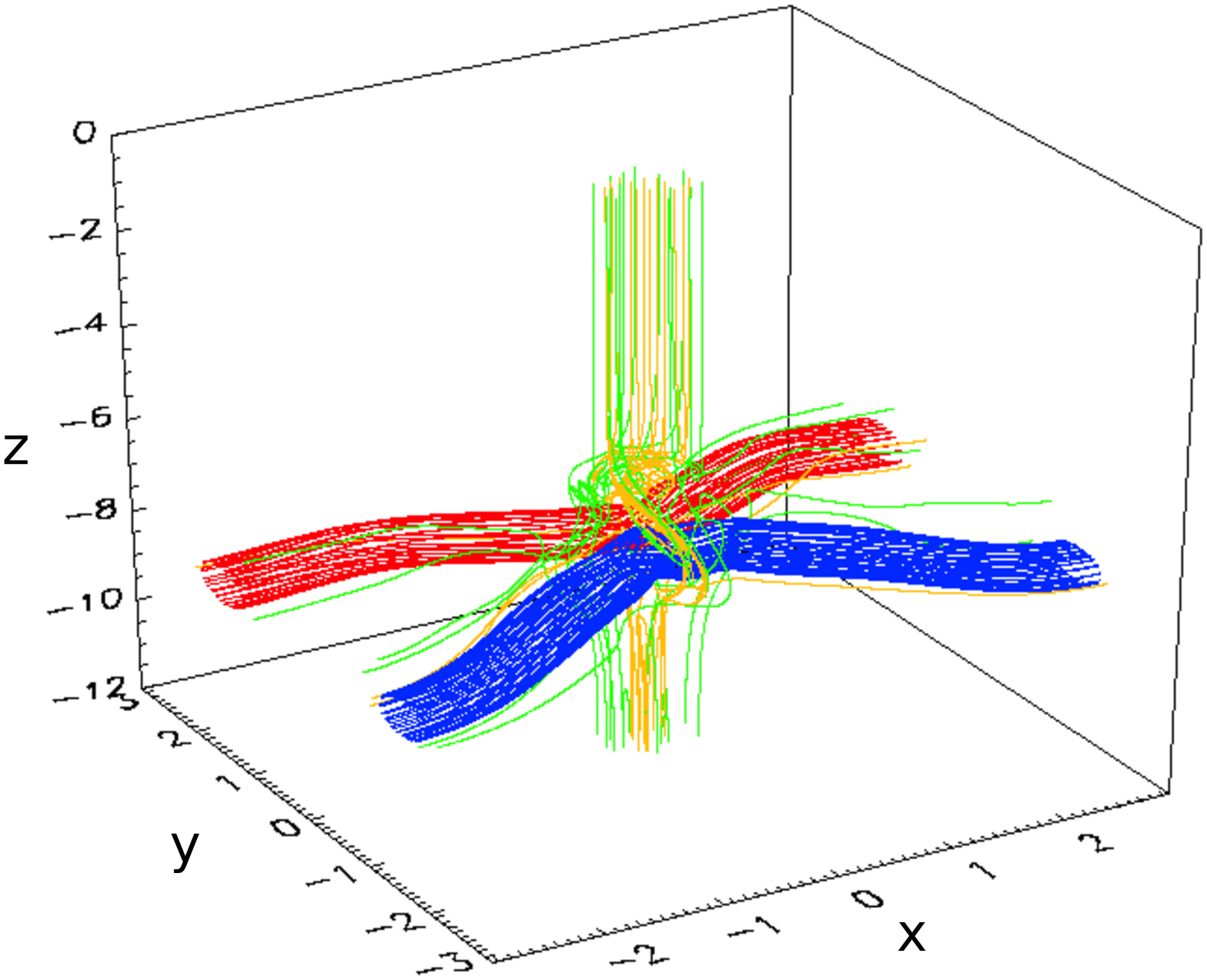}
(d)\includegraphics[width=7cm]{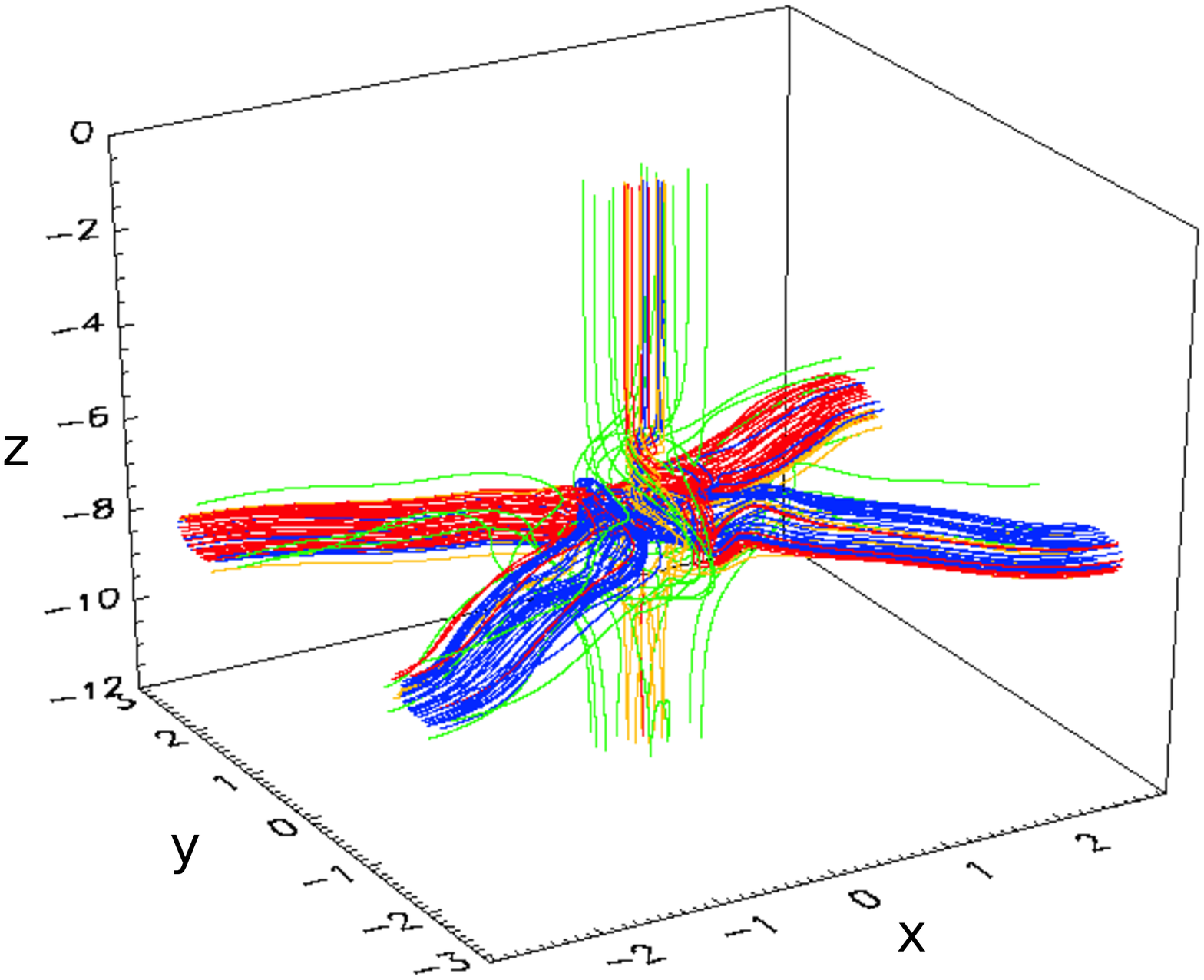}
(e)\includegraphics[width=7cm]{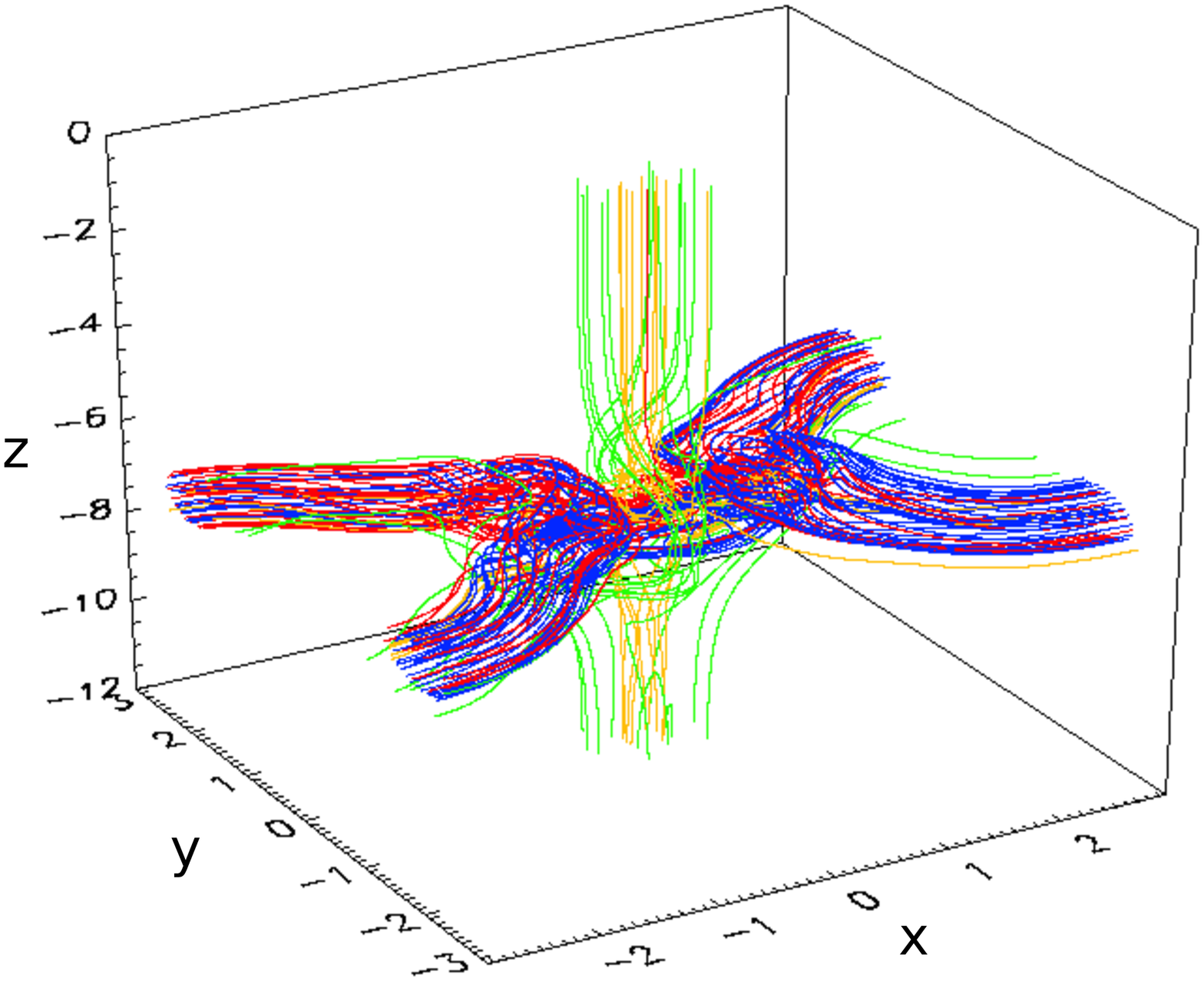}
(f)\includegraphics[width=7cm]{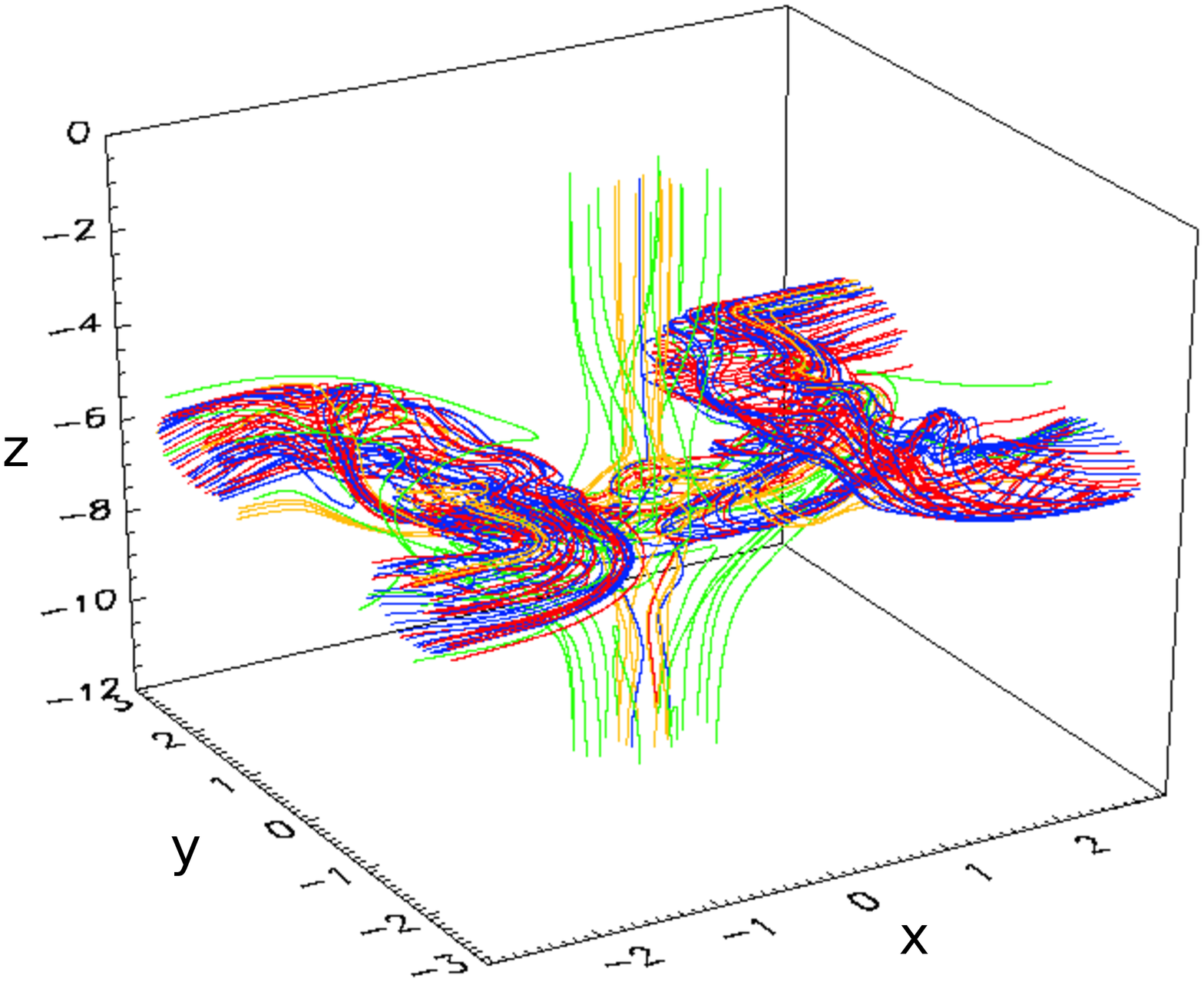}
\caption{$30\%$ of maximum vorticity fieldlines at $x=-3,3$ boundaries (red and blue) and $z=-12,0$ boundaries (green ($\shell$) and yellow ($\core$)) at (a) $t=0$, (b) $t=15$, (c) $t=30$, (d) $t=45$, (e) $t=60$ and (f) $t=90$ for $Re=2000$.}
\label{fig:pthighres30fieldlines}
\end{center}
\end{figure*}

\subsection{Qualitative Evolution and Vorticity Isosurfaces}

The system begins with a surface of zero vorticity between $\core$ and $\shell$ on which $\nabla\times\vort\neq {\bf 0}$, causing the flux of $\ctube$ to begin annihilating, across the null surface between $\core$ and $\shell$ (note that the magnitude of $\ctube$ is sufficiently large that an appreciable $\omega_z$ remains on the $z$-axis when the anti-parallel tubes impinge on that region). As $\atube_1$ and $\atube_2$ rotate towards each other $\ctube$ is squeezed. This breaks the symmetry required for flux annihilation, and instead we have a reconnection process that creates vortex lines connecting between $\core$ and $\shell$, this reconnection occurring at a ring-shaped null line where $\ctube$ is squeezed. However, we will focus our attention on reconnection between $\atube_1$, $\atube_2$ and $\ctube$.
\begin{figure*}
\begin{center}
(a)\includegraphics[width=6.5cm]{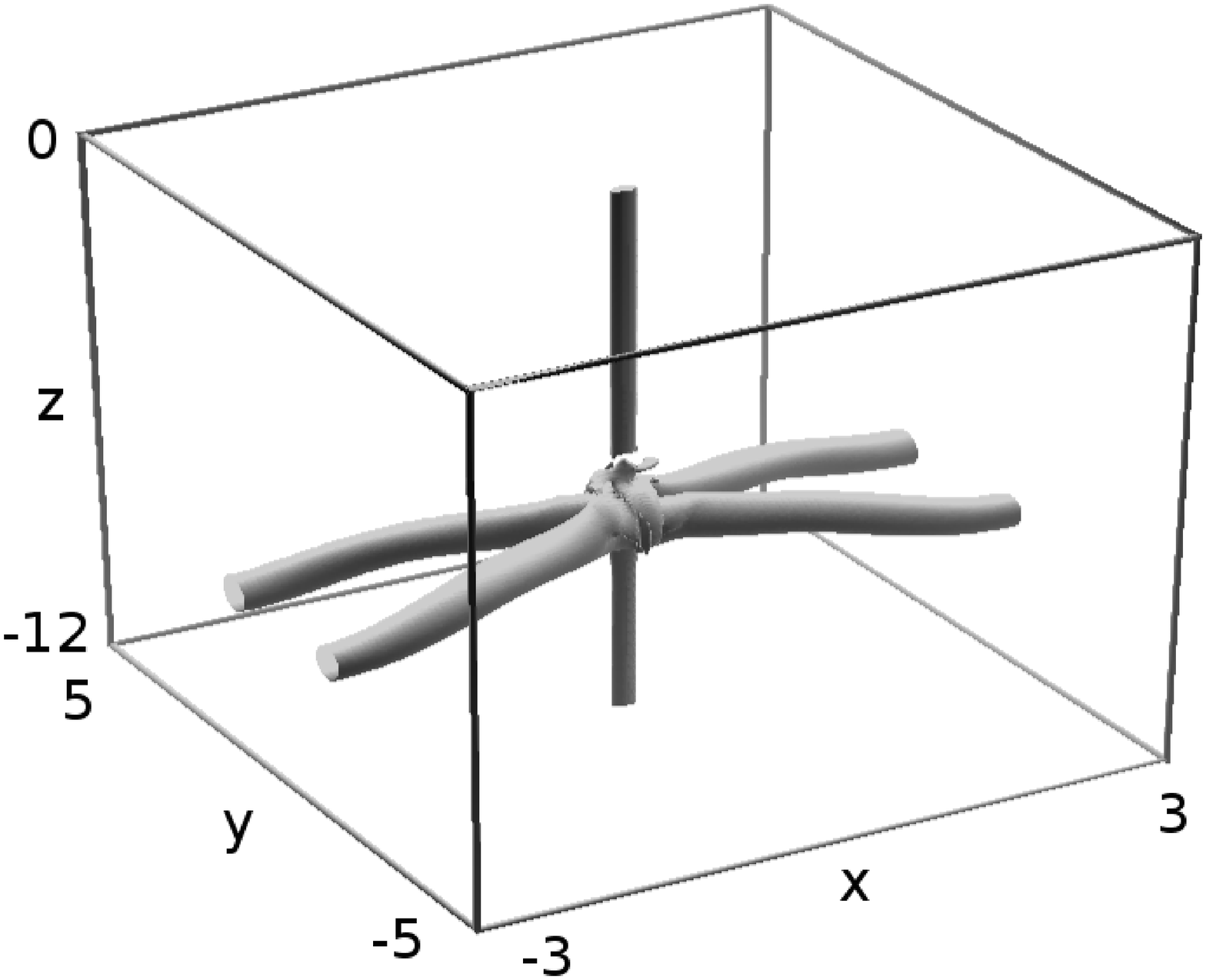}
(b)\includegraphics[width=8.5cm]{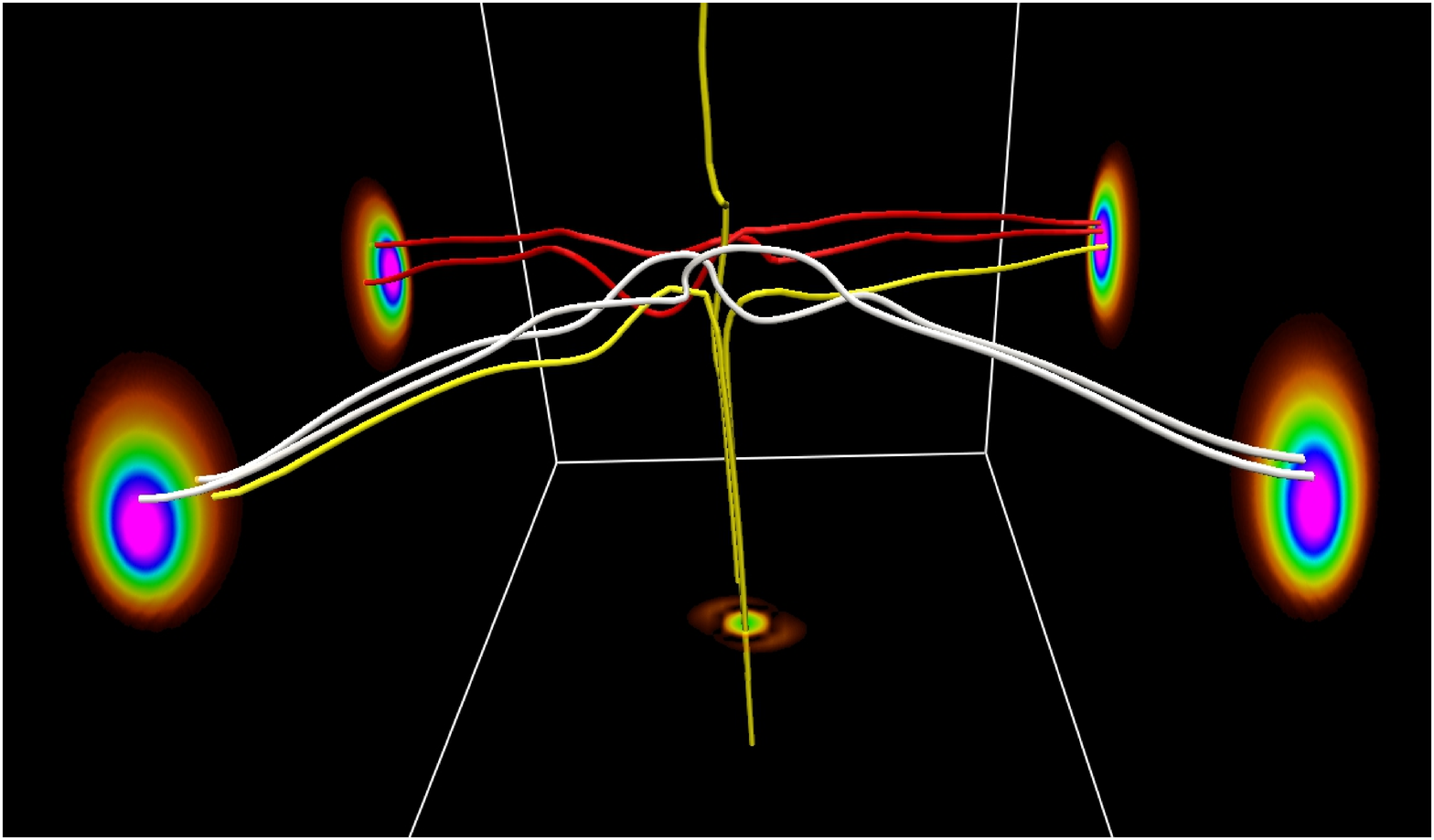}\\
(c)\includegraphics[width=8cm]{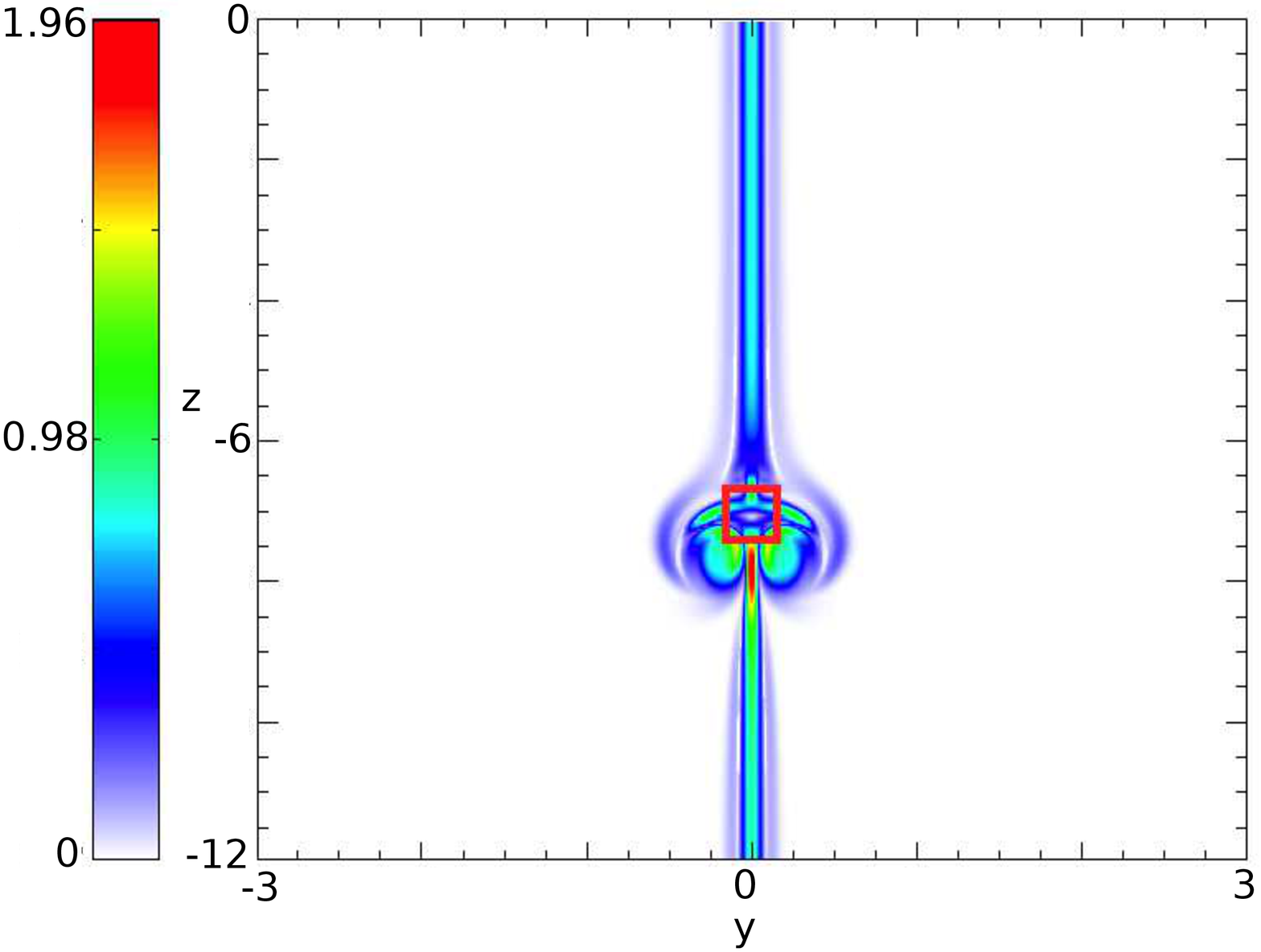}
(d)\includegraphics[width=8cm]{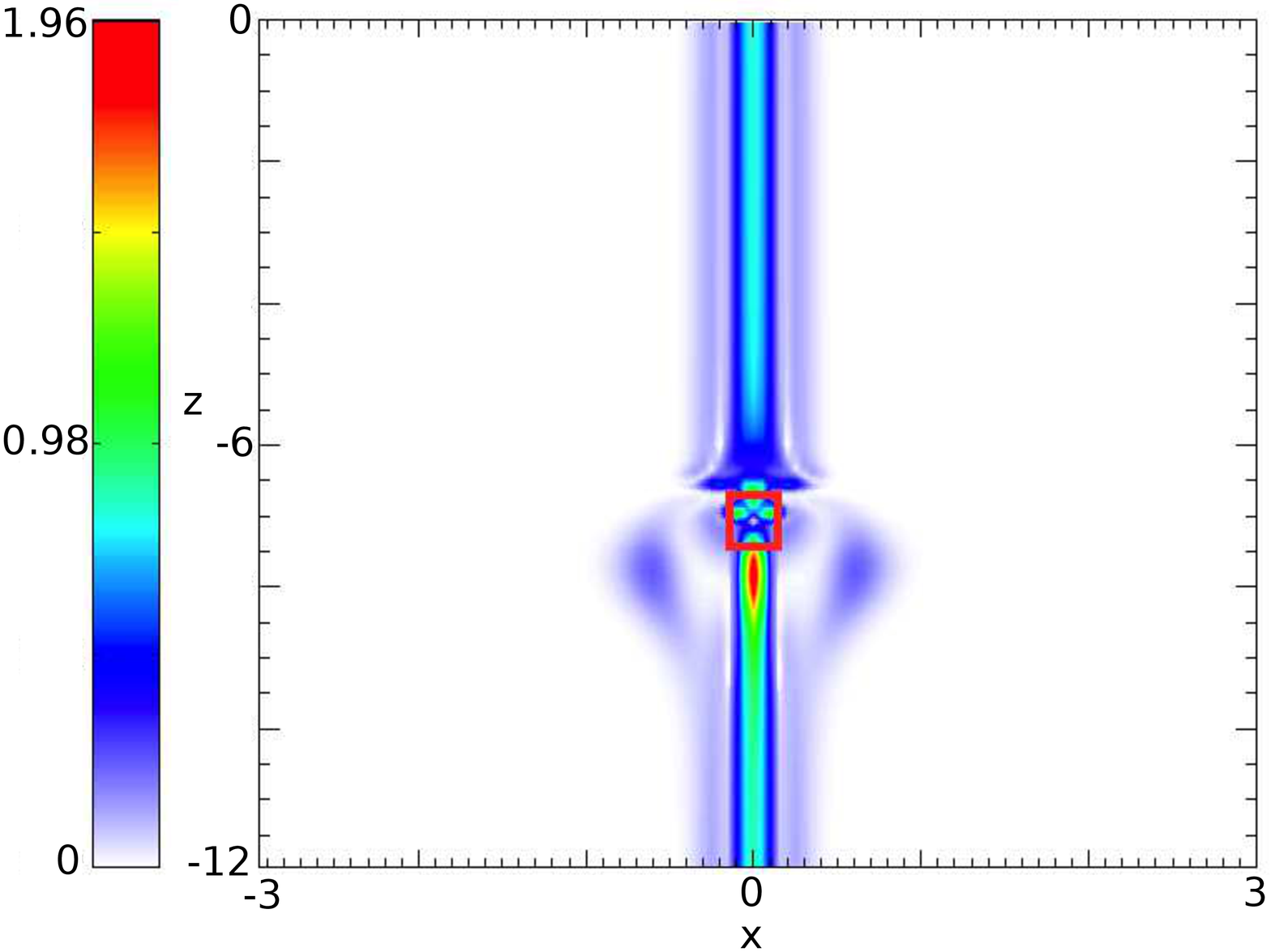}
\caption{(a) Vorticity isosurface at $30\%$ of the maximum vorticity at the $z=0$ boundary at $t=30$. (b) Sample field lines at $t=45$, demonstrating that field lines are not antiparallel prior to reconnection -- boundary shading shows $|\vort|$.
(c) Contours of absolute vorticity in the symmetry plane $x=0$ at $t=24$, and (d) the same plot in the dividing plane. The red square marks the location of the identified vortex nulls.}
\label{fig:ptres30isosurfct}
\end{center}
\end{figure*}
To study the reconnection process we concentrate below on visualising the vortex lines. However, one interesting feature is most clearly seen by looking at the $|\vort|$ isosurfaces. Specifically, the velocity associated with $\ctube$ leads to an asymmetry in $\atube_1$ and $\atube_2$ as they approach, meaning that the vortex sheet will not form exactly in the dividing plane ($y=0$), as shown in Figure~\ref{fig:ptres30isosurfct}(a). In addition, bridge-like structures are formed between $\atube_1$ and $\atube_2$ and $\ctube$.

Further insight can be gained by examining the distribution of $|\vort|$ in the $x=0$ and $y=0$ planes (though note that these are no longer symmetry planes), in Figures \ref{fig:ptres30isosurfct}(c,d). The presence of each of $\atube_1$, $\atube_2$ and $\ctube$ is clear to identify. Also,
the curve in $\shell$ around $z=-7$ explains the bridges seen in Figure~\ref{fig:ptres30isosurfct}(a). The most important feature observed in these plots is a region of very weak vorticity just above (in $z$) $\atube_1$ and $\atube_2$ (marked by a red box), which as we will see below contains a vortex null point. This has implications for the topology of the vorticity field in the vicinity of the reconnection site, see below.

\subsection{Visualising the Reconnection Process - Vorticity Fieldlines}\label{subsec:ptresfieldlines}

In order to visualise the reconnection process we plot vortex lines originating from contours of $|\vort|$ on the $x=\pm3$, $z=-12$ and $z=0$ planes -- Figure \ref{fig:pthighres30fieldlines}. The vortex lines are plotted from contours at 30$\%$ of the individual maximum $|\vort|$ for each of $\atube_1,\atube_2,\core,\shell$.  
By $t=15$ in (b) we see $\ctube$ twisting so that it can reconnect with $\atube_1$ and $\atube_2$ in a configuration with anti-parallel vortex lines \citep{1992PhFlA...4..581B}. At $t=30$ in (c) the fieldlines from $\atube_1$ and $\atube_2$ and $\ctube$ have begun reconnecting but we also see some field lines from $\ctube$  that have reconnected twice, such that they now wrap on the outside of $\atube_1$ and $\atube_2$ (so that they penetrate both top and bottom $z$ boundaries). This double reconnection essentially allows $\atube_1$ and $\atube_2$ to pass through $\ctube$ (in a manner similar to the ``tunnel" interaction of magnetic flux tubes described by \cite{2001ApJ...553..905L}), and allows $\atube_1$ and $\atube_2$ to proceed towards one another. In (d-f) we observe vortex lines from $\atube_1$ and $\atube_2$ connecting to the top of the box close to the central axis: $\atube_1$ and $\atube_2$ are reconnecting with one another, but the vortex lines connect in an intermediate step to $\core$. 

Interestingly, vortex lines no longer lie locally in a plane when they reconnect, but rather are seen to exhibit a finite angle of inclination across the vortex sheet -- see Figure \ref{fig:ptres30isosurfct}(b). This means that the reconnection process is fully 3D, thus implying that vortex lines do not reconnect pairwise along a single line, but rather reconnect throughout a finite volume, defined by $(\nabla\times\vort)\cdot\vort\neq 0$ \citep{2003JGRA..108.1285P}. 

\subsection{Generation of null points}

\begin{figure}
\begin{center}
(a)\includegraphics[width=8cm]{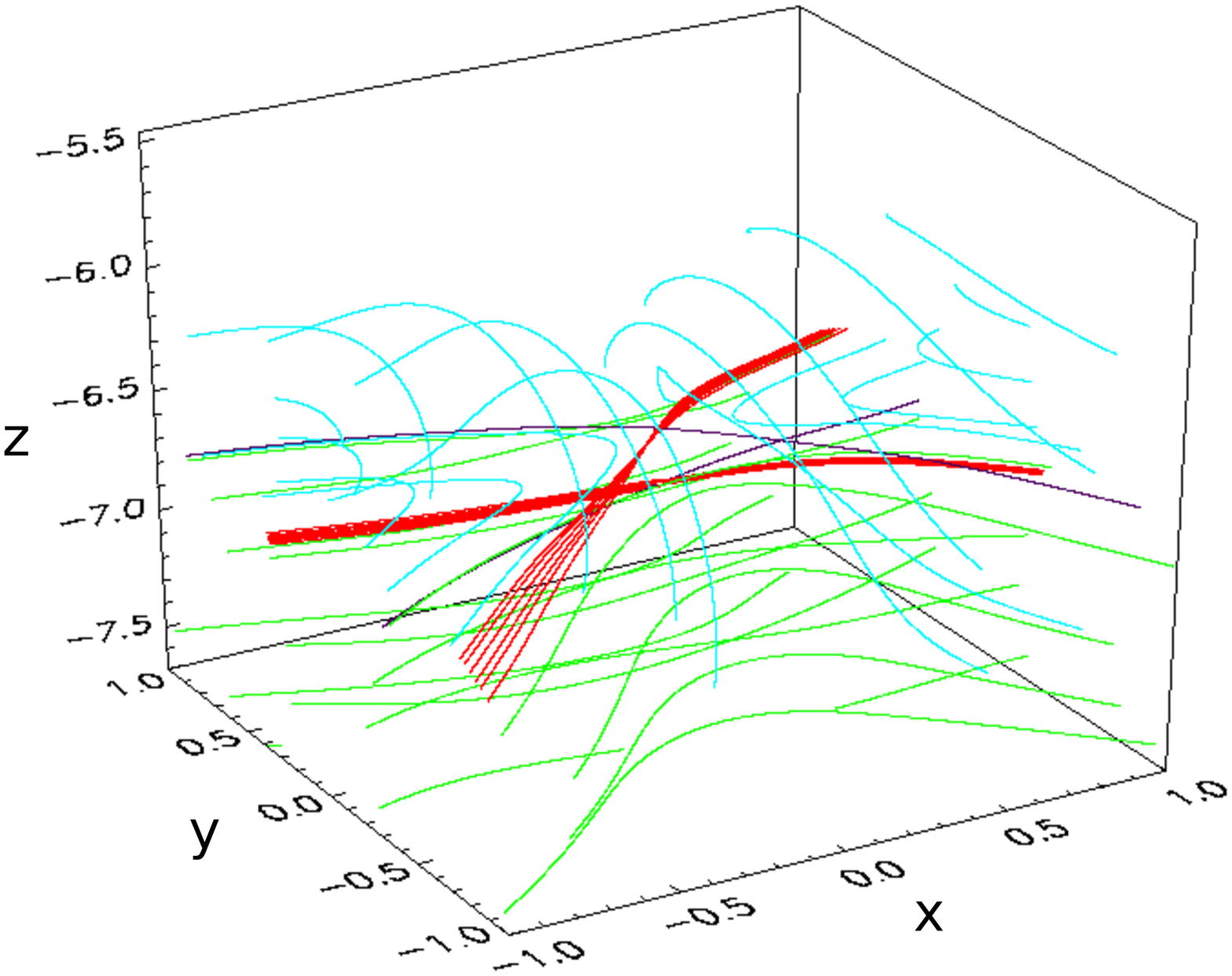}
(b)\includegraphics[width=8.3cm]{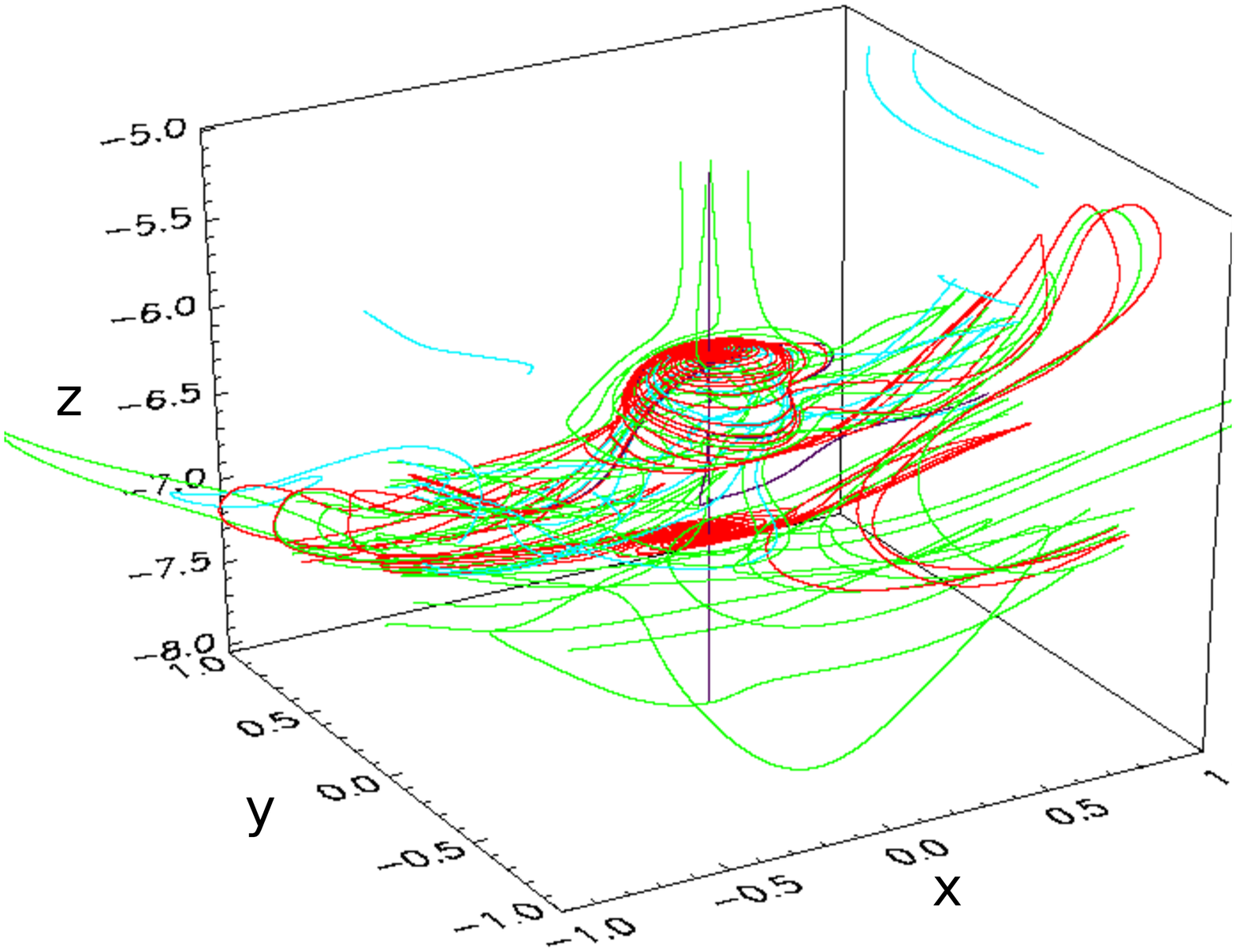}
\caption{$30\%$ of maximum vorticity fieldlines at $x=0$ (green) and $y=0$ (blue) with fans (red) and spines (black) plotted from null points for (a) $Re=800, t=48$, and (b) $Re=2000, t=93$.}
\label{fig:ptresnullpoint16}
\end{center}
\end{figure}

As $\atube_1$ and $\atube_2$ reconnect sequentially with vortex lines of $\ctube$, the thread field lines of $\atube_1$ and $\atube_2$ eventually meet at the $z$-axis. Subsequently, $\atube_1$ and $\atube_2$ begin reconnecting with one another through this central axis. At $t=0$, $\omega_z<0$ along the entire $z$-axis. However, we observe that when $\atube_1$ and $\atube_2$ impinge on the $z$-axis, it leads to the generation of region of $\omega_z>0$ on a segment of the $z$-axis. This coincides with the generation of a pair of vorticity null points of opposite topological degree, consistent with the discussion in Section~\ref{sec:rectheory}. These vorticity nulls are located by applying a numerical implementation of the method described by \cite{haynes2007}. To understand the nature of the vorticity field around the null point pairs we plot their fans and spines (see Section~\ref{sec:rectheory}) together with some neighbouring vortex lines in Figure~\ref{fig:ptresnullpoint16}. 

For the simulation with $Re=800$, the fans and spines are seen to follow the geometry of the reconnecting anti-parallel tubes (Figure \ref{fig:ptresnullpoint16}a). In particular, they lie between the threads and the reconnected vortex rings, and are oriented such that there is a topologically stable separator formed by the intersection of the fan surfaces that connects the nulls (see Section~\ref{sec:rectheory}). With this orientation, there is reconnection occurring at the separator, and what is more it involves the transfer of flux between the flux domains delineated by the associated separatrix surfaces, exactly as in the prototype separator reconnection models for magnetic fields \citep[e.g.][]{lau1990,1993PhFlB...5.2355G}. We also note, however, that the separator field line is relatively short, and that only a fraction of the flux is reconnected through the separator.  

Turning now to the simulation with $Re=2000$, we find a completely different configuration of the vortex lines in the vicinity of the null point pair created (Figure \ref{fig:ptresnullpoint16}b). In particular, calculation of the eignevectors of $\nabla\vort$ at the null reveals that, in contrast to the lower $Re$ simulations, the spines align themselves approximately with the $z$-axis and the fan surfaces form into a nearly closed configuration reminiscent of a `spheromak' geometry \citep{1979NucFu..19..489R} -- though note that they cannot form a completely closed flux surface for any finite time \citep{hu2004,wyper2014b}. It is also worth noting that there are indications of additional pairs of nulls forming at higher values of $Re$.

\subsection{Flux Evolution}

\begin{figure}
\begin{center}
(a)\includegraphics[width=0.45\textwidth]{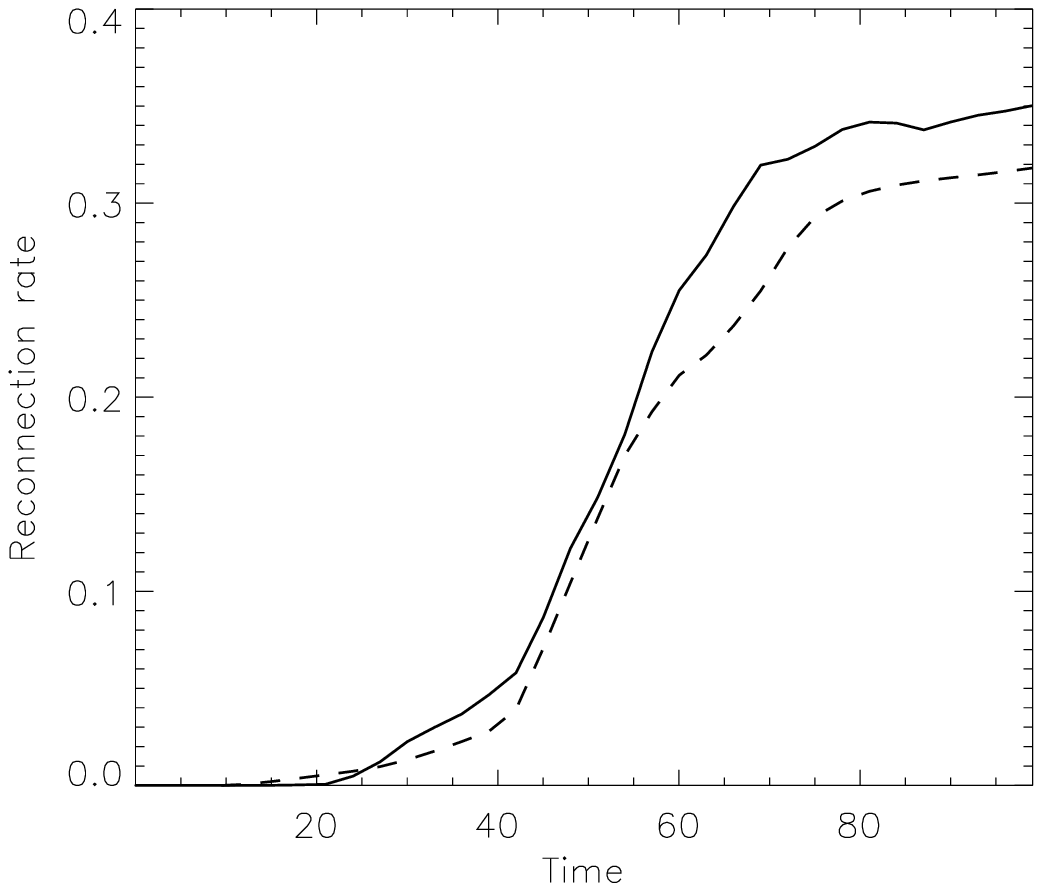}
(b)\includegraphics[width=0.45\textwidth]{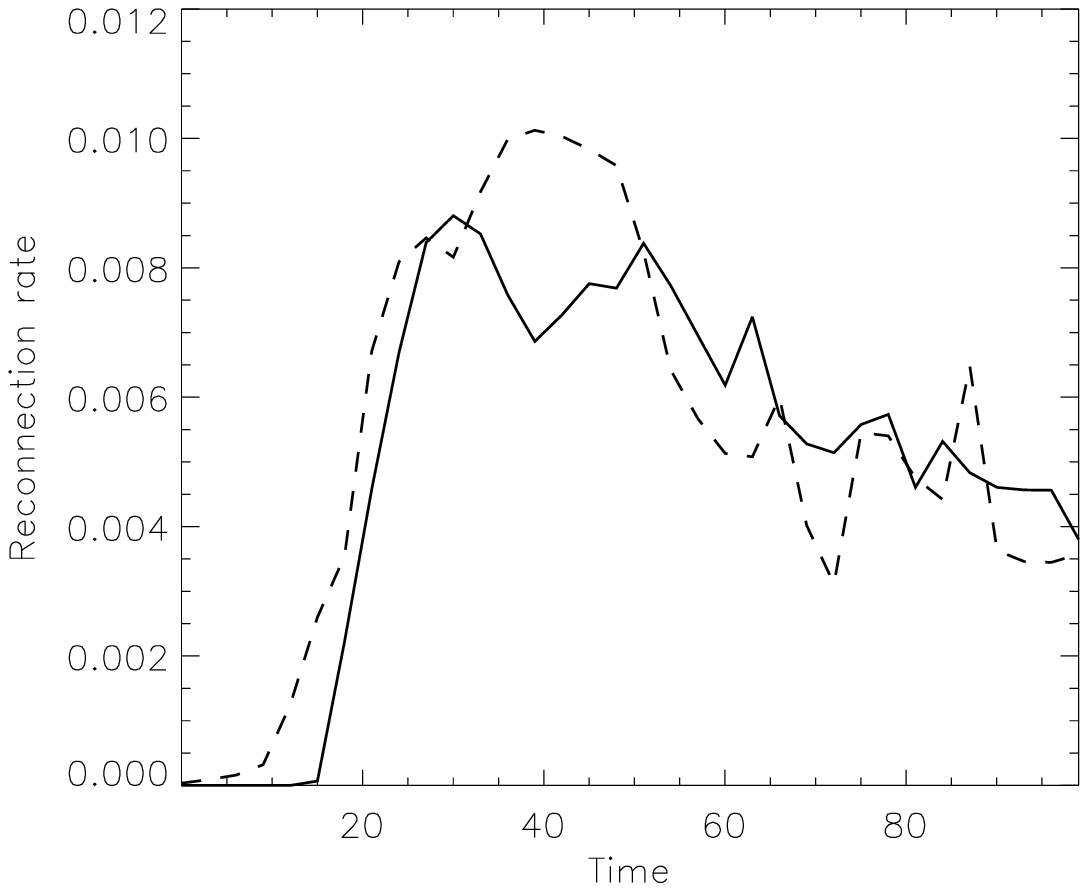}
\caption{(a) Estimated change in flux between $\atube_1$ and $\atube_2$ by integrating $\nu(\nabla\times\vort)_z$ along the central axis (dashed), and direct flux counting (solid line). (b) change in flux between $\atube_1$ and $\core$ (solid line) and between $\atube_1$ and the $\shell$ (dashed).}
\label{fig:ptresvortflux}
\end{center}
\end{figure}

Due to the absence of symmetry, flux measurements in the $x=0$ and $y=0$ planes cannot be used to measure the rate of reconnection. Instead, to measure the flux reconnected between each tube we use a brute force method; we integrate a large number of vortex lines from the footprint of $\atube_1$ on the $x=3$ plane, and determine to which boundary they connect. These are then assigned as having reconnected to $\atube_2$, $\shell$, or $\core$ if they connect to  $x=-3$, $z=0$, $z=-12$, respectively. Fluxes are then estimated by counting the numbers of field lines in each category, weighted by the corresponding local area and $|\vort|$ at $x=3$. The flux between $\atube_1$ and $\atube_2$ is shown in Figure~\ref{fig:ptresvortflux}(a), and compared with the estimated flux reconnected based on the integration of $(\nabla\times\vort)_\|$ along the $z$-axis (dashed line). The discrepancy is due to instabilities that break the symmetry. The curves describing the flux connecting $\atube_1$ to $\core$ and $\shell$ (Figure~\ref{fig:ptresvortflux}b) follow a similar profile after their respective peaks due to the annihilation of $\ctube$. The delay in reconnection between $\atube_1$ and $\core$ is clearly evident.

\section{Conclusions}

In this paper we have studied examples of both 2D and 3D reconnection processes, occuring between vortex tubes. The focus was on the topology of vortex lines during the reconnection process (in contrast to previous studies that analysed mainly isosurfaces of the vorticity magnitude). Our analyses provide new insights into the interaction of vortex tubes, and in particular reveal that the topology of the vorticity field during the process is more complex than originally appreciated.  

Considering first the interaction of an isolated pair of anti-parallel vortex tubes, we noted features observed previously by various authors, including the onset of the Kelvin-Helmholtz instability, and the choking off of the reconnection by thread curvature. The main new results are as follows.
\begin{enumerate}
\item
We have identified the generation of many small flux rings, formed when reconnection occurs simultaneously at multiple locations in the vortex sheet between the tubes. The rings form with varying sizes during the interaction, and are more numerous at higher $Re$. While such rings have been identified in the past, principally from isosurface plots of $|\vort|$ \cite{zabusky1991,1995AcMSn..11..209W}, this is to the best of our knowledge the first time their flux has been quantified.
\item
We demonstrated the link between regions of localised twist of the vortex lines and kinetic helicity density oscillations.
\item
Consideration of three-dimensional reconnection principles leads us to describe how to correctly measure the reconnection rate, even once instabilities break the symmetry. It also allows us to identify for the first time internal `slipping' reconnection within the vortex tubes due to buildup of twist gradients. At $Re=4000$, we find that on average all flux is `internally' reconnected once within each tube.
\end{enumerate}
These results permit a deeper understanding of the reconnection process, and a systematic study at higher Reynolds numbers would be an interesting future extension.

The introduction of a third vortex tube perpendicular to the anti-parallel tubes was found to render the vorticity field in the vicinity of the reconnection site fully three-dimensional. The main results from these simulations are as follows. 
\begin{enumerate}
\item
Vortex lines no longer lie locally in a plane when they reconnect, but rather exhibit some finite angle of inclination across the vortex sheet. This implies that vortex lines no longer reconnect pairwise along a single line, but rather reconnect throughout some finite volume, defined by $(\nabla\times\vort)\cdot\vort\neq 0$ \citep{2003JGRA..108.1285P}.
\item
We noted the generation of pairs of vorticity null points during the reconnection process, together with associated separator field lines. The generation of vorticity null points is of interest given their relevance for the problem of finite-time blow-up \citep[e.g.][]{bhattacharjee1992,pelz2002}.
\end{enumerate}
In future it would be of interest to study a configuration in which the perpendicular tube has no `return vorticity' shell (this requiring a change to the boundary conditions), and to investigate the formation of vortex nulls for different parameters. The exchange of kinetic helicity between the vortex tubes, as well as the dependence on $Re$ should also be addressed in future. 


\begin{acknowledgments}
The authors gratefully acknowledge financial support from the Leverhulme Trust and EPSRC, and helpful discussions with G.~Hornig.
\end{acknowledgments}


\begin{thebibliography}{56}%
\makeatletter
\providecommand \@ifxundefined [1]{%
 \@ifx{#1\undefined}
}%
\providecommand \@ifnum [1]{%
 \ifnum #1\expandafter \@firstoftwo
 \else \expandafter \@secondoftwo
 \fi
}%
\providecommand \@ifx [1]{%
 \ifx #1\expandafter \@firstoftwo
 \else \expandafter \@secondoftwo
 \fi
}%
\providecommand \natexlab [1]{#1}%
\providecommand \enquote  [1]{``#1''}%
\providecommand \bibnamefont  [1]{#1}%
\providecommand \bibfnamefont [1]{#1}%
\providecommand \citenamefont [1]{#1}%
\providecommand \href@noop [0]{\@secondoftwo}%
\providecommand \href [0]{\begingroup \@sanitize@url \@href}%
\providecommand \@href[1]{\@@startlink{#1}\@@href}%
\providecommand \@@href[1]{\endgroup#1\@@endlink}%
\providecommand \@sanitize@url [0]{\catcode `\\12\catcode `\$12\catcode
  `\&12\catcode `\#12\catcode `\^12\catcode `\_12\catcode `\%12\relax}%
\providecommand \@@startlink[1]{}%
\providecommand \@@endlink[0]{}%
\providecommand \url  [0]{\begingroup\@sanitize@url \@url }%
\providecommand \@url [1]{\endgroup\@href {#1}{\urlprefix }}%
\providecommand \urlprefix  [0]{URL }%
\providecommand \Eprint [0]{\href }%
\providecommand \doibase [0]{http://dx.doi.org/}%
\providecommand \selectlanguage [0]{\@gobble}%
\providecommand \bibinfo  [0]{\@secondoftwo}%
\providecommand \bibfield  [0]{\@secondoftwo}%
\providecommand \translation [1]{[#1]}%
\providecommand \BibitemOpen [0]{}%
\providecommand \bibitemStop [0]{}%
\providecommand \bibitemNoStop [0]{.\EOS\space}%
\providecommand \EOS [0]{\spacefactor3000\relax}%
\providecommand \BibitemShut  [1]{\csname bibitem#1\endcsname}%
\let\auto@bib@innerbib\@empty
\bibitem [{\citenamefont {{Crow}}(1970)}]{1970AIAAJ...8.2172C}%
  \BibitemOpen
  \bibfield  {author} {\bibinfo {author} {\bibfnamefont {S.~C.}\ \bibnamefont
  {{Crow}}},\ }\bibfield  {title} {\enquote {\bibinfo {title} {{Stability
  theory for a pair of trailing vortices}},}\ }\href {\doibase 10.2514/3.6083}
  {\bibfield  {journal} {\bibinfo  {journal} {AIAA Journal}\ }\textbf {\bibinfo
  {volume} {8}},\ \bibinfo {pages} {2172--2179} (\bibinfo {year}
  {1970})}\BibitemShut {NoStop}%
\bibitem [{\citenamefont {{Spalart}}(1998)}]{1998AnRFM..30..107S}%
  \BibitemOpen
  \bibfield  {author} {\bibinfo {author} {\bibfnamefont {P.~R.}\ \bibnamefont
  {{Spalart}}},\ }\bibfield  {title} {\enquote {\bibinfo {title} {{Airplane
  Trailing Vortices}},}\ }\href {\doibase 10.1146/annurev.fluid.30.1.107}
  {\bibfield  {journal} {\bibinfo  {journal} {Annual Review of Fluid
  Mechanics}\ }\textbf {\bibinfo {volume} {30}},\ \bibinfo {pages} {107--138}
  (\bibinfo {year} {1998})}\BibitemShut {NoStop}%
\bibitem [{\citenamefont {{Kida}}\ \emph {et~al.}(1989)\citenamefont {{Kida}},
  \citenamefont {{Takaoka}},\ and\ \citenamefont
  {{Hussain}}}]{1989PhFlA...1..630K}%
  \BibitemOpen
  \bibfield  {author} {\bibinfo {author} {\bibfnamefont {S.}~\bibnamefont
  {{Kida}}}, \bibinfo {author} {\bibfnamefont {M.}~\bibnamefont {{Takaoka}}}, \
  and\ \bibinfo {author} {\bibfnamefont {F.}~\bibnamefont {{Hussain}}},\
  }\bibfield  {title} {\enquote {\bibinfo {title} {{Reconnection of two vortex
  rings}},}\ }\href {\doibase 10.1063/1.857436} {\bibfield  {journal} {\bibinfo
   {journal} {Physics of Fluids A}\ }\textbf {\bibinfo {volume} {1}},\ \bibinfo
  {pages} {630--632} (\bibinfo {year} {1989})}\BibitemShut {NoStop}%
\bibitem [{\citenamefont {{Hussain}}\ and\ \citenamefont
  {{Duraisamy}}(2011)}]{2011PhFl...23b1701H}%
  \BibitemOpen
  \bibfield  {author} {\bibinfo {author} {\bibfnamefont {F.}~\bibnamefont
  {{Hussain}}}\ and\ \bibinfo {author} {\bibfnamefont {K.}~\bibnamefont
  {{Duraisamy}}},\ }\bibfield  {title} {\enquote {\bibinfo {title} {{Mechanics
  of viscous vortex reconnection}},}\ }\href {\doibase 10.1063/1.3532039}
  {\bibfield  {journal} {\bibinfo  {journal} {Physics of Fluids}\ }\textbf
  {\bibinfo {volume} {23}},\ \bibinfo {pages} {021701--021701} (\bibinfo {year}
  {2011})}\BibitemShut {NoStop}%
\bibitem [{\citenamefont {{Kleckner}}\ and\ \citenamefont
  {{Irvine}}(2013)}]{2013NatPh...9..253K}%
  \BibitemOpen
  \bibfield  {author} {\bibinfo {author} {\bibfnamefont {D.}~\bibnamefont
  {{Kleckner}}}\ and\ \bibinfo {author} {\bibfnamefont {W.~T.~M.}\ \bibnamefont
  {{Irvine}}},\ }\bibfield  {title} {\enquote {\bibinfo {title} {{Creation and
  dynamics of knotted vortices}},}\ }\href {\doibase 10.1038/nphys2560}
  {\bibfield  {journal} {\bibinfo  {journal} {Nature Physics}\ }\textbf
  {\bibinfo {volume} {9}},\ \bibinfo {pages} {253--258} (\bibinfo {year}
  {2013})}\BibitemShut {NoStop}%
\bibitem [{\citenamefont {{Pumir}}\ and\ \citenamefont
  {{Kerr}}(1987)}]{1987PhRvL..58.1636P}%
  \BibitemOpen
  \bibfield  {author} {\bibinfo {author} {\bibfnamefont {A.}~\bibnamefont
  {{Pumir}}}\ and\ \bibinfo {author} {\bibfnamefont {R.~M.}\ \bibnamefont
  {{Kerr}}},\ }\bibfield  {title} {\enquote {\bibinfo {title} {{Numerical
  simulation of interacting vortex tubes}},}\ }\href {\doibase
  10.1103/PhysRevLett.58.1636} {\bibfield  {journal} {\bibinfo  {journal}
  {Physical Review Letters}\ }\textbf {\bibinfo {volume} {58}},\ \bibinfo
  {pages} {1636--1639} (\bibinfo {year} {1987})}\BibitemShut {NoStop}%
\bibitem [{\citenamefont {{Melander}}\ and\ \citenamefont
  {{Hussain}}(1989{\natexlab{a}})}]{1989PhFl....1..633M}%
  \BibitemOpen
  \bibfield  {author} {\bibinfo {author} {\bibfnamefont {M.~V.}\ \bibnamefont
  {{Melander}}}\ and\ \bibinfo {author} {\bibfnamefont {F.}~\bibnamefont
  {{Hussain}}},\ }\bibfield  {title} {\enquote {\bibinfo {title}
  {{Cross-linking of two antiparallel vortex tubes}},}\ }\href {\doibase
  10.1063/1.857437} {\bibfield  {journal} {\bibinfo  {journal} {Physics of
  Fluids}\ }\textbf {\bibinfo {volume} {1}},\ \bibinfo {pages} {633--636}
  (\bibinfo {year} {1989}{\natexlab{a}})}\BibitemShut {NoStop}%
\bibitem [{\citenamefont {{Melander}}\ and\ \citenamefont
  {{Hussain}}(1989{\natexlab{b}})}]{iutam1989melander}%
  \BibitemOpen
  \bibfield  {author} {\bibinfo {author} {\bibfnamefont {M.~V.}\ \bibnamefont
  {{Melander}}}\ and\ \bibinfo {author} {\bibfnamefont {F.}~\bibnamefont
  {{Hussain}}},\ }\bibfield  {title} {\enquote {\bibinfo {title} {{Topological
  Aspects of Vortex Reconnection}},}\ }in\ \href@noop {} {\emph {\bibinfo
  {booktitle} {Topological Fluid Mechanics: Proceedings of the IUTAM
  Symposium}}},\ \bibinfo {editor} {edited by\ \bibinfo {editor} {\bibfnamefont
  {H.K.}\ \bibnamefont {Moffatt}}\ and\ \bibinfo {editor} {\bibfnamefont
  {A.}~\bibnamefont {Tsinober}}}\ (\bibinfo {year} {1989})\ pp.\ \bibinfo
  {pages} {485--499}\BibitemShut {NoStop}%
\bibitem [{\citenamefont {Kida}\ and\ \citenamefont
  {Takaoka}(1987)}]{kida1987}%
  \BibitemOpen
  \bibfield  {author} {\bibinfo {author} {\bibfnamefont {S}~\bibnamefont
  {Kida}}\ and\ \bibinfo {author} {\bibfnamefont {M}~\bibnamefont {Takaoka}},\
  }\bibfield  {title} {\enquote {\bibinfo {title} {{Bridging in vortex
  reconnection}},}\ }\href@noop {} {\bibfield  {journal} {\bibinfo  {journal}
  {Phys Fluids}\ }\textbf {\bibinfo {volume} {30}},\ \bibinfo {pages} {2911--5}
  (\bibinfo {year} {1987})}\BibitemShut {NoStop}%
\bibitem [{\citenamefont {{Virk}}\ \emph {et~al.}(1995)\citenamefont {{Virk}},
  \citenamefont {{Hussain}},\ and\ \citenamefont
  {{Kerr}}}]{1995JFM...304...47V}%
  \BibitemOpen
  \bibfield  {author} {\bibinfo {author} {\bibfnamefont {D.}~\bibnamefont
  {{Virk}}}, \bibinfo {author} {\bibfnamefont {F.}~\bibnamefont {{Hussain}}}, \
  and\ \bibinfo {author} {\bibfnamefont {R.~M.}\ \bibnamefont {{Kerr}}},\
  }\bibfield  {title} {\enquote {\bibinfo {title} {{Compressible vortex
  reconnection}},}\ }\href {\doibase 10.1017/S0022112095004344} {\bibfield
  {journal} {\bibinfo  {journal} {Journal of Fluid Mechanics}\ }\textbf
  {\bibinfo {volume} {304}},\ \bibinfo {pages} {47--86} (\bibinfo {year}
  {1995})}\BibitemShut {NoStop}%
\bibitem [{\citenamefont {{van Rees}}\ \emph {et~al.}(2012)\citenamefont {{van
  Rees}}, \citenamefont {{Hussain}},\ and\ \citenamefont
  {{Koumoutsakos}}}]{2012PhFl...24g5105V}%
  \BibitemOpen
  \bibfield  {author} {\bibinfo {author} {\bibfnamefont {W.~M.}\ \bibnamefont
  {{van Rees}}}, \bibinfo {author} {\bibfnamefont {F.}~\bibnamefont
  {{Hussain}}}, \ and\ \bibinfo {author} {\bibfnamefont {P.}~\bibnamefont
  {{Koumoutsakos}}},\ }\bibfield  {title} {\enquote {\bibinfo {title} {{Vortex
  tube reconnection at Re = 10$^{4}$}},}\ }\href {\doibase 10.1063/1.4731809}
  {\bibfield  {journal} {\bibinfo  {journal} {Physics of Fluids}\ }\textbf
  {\bibinfo {volume} {24}},\ \bibinfo {pages} {075105--075105} (\bibinfo {year}
  {2012})}\BibitemShut {NoStop}%
\bibitem [{\citenamefont {{Ashurst}}\ and\ \citenamefont
  {{Meiron}}(1987)}]{1987PhRvL..58.1632A}%
  \BibitemOpen
  \bibfield  {author} {\bibinfo {author} {\bibfnamefont {W.~T.}\ \bibnamefont
  {{Ashurst}}}\ and\ \bibinfo {author} {\bibfnamefont {D.~I.}\ \bibnamefont
  {{Meiron}}},\ }\bibfield  {title} {\enquote {\bibinfo {title} {{Numerical
  study of vortex reconnection}},}\ }\href {\doibase
  10.1103/PhysRevLett.58.1632} {\bibfield  {journal} {\bibinfo  {journal}
  {Physical Review Letters}\ }\textbf {\bibinfo {volume} {58}},\ \bibinfo
  {pages} {1632--1635} (\bibinfo {year} {1987})}\BibitemShut {NoStop}%
\bibitem [{\citenamefont {Scheeler}\ \emph {et~al.}(2017)\citenamefont
  {Scheeler}, \citenamefont {van Rees}, \citenamefont {Kedia}, \citenamefont
  {Kleckner},\ and\ \citenamefont {Irvine}}]{scheeler2017}%
  \BibitemOpen
  \bibfield  {author} {\bibinfo {author} {\bibfnamefont {Martin~W.}\
  \bibnamefont {Scheeler}}, \bibinfo {author} {\bibfnamefont {Wim~M.}\
  \bibnamefont {van Rees}}, \bibinfo {author} {\bibfnamefont {Hridesh}\
  \bibnamefont {Kedia}}, \bibinfo {author} {\bibfnamefont {Dustin}\
  \bibnamefont {Kleckner}}, \ and\ \bibinfo {author} {\bibfnamefont {William
  T.~M.}\ \bibnamefont {Irvine}},\ }\bibfield  {title} {\enquote {\bibinfo
  {title} {Complete measurement of helicity and its dynamics in vortex
  tubes},}\ }\href {\doibase 10.1126/science.aam6897} {\bibfield  {journal}
  {\bibinfo  {journal} {Science}\ }\textbf {\bibinfo {volume} {357}},\ \bibinfo
  {pages} {487--491} (\bibinfo {year} {2017})}\BibitemShut {NoStop}%
\bibitem [{\citenamefont {{Priest}}\ \emph {et~al.}(2003)\citenamefont
  {{Priest}}, \citenamefont {{Hornig}},\ and\ \citenamefont
  {{Pontin}}}]{2003JGRA..108.1285P}%
  \BibitemOpen
  \bibfield  {author} {\bibinfo {author} {\bibfnamefont {E.~R.}\ \bibnamefont
  {{Priest}}}, \bibinfo {author} {\bibfnamefont {G.}~\bibnamefont {{Hornig}}},
  \ and\ \bibinfo {author} {\bibfnamefont {D.~I.}\ \bibnamefont {{Pontin}}},\
  }\bibfield  {title} {\enquote {\bibinfo {title} {{On the nature of
  three-dimensional magnetic reconnection}},}\ }\href {\doibase
  10.1029/2002JA009812} {\bibfield  {journal} {\bibinfo  {journal} {J.
  Geophys. Res. (Space Phys)}\ }\textbf {\bibinfo {volume} {108}},\
  \bibinfo {eid} {1285} (\bibinfo {year} {2003})}\BibitemShut {NoStop}%
\bibitem [{\citenamefont {Birn}\ and\ \citenamefont {Priest}(2007)}]{birn2007}%
  \BibitemOpen
  \bibinfo {editor} {\bibfnamefont {J.}~\bibnamefont {Birn}}\ and\ \bibinfo
  {editor} {\bibfnamefont {E.~R.}\ \bibnamefont {Priest}},\ eds.,\ \href@noop
  {} {\emph {\bibinfo {title} {Reconnection of Magnetic Fields :
  Magnetohydrodynamics and Collisionless Theory and Observations}}}\ (\bibinfo
  {publisher} {Cambridge University Press},\ \bibinfo {year}
  {2007})\BibitemShut {NoStop}%
\bibitem [{\citenamefont {Moffatt}(1978)}]{1978magnetic}%
  \BibitemOpen
  \bibfield  {author} {\bibinfo {author} {\bibfnamefont {K.}~\bibnamefont
  {Moffatt}},\ }\href {https://books.google.co.uk/books?id=cAo4AAAAIAAJ} {\emph
  {\bibinfo {title} {Magnetic Field Generation in Electrically Conducting
  Fluids}}},\ Cambridge Monographs on Mechanics\ (\bibinfo  {publisher}
  {Cambridge University Press},\ \bibinfo {year} {1978})\BibitemShut {NoStop}%
\bibitem [{\citenamefont {{Kida}}\ and\ \citenamefont
  {{Takaoka}}(1994)}]{1994AnRFM..26..169K}%
  \BibitemOpen
  \bibfield  {author} {\bibinfo {author} {\bibfnamefont {S.}~\bibnamefont
  {{Kida}}}\ and\ \bibinfo {author} {\bibfnamefont {M.}~\bibnamefont
  {{Takaoka}}},\ }\bibfield  {title} {\enquote {\bibinfo {title} {{Vortex
  reconnection}},}\ }\href {\doibase 10.1146/annurev.fl.26.010194.001125}
  {\bibfield  {journal} {\bibinfo  {journal} {Annual Review of Fluid
  Mechanics}\ }\textbf {\bibinfo {volume} {26}},\ \bibinfo {pages} {169--189}
  (\bibinfo {year} {1994})}\BibitemShut {NoStop}%
\bibitem [{\citenamefont {{Hornig}}(2001)}]{2001LNP...571..373H}%
  \BibitemOpen
  \bibfield  {author} {\bibinfo {author} {\bibfnamefont {G.}~\bibnamefont
  {{Hornig}}},\ }\bibfield  {title} {\enquote {\bibinfo {title} {{The Geometry
  of Magnetic and Vortex Reconnection}},}\ }in\ \href@noop {} {\emph {\bibinfo
  {booktitle} {Quantized Vortex Dynamics and Superfluid Turbulence}}},\
  \bibinfo {series} {Lecture Notes in Physics, Berlin Springer Verlag}, Vol.\
  \bibinfo {volume} {571},\ \bibinfo {editor} {edited by\ \bibinfo {editor}
  {\bibfnamefont {C.~F.}\ \bibnamefont {{Barenghi}}}, \bibinfo {editor}
  {\bibfnamefont {R.~J.}\ \bibnamefont {{Donnelly}}}, \ and\ \bibinfo {editor}
  {\bibfnamefont {W.~F.}\ \bibnamefont {{Vinen}}}}\ (\bibinfo {year} {2001})\
  p.\ \bibinfo {pages} {373}\BibitemShut {NoStop}%
\bibitem [{\citenamefont {{Greene}}(1993)}]{1993PhFlB...5.2355G}%
  \BibitemOpen
  \bibfield  {author} {\bibinfo {author} {\bibfnamefont {J.~M.}\ \bibnamefont
  {{Greene}}},\ }\bibfield  {title} {\enquote {\bibinfo {title} {{Reconnection
  of vorticity lines and magnetic lines}},}\ }\href {\doibase 10.1063/1.860718}
  {\bibfield  {journal} {\bibinfo  {journal} {Physics of Fluids B}\ }\textbf
  {\bibinfo {volume} {5}},\ \bibinfo {pages} {2355--2362} (\bibinfo {year}
  {1993})}\BibitemShut {NoStop}%
\bibitem [{\citenamefont {Kida}\ and\ \citenamefont
  {Takaoka}(1991)}]{kida1991}%
  \BibitemOpen
  \bibfield  {author} {\bibinfo {author} {\bibfnamefont {Shigeo}\ \bibnamefont
  {Kida}}\ and\ \bibinfo {author} {\bibfnamefont {Masanori}\ \bibnamefont
  {Takaoka}},\ }\bibfield  {title} {\enquote {\bibinfo {title} {{Breakdown of
  Frozen Motion of Vorticity Fieldand Vorticity Reconnection}},}\ }\href@noop
  {} {\bibfield  {journal} {\bibinfo  {journal} {Journal of the Physical
  Society of Japan}\ }\textbf {\bibinfo {volume} {60}},\ \bibinfo {pages}
  {2184--2196} (\bibinfo {year} {1991})}\BibitemShut {NoStop}%
\bibitem [{\citenamefont {Takaoka}(1996)}]{takaoka1996}%
  \BibitemOpen
  \bibfield  {author} {\bibinfo {author} {\bibfnamefont {M}~\bibnamefont
  {Takaoka}},\ }\bibfield  {title} {\enquote {\bibinfo {title} {{Helicity
  generation and vorticity dynamics in helically symmetric flow}},}\
  }\href@noop {} {\bibfield  {journal} {\bibinfo  {journal} {Journal of Fluid
  Mechanics}\ }\textbf {\bibinfo {volume} {319}},\ \bibinfo {pages} {125--149}
  (\bibinfo {year} {1996})}\BibitemShut {NoStop}%
\bibitem [{\citenamefont {{Schindler}}\ \emph {et~al.}(1988)\citenamefont
  {{Schindler}}, \citenamefont {{Hesse}},\ and\ \citenamefont
  {{Birn}}}]{1988JGR....93.5547S}%
  \BibitemOpen
  \bibfield  {author} {\bibinfo {author} {\bibfnamefont {K.}~\bibnamefont
  {{Schindler}}}, \bibinfo {author} {\bibfnamefont {M.}~\bibnamefont
  {{Hesse}}}, \ and\ \bibinfo {author} {\bibfnamefont {J.}~\bibnamefont
  {{Birn}}},\ }\bibfield  {title} {\enquote {\bibinfo {title} {{General
  magnetic reconnection, parallel electric fields, and helicity}},}\ }\href
  {\doibase 10.1029/JA093iA06p05547} {\bibfield  {journal} {\bibinfo  {journal}
  {Journal of Geophysics Research}\ }\textbf {\bibinfo {volume} {93}},\
  \bibinfo {pages} {5547--5557} (\bibinfo {year} {1988})}\BibitemShut {NoStop}%
\bibitem [{\citenamefont {Hornig}\ and\ \citenamefont
  {Schindler}(1996)}]{hornig1996}%
  \BibitemOpen
  \bibfield  {author} {\bibinfo {author} {\bibfnamefont {G.}~\bibnamefont
  {Hornig}}\ and\ \bibinfo {author} {\bibfnamefont {K.}~\bibnamefont
  {Schindler}},\ }\bibfield  {title} {\enquote {\bibinfo {title} {Magnetic
  topology and the problem of its invariant definition},}\ }\href@noop {}
  {\bibfield  {journal} {\bibinfo  {journal} {Phys. Plasmas}\ }\textbf
  {\bibinfo {volume} {3}},\ \bibinfo {pages} {781--791} (\bibinfo {year}
  {1996})}\BibitemShut {NoStop}%
\bibitem [{\citenamefont {Melander}\ and\ \citenamefont
  {Hussain}(1994)}]{melander1994}%
  \BibitemOpen
  \bibfield  {author} {\bibinfo {author} {\bibfnamefont {Mogens~V}\
  \bibnamefont {Melander}}\ and\ \bibinfo {author} {\bibfnamefont {Fazle}\
  \bibnamefont {Hussain}},\ }\bibfield  {title} {\enquote {\bibinfo {title}
  {{Topological vortex dynamics in axisymmetric viscous flows}},}\ }\href@noop
  {} {\bibfield  {journal} {\bibinfo  {journal} {Journal of Fluid Mechanics}\
  }\textbf {\bibinfo {volume} {260}},\ \bibinfo {pages} {57--80} (\bibinfo
  {year} {1994})}\BibitemShut {NoStop}%
\bibitem [{\citenamefont {{Eyink}}(2015)}]{eyink2015}%
  \BibitemOpen
  \bibfield  {author} {\bibinfo {author} {\bibfnamefont {G.~L.}\ \bibnamefont
  {{Eyink}}},\ }\bibfield  {title} {\enquote {\bibinfo {title} {{Turbulent
  General Magnetic Reconnection}},}\ }\href {\doibase
  10.1088/0004-637X/807/2/137} {\bibfield  {journal} {\bibinfo  {journal}
  {Astrophys.~J.}\ }\textbf {\bibinfo {volume} {807}},\ \bibinfo {eid} {137}
  (\bibinfo {year} {2015})}\BibitemShut {NoStop}%
\bibitem [{\citenamefont {{Buntine}}\ and\ \citenamefont
  {{Pullin}}(1989)}]{1989JFM...205..263B}%
  \BibitemOpen
  \bibfield  {author} {\bibinfo {author} {\bibfnamefont {J.~D.}\ \bibnamefont
  {{Buntine}}}\ and\ \bibinfo {author} {\bibfnamefont {D.~I.}\ \bibnamefont
  {{Pullin}}},\ }\bibfield  {title} {\enquote {\bibinfo {title} {{Merger and
  cancellation of strained vortices}},}\ }\href {\doibase
  10.1017/S002211208900203X} {\bibfield  {journal} {\bibinfo  {journal}
  {Journal of Fluid Mechanics}\ }\textbf {\bibinfo {volume} {205}},\ \bibinfo
  {pages} {263--295} (\bibinfo {year} {1989})}\BibitemShut {NoStop}%
\bibitem [{\citenamefont {Kida}\ and\ \citenamefont
  {Takaoka}(1988)}]{kida1988}%
  \BibitemOpen
  \bibfield  {author} {\bibinfo {author} {\bibfnamefont {S}~\bibnamefont
  {Kida}}\ and\ \bibinfo {author} {\bibfnamefont {M}~\bibnamefont {Takaoka}},\
  }\bibfield  {title} {\enquote {\bibinfo {title} {{Reconnection of vortex
  tubes}},}\ }\href@noop {} {\bibfield  {journal} {\bibinfo  {journal} {Fluid
  Dynamics Research}\ }\textbf {\bibinfo {volume} {3}},\ \bibinfo {pages}
  {257--261} (\bibinfo {year} {1988})}\BibitemShut {NoStop}%
\bibitem [{\citenamefont {{Boratav}}\ \emph {et~al.}(1992)\citenamefont
  {{Boratav}}, \citenamefont {{Pelz}},\ and\ \citenamefont
  {{Zabusky}}}]{1992PhFlA...4..581B}%
  \BibitemOpen
  \bibfield  {author} {\bibinfo {author} {\bibfnamefont {O.~N.}\ \bibnamefont
  {{Boratav}}}, \bibinfo {author} {\bibfnamefont {R.~B.}\ \bibnamefont
  {{Pelz}}}, \ and\ \bibinfo {author} {\bibfnamefont {N.~J.}\ \bibnamefont
  {{Zabusky}}},\ }\bibfield  {title} {\enquote {\bibinfo {title} {{Reconnection
  in orthogonally interacting vortex tubes: Direct numerical simulations and
  quantifications}},}\ }\href {\doibase 10.1063/1.858329} {\bibfield  {journal}
  {\bibinfo  {journal} {Physics of Fluids A}\ }\textbf {\bibinfo {volume}
  {4}},\ \bibinfo {pages} {581--605} (\bibinfo {year} {1992})}\BibitemShut
  {NoStop}%
\bibitem [{\citenamefont {Greene}(1993)}]{greene1993}%
  \BibitemOpen
  \bibfield  {author} {\bibinfo {author} {\bibfnamefont {J.~M.}\ \bibnamefont
  {Greene}},\ }\bibfield  {title} {\enquote {\bibinfo {title} {Reconnection of
  vorticity lines and magnetic lines},}\ }\href@noop {} {\bibfield  {journal}
  {\bibinfo  {journal} {Phys. Fluids B}\ }\textbf {\bibinfo {volume} {5}},\
  \bibinfo {pages} {2355--2362} (\bibinfo {year} {1993})}\BibitemShut {NoStop}%
\bibitem [{\citenamefont {{Hesse}}\ and\ \citenamefont
  {{Schindler}}(1988)}]{1988JGR....93.5559H}%
  \BibitemOpen
  \bibfield  {author} {\bibinfo {author} {\bibfnamefont {M.}~\bibnamefont
  {{Hesse}}}\ and\ \bibinfo {author} {\bibfnamefont {K.}~\bibnamefont
  {{Schindler}}},\ }\bibfield  {title} {\enquote {\bibinfo {title} {{A
  theoretical foundation of general magnetic reconnection}},}\ }\href {\doibase
  10.1029/JA093iA06p05559} {\bibfield  {journal} {\bibinfo  {journal} {Journal
  of Geophysics Research}\ }\textbf {\bibinfo {volume} {93}},\ \bibinfo {pages}
  {5559--5567} (\bibinfo {year} {1988})}\BibitemShut {NoStop}%
\bibitem [{\citenamefont {{Hornig}}\ and\ \citenamefont
  {{Priest}}(2003)}]{hornig2003}%
  \BibitemOpen
  \bibfield  {author} {\bibinfo {author} {\bibfnamefont {G.}~\bibnamefont
  {{Hornig}}}\ and\ \bibinfo {author} {\bibfnamefont {E.}~\bibnamefont
  {{Priest}}},\ }\bibfield  {title} {\enquote {\bibinfo {title} {{Evolution of
  magnetic flux in an isolated reconnection process}},}\ }\href {\doibase
  10.1063/1.1580120} {\bibfield  {journal} {\bibinfo  {journal} {Physics of
  Plasmas}\ }\textbf {\bibinfo {volume} {10}},\ \bibinfo {pages} {2712--2721}
  (\bibinfo {year} {2003})}\BibitemShut {NoStop}%
\bibitem [{\citenamefont {{Hesse}}\ \emph {et~al.}(2005)\citenamefont
  {{Hesse}}, \citenamefont {{Forbes}},\ and\ \citenamefont
  {{Birn}}}]{2005ApJ...631.1227H}%
  \BibitemOpen
  \bibfield  {author} {\bibinfo {author} {\bibfnamefont {M.}~\bibnamefont
  {{Hesse}}}, \bibinfo {author} {\bibfnamefont {T.~G.}\ \bibnamefont
  {{Forbes}}}, \ and\ \bibinfo {author} {\bibfnamefont {J.}~\bibnamefont
  {{Birn}}},\ }\bibfield  {title} {\enquote {\bibinfo {title} {{On the Relation
  between Reconnected Magnetic Flux and Parallel Electric Fields in the Solar
  Corona}},}\ }\href {\doibase 10.1086/432677} {\bibfield  {journal} {\bibinfo
  {journal} {The Astrophysical Journal}\ }\textbf {\bibinfo {volume} {631}},\
  \bibinfo {pages} {1227--1238} (\bibinfo {year} {2005})}\BibitemShut {NoStop}%
\bibitem [{\citenamefont {Pontin}(2011)}]{pontin2011b}%
  \BibitemOpen
  \bibfield  {author} {\bibinfo {author} {\bibfnamefont {D.~I.}\ \bibnamefont
  {Pontin}},\ }\bibfield  {title} {\enquote {\bibinfo {title}
  {Three-dimensional magnetic reconnection regimes: A review},}\ }\href
  {\doibase 10.1016/j.asr.2010.12.022} {\bibfield  {journal} {\bibinfo
  {journal} {Adv.~Space Res.}\ }\textbf {\bibinfo {volume} {47}},\ \bibinfo
  {pages} {1508--1522} (\bibinfo {year} {2011})}\BibitemShut {NoStop}%
\bibitem [{\citenamefont {Pontin}\ \emph {et~al.}(2005)\citenamefont {Pontin},
  \citenamefont {Hornig},\ and\ \citenamefont {Priest}}]{pontin2005a}%
  \BibitemOpen
  \bibfield  {author} {\bibinfo {author} {\bibfnamefont {D.~I.}\ \bibnamefont
  {Pontin}}, \bibinfo {author} {\bibfnamefont {G.}~\bibnamefont {Hornig}}, \
  and\ \bibinfo {author} {\bibfnamefont {E.~R.}\ \bibnamefont {Priest}},\
  }\bibfield  {title} {\enquote {\bibinfo {title} {Kinematic reconnection at a
  magnetic null point: Fan-aligned current},}\ }\href {\doibase
  10.1080/03091920512331328071} {\bibfield  {journal} {\bibinfo  {journal}
  {Geophys. Astrophys. Fluid Dynamics}\ }\textbf {\bibinfo {volume} {99}},\
  \bibinfo {pages} {77--93} (\bibinfo {year} {2005})}\BibitemShut {NoStop}%
\bibitem [{\citenamefont {Parnell}\ \emph {et~al.}(1996)\citenamefont
  {Parnell}, \citenamefont {Smith}, \citenamefont {Neukirch},\ and\
  \citenamefont {Priest}}]{parnell1996}%
  \BibitemOpen
  \bibfield  {author} {\bibinfo {author} {\bibfnamefont {C.~E.}\ \bibnamefont
  {Parnell}}, \bibinfo {author} {\bibfnamefont {J.~M.}\ \bibnamefont {Smith}},
  \bibinfo {author} {\bibfnamefont {T.}~\bibnamefont {Neukirch}}, \ and\
  \bibinfo {author} {\bibfnamefont {E.~R.}\ \bibnamefont {Priest}},\ }\bibfield
   {title} {\enquote {\bibinfo {title} {The structure of three-dimensional
  magnetic neutral points},}\ }\href@noop {} {\bibfield  {journal} {\bibinfo
  {journal} {Phys.~Plasmas}\ }\textbf {\bibinfo {volume} {3}},\ \bibinfo
  {pages} {759--770} (\bibinfo {year} {1996})}\BibitemShut {NoStop}%
\bibitem [{\citenamefont {Frankel}(2004)}]{frankel2004geometry}%
  \BibitemOpen
  \bibfield  {author} {\bibinfo {author} {\bibfnamefont {T.}~\bibnamefont
  {Frankel}},\ }\href {https://books.google.co.uk/books?id=2iq2EGNgX24C} {\emph
  {\bibinfo {title} {The Geometry of Physics: An Introduction}}}\ (\bibinfo
  {publisher} {Cambridge University Press},\ \bibinfo {year}
  {2004})\BibitemShut {NoStop}%
\bibitem [{\citenamefont {{Bustamante}}\ and\ \citenamefont
  {{Kerr}}(2008)}]{2008PhyD..237.1912B}%
  \BibitemOpen
  \bibfield  {author} {\bibinfo {author} {\bibfnamefont {M.~D.}\ \bibnamefont
  {{Bustamante}}}\ and\ \bibinfo {author} {\bibfnamefont {R.~M.}\ \bibnamefont
  {{Kerr}}},\ }\bibfield  {title} {\enquote {\bibinfo {title} {{3D Euler about
  a 2D symmetry plane}},}\ }\href {\doibase 10.1016/j.physd.2008.02.007}
  {\bibfield  {journal} {\bibinfo  {journal} {Physica D Nonlinear Phenomena}\
  }\textbf {\bibinfo {volume} {237}},\ \bibinfo {pages} {1912--1920} (\bibinfo
  {year} {2008})},\ \Eprint {http://arxiv.org/abs/0802.3369} {arXiv:0802.3369
  [physics.flu-dyn]} \BibitemShut {NoStop}%
\bibitem [{\citenamefont {{Nordlund}}\ and\ \citenamefont
  {{Galsgaard}}(1995)}]{nordlund19953d}%
  \BibitemOpen
  \bibfield  {author} {\bibinfo {author} {\bibfnamefont {{\AA}.}~\bibnamefont
  {{Nordlund}}}\ and\ \bibinfo {author} {\bibfnamefont {K.}~\bibnamefont
  {{Galsgaard}}},\ }\href@noop {} {\emph {\bibinfo {title} {{A 3D MHD code for
  Parallel Computers}}}},\ \bibinfo {type} {Tech. Rep.}\ (\bibinfo
  {institution} {Niels Bohr Institute for Astronomy, Physics and Geophysics},\
  \bibinfo {year} {1995})\BibitemShut {NoStop}%
\bibitem [{\citenamefont {Galsgaard}\ and\ \citenamefont
  {Nordlund}(1996)}]{galsgaard1996}%
  \BibitemOpen
  \bibfield  {author} {\bibinfo {author} {\bibfnamefont {K.}~\bibnamefont
  {Galsgaard}}\ and\ \bibinfo {author} {\bibfnamefont {A.}~\bibnamefont
  {Nordlund}},\ }\bibfield  {title} {\enquote {\bibinfo {title} {{Heating and
  activity of the solar corona: 1. Boundary shearing of an initially
  homogeneous magnetic field}},}\ }\href@noop {} {\bibfield  {journal}
  {\bibinfo  {journal} {J. Geophys. Res.}\ }\textbf {\bibinfo {volume} {101}},\
  \bibinfo {pages} {13445--13460} (\bibinfo {year} {1996})}\BibitemShut
  {NoStop}%
\bibitem [{\citenamefont {Chandrasekhar}(1981)}]{opac-b1090797}%
  \BibitemOpen
  \bibfield  {author} {\bibinfo {author} {\bibfnamefont {S.}~\bibnamefont
  {Chandrasekhar}},\ }\href {http://opac.inria.fr/record=b1090797} {\emph
  {\bibinfo {title} {Hydrodynamic and hydromagnetic stability}}},\ Dover
  classics of science and mathematics\ (\bibinfo  {publisher} {Dover
  Publications, cop. 1961},\ \bibinfo {address} {New York},\ \bibinfo {year}
  {1981})\ \bibinfo {note} {publiŽ pour la 1re fois en 1961 chez Clarendon
  Press}\BibitemShut {NoStop}%
\bibitem [{\citenamefont {McGavin}(2017)}]{mcgavin2017}%
  \BibitemOpen
  \bibfield  {author} {\bibinfo {author} {\bibfnamefont {P.}~\bibnamefont
  {McGavin}},\ }\emph {\bibinfo {title} {Understanding vortex reconnection in
  complex fluid flows}},\ \href@noop {} {Ph.D. thesis},\ \bibinfo  {school}
  {University of Dundee} (\bibinfo {year} {2017})\BibitemShut {NoStop}%
\bibitem [{\citenamefont {{Weiguo}}\ \emph {et~al.}(1995)\citenamefont
  {{Weiguo}}, \citenamefont {{Changchun}},\ and\ \citenamefont
  {{Yaosong}}}]{1995AcMSn..11..209W}%
  \BibitemOpen
  \bibfield  {author} {\bibinfo {author} {\bibfnamefont {W.}~\bibnamefont
  {{Weiguo}}}, \bibinfo {author} {\bibfnamefont {S.}~\bibnamefont
  {{Changchun}}}, \ and\ \bibinfo {author} {\bibfnamefont {C.}~\bibnamefont
  {{Yaosong}}},\ }\bibfield  {title} {\enquote {\bibinfo {title} {{Numerical
  study of vortex reconnection for two anti-parallel vortex tubes}},}\ }\href
  {\doibase 10.1007/BF02487724} {\bibfield  {journal} {\bibinfo  {journal}
  {Acta Mechanica Sinica}\ }\textbf {\bibinfo {volume} {11}},\ \bibinfo {pages}
  {209--218} (\bibinfo {year} {1995})}\BibitemShut {NoStop}%
\bibitem [{\citenamefont {Moffatt}(1969)}]{moffatt1969}%
  \BibitemOpen
  \bibfield  {author} {\bibinfo {author} {\bibfnamefont {H~K}\ \bibnamefont
  {Moffatt}},\ }\bibfield  {title} {\enquote {\bibinfo {title} {{The degree of
  knottedness of tangled vortex lines}},}\ }\href@noop {} {\bibfield  {journal}
  {\bibinfo  {journal} {Journal of Fluid Mechanics}\ }\textbf {\bibinfo
  {volume} {35}},\ \bibinfo {pages} {117--129} (\bibinfo {year}
  {1969})}\BibitemShut {NoStop}%
\bibitem [{\citenamefont {Berger}(1988)}]{berger1988}%
  \BibitemOpen
  \bibfield  {author} {\bibinfo {author} {\bibfnamefont {M~A}\ \bibnamefont
  {Berger}},\ }\bibfield  {title} {\enquote {\bibinfo {title} {{An energy
  formula for nonlinear force-free magnetic fields}},}\ }\href@noop {}
  {\bibfield  {journal} {\bibinfo  {journal} {Astronomy and Astrophysics}\
  }\textbf {\bibinfo {volume} {201}},\ \bibinfo {pages} {355} (\bibinfo {year}
  {1988})}\BibitemShut {NoStop}%
\bibitem [{\citenamefont {{Betchov}}(1965)}]{1965JFM....22..471B}%
  \BibitemOpen
  \bibfield  {author} {\bibinfo {author} {\bibfnamefont {R.}~\bibnamefont
  {{Betchov}}},\ }\bibfield  {title} {\enquote {\bibinfo {title} {{On the
  curvature and torsion of an isolated vortex filament}},}\ }\href {\doibase
  10.1017/S0022112065000915} {\bibfield  {journal} {\bibinfo  {journal}
  {Journal of Fluid Mechanics}\ }\textbf {\bibinfo {volume} {22}},\ \bibinfo
  {pages} {471--479} (\bibinfo {year} {1965})}\BibitemShut {NoStop}%
\bibitem [{\citenamefont {Kida}\ \emph {et~al.}(1991)\citenamefont {Kida},
  \citenamefont {Takaoka},\ and\ \citenamefont {Hussain}}]{kida1991b}%
  \BibitemOpen
  \bibfield  {author} {\bibinfo {author} {\bibfnamefont {S}~\bibnamefont
  {Kida}}, \bibinfo {author} {\bibfnamefont {M}~\bibnamefont {Takaoka}}, \ and\
  \bibinfo {author} {\bibfnamefont {F}~\bibnamefont {Hussain}},\ }\bibfield
  {title} {\enquote {\bibinfo {title} {{Collision of two vortex rings}},}\
  }\href@noop {} {\bibfield  {journal} {\bibinfo  {journal} {Journal of Fluid
  Mechanics (ISSN 0022-1120)}\ }\textbf {\bibinfo {volume} {230}},\ \bibinfo
  {pages} {583--646} (\bibinfo {year} {1991})}\BibitemShut {NoStop}%
\bibitem [{\citenamefont {{Wyper}}\ and\ \citenamefont
  {{Hesse}}(2015)}]{2015PhPl...22d2117W}%
  \BibitemOpen
  \bibfield  {author} {\bibinfo {author} {\bibfnamefont {P.~F.}\ \bibnamefont
  {{Wyper}}}\ and\ \bibinfo {author} {\bibfnamefont {M.}~\bibnamefont
  {{Hesse}}},\ }\bibfield  {title} {\enquote {\bibinfo {title} {{Quantifying
  three dimensional reconnection in fragmented current layers}},}\ }\href
  {\doibase 10.1063/1.4918335} {\bibfield  {journal} {\bibinfo  {journal}
  {Physics of Plasmas}\ }\textbf {\bibinfo {volume} {22}},\ \bibinfo {eid}
  {042117} (\bibinfo {year} {2015})},\ \Eprint
  {http://arxiv.org/abs/1502.00654} {arXiv:1502.00654 [astro-ph.SR]}
  \BibitemShut {NoStop}%
\bibitem [{\citenamefont {{Linton}}\ \emph {et~al.}(2001)\citenamefont
  {{Linton}}, \citenamefont {{Dahlburg}},\ and\ \citenamefont
  {{Antiochos}}}]{2001ApJ...553..905L}%
  \BibitemOpen
  \bibfield  {author} {\bibinfo {author} {\bibfnamefont {M.~G.}\ \bibnamefont
  {{Linton}}}, \bibinfo {author} {\bibfnamefont {R.~B.}\ \bibnamefont
  {{Dahlburg}}}, \ and\ \bibinfo {author} {\bibfnamefont {S.~K.}\ \bibnamefont
  {{Antiochos}}},\ }\bibfield  {title} {\enquote {\bibinfo {title}
  {{Reconnection of Twisted Flux Tubes as a Function of Contact Angle}},}\
  }\href {\doibase 10.1086/320974} {\bibfield  {journal} {\bibinfo  {journal}
  {The Astrophysical Journal}\ }\textbf {\bibinfo {volume} {553}},\ \bibinfo
  {pages} {905--921} (\bibinfo {year} {2001})}\BibitemShut {NoStop}%
\bibitem [{\citenamefont {{Haynes}}\ and\ \citenamefont
  {{Parnell}}(2007)}]{haynes2007}%
  \BibitemOpen
  \bibfield  {author} {\bibinfo {author} {\bibfnamefont {A.~L.}\ \bibnamefont
  {{Haynes}}}\ and\ \bibinfo {author} {\bibfnamefont {C.~E.}\ \bibnamefont
  {{Parnell}}},\ }\bibfield  {title} {\enquote {\bibinfo {title} {{A trilinear
  method for finding null points in a three-dimensional vector space}},}\
  }\href {\doibase 10.1063/1.2756751} {\bibfield  {journal} {\bibinfo
  {journal} {Phys.~Plasmas}\ }\textbf {\bibinfo {volume} {14}},\ \bibinfo
  {pages} {082107--082107} (\bibinfo {year} {2007})}\BibitemShut {NoStop}%
\bibitem [{\citenamefont {Lau}\ and\ \citenamefont {Finn}(1990)}]{lau1990}%
  \BibitemOpen
  \bibfield  {author} {\bibinfo {author} {\bibfnamefont {Y.~T.}\ \bibnamefont
  {Lau}}\ and\ \bibinfo {author} {\bibfnamefont {J.~M.}\ \bibnamefont {Finn}},\
  }\bibfield  {title} {\enquote {\bibinfo {title} {Three dimensional kinematic
  reconnection in the presence of field nulls and closed field lines},}\
  }\href@noop {} {\bibfield  {journal} {\bibinfo  {journal} {Astrophys. J.}\
  }\textbf {\bibinfo {volume} {350}},\ \bibinfo {pages} {672--691} (\bibinfo
  {year} {1990})}\BibitemShut {NoStop}%
\bibitem [{\citenamefont {{Rosenbluth}}\ and\ \citenamefont
  {{Bussac}}(1979)}]{1979NucFu..19..489R}%
  \BibitemOpen
  \bibfield  {author} {\bibinfo {author} {\bibfnamefont {M.~N.}\ \bibnamefont
  {{Rosenbluth}}}\ and\ \bibinfo {author} {\bibfnamefont {M.~N.}\ \bibnamefont
  {{Bussac}}},\ }\bibfield  {title} {\enquote {\bibinfo {title} {{MHD stability
  of Spheromak}},}\ }\href@noop {} {\bibfield  {journal} {\bibinfo  {journal}
  {Nuclear Fusion}\ }\textbf {\bibinfo {volume} {19}},\ \bibinfo {pages}
  {489--498} (\bibinfo {year} {1979})}\BibitemShut {NoStop}%
\bibitem [{\citenamefont {Hu}\ \emph {et~al.}(2004)\citenamefont {Hu},
  \citenamefont {Bhattacharjee}, \citenamefont {Dorelli},\ and\ \citenamefont
  {Greene}}]{hu2004}%
  \BibitemOpen
  \bibfield  {author} {\bibinfo {author} {\bibfnamefont {S}~\bibnamefont {Hu}},
  \bibinfo {author} {\bibfnamefont {A}~\bibnamefont {Bhattacharjee}}, \bibinfo
  {author} {\bibfnamefont {J}~\bibnamefont {Dorelli}}, \ and\ \bibinfo {author}
  {\bibfnamefont {J~M}\ \bibnamefont {Greene}},\ }\bibfield  {title} {\enquote
  {\bibinfo {title} {{The spherical tearing mode}},}\ }\href@noop {} {\bibfield
   {journal} {\bibinfo  {journal} {Geophysical Research Letters}\ }\textbf
  {\bibinfo {volume} {31}},\ \bibinfo {pages} {19806} (\bibinfo {year}
  {2004})}\BibitemShut {NoStop}%
\bibitem [{\citenamefont {{Wyper}}\ and\ \citenamefont
  {{Pontin}}(2014)}]{wyper2014b}%
  \BibitemOpen
  \bibfield  {author} {\bibinfo {author} {\bibfnamefont {P.~F.}\ \bibnamefont
  {{Wyper}}}\ and\ \bibinfo {author} {\bibfnamefont {D.~I.}\ \bibnamefont
  {{Pontin}}},\ }\bibfield  {title} {\enquote {\bibinfo {title} {{Dynamic
  topology and flux rope evolution during non-linear tearing of 3D null point
  current sheets}},}\ }\href {\doibase 10.1063/1.4896060]} {\bibfield
  {journal} {\bibinfo  {journal} {Phys.~Plasmas}\ }\textbf {\bibinfo {volume}
  {21}},\ \bibinfo {eid} {102102} (\bibinfo {year} {2014})}\BibitemShut
  {NoStop}%
\bibitem [{\citenamefont {Zabusky}\ \emph {et~al.}(1991)\citenamefont
  {Zabusky}, \citenamefont {Boratav}, \citenamefont {Pelz}, \citenamefont
  {Gao}, \citenamefont {Silver},\ and\ \citenamefont {Cooper}}]{zabusky1991}%
  \BibitemOpen
  \bibfield  {author} {\bibinfo {author} {\bibfnamefont {N~J}\ \bibnamefont
  {Zabusky}}, \bibinfo {author} {\bibfnamefont {O~N}\ \bibnamefont {Boratav}},
  \bibinfo {author} {\bibfnamefont {R~B}\ \bibnamefont {Pelz}}, \bibinfo
  {author} {\bibfnamefont {M}~\bibnamefont {Gao}}, \bibinfo {author}
  {\bibfnamefont {D}~\bibnamefont {Silver}}, \ and\ \bibinfo {author}
  {\bibfnamefont {S~P}\ \bibnamefont {Cooper}},\ }\bibfield  {title} {\enquote
  {\bibinfo {title} {{Emergence of coherent patterns of vortex stretching
  during reconnection - A scattering paradigm}},}\ }\href@noop {} {\bibfield
  {journal} {\bibinfo  {journal} {Phys.~Rev.~Lett.}\ }\textbf {\bibinfo
  {volume} {67}},\ \bibinfo {pages} {2469--2472} (\bibinfo {year}
  {1991})}\BibitemShut {NoStop}%
\bibitem [{\citenamefont {Bhattacharjee}\ and\ \citenamefont
  {Wang}(1992)}]{bhattacharjee1992}%
  \BibitemOpen
  \bibfield  {author} {\bibinfo {author} {\bibfnamefont {A}~\bibnamefont
  {Bhattacharjee}}\ and\ \bibinfo {author} {\bibfnamefont {X}~\bibnamefont
  {Wang}},\ }\bibfield  {title} {\enquote {\bibinfo {title} {{Finite-time
  vortex singularity in a Model of Three-dimensional Euler flows}},}\
  }\href@noop {} {\bibfield  {journal} {\bibinfo  {journal} {Phys Rev. Lett.}\
  }\textbf {\bibinfo {volume} {69}},\ \bibinfo {pages} {2196--2199} (\bibinfo
  {year} {1992})}\BibitemShut {NoStop}%
\bibitem [{\citenamefont {Pelz}(2002)}]{pelz2002}%
  \BibitemOpen
  \bibfield  {author} {\bibinfo {author} {\bibfnamefont {Richard~B.}\
  \bibnamefont {Pelz}},\ }\enquote {\bibinfo {title} {Discrete groups,
  symmetric flows and hydrodynamic blowup},}\ in\ \href {\doibase
  10.1007/0-306-48420-X_34} {\emph {\bibinfo {booktitle} {Tubes, Sheets and
  Singularities in Fluid Dynamics: Proceedings of the NATO RAW held in
  Zakopane, Poland, 2--7 September 2001, Sponsored as an IUTAM Symposium by the
  International Union of Theoretical and Applied Mechanics}}},\ \bibinfo
  {editor} {edited by\ \bibinfo {editor} {\bibfnamefont {K.}~\bibnamefont
  {Bajer}}\ and\ \bibinfo {editor} {\bibfnamefont {H.~K.}\ \bibnamefont
  {Moffatt}}}\ (\bibinfo  {publisher} {Springer Netherlands},\ \bibinfo
  {address} {Dordrecht},\ \bibinfo {year} {2002})\ pp.\ \bibinfo {pages}
  {269--283}\BibitemShut {NoStop}%
\end{thebibliography}

%

\end{document}